\input gt.sty
\input epsf.sty
\input psfig.sty
\magnification1050

 \def\leaderfill{\kern0.5em\leaders\hbox to 0.5em{\hss.\hss}\hfill\kern
      0.5em}

\def\notightboxit{\relax}
\def\otightboxit{\tightboxit}

\nopagenumbers
\rightline\timestamp

\bigskip
\hrule height .3mm
\vskip.6cm
\centerline{\bigfib \bigfib Elements of Group Theory}
\medskip
\centerrule{.7cm}
\vskip1.7cm
\setbox9=\vbox{\hsize65mm {\noindent\fib F. J. 
Yndur\'ain} 
\vskip .1cm
\noindent{\addressfont Departamento de F\'{\i}sica Te\'orica, C-XI\hb
 Universidad Aut\'onoma de Madrid,\hb
 Canto Blanco,\hb
E-28049, Madrid, Spain.}\hb
{\petit e-mail: fjy@delta.ft.uam.es}}
\smallskip
\centerline{\box9}
\bigskip
\vskip2truecm
{\abstracttype{Abstract}
\noindent 1.  Generalities\hfil\break
2. Lie groups and Lie algebras\hfil\break
 3.  The unitary groups\hfil\break
 4.  Representations of the SU(n) groups (and of their algebras)\hfil\break 
 5.  The tensor method for unitary groups, and\hb the permutation group\hfil\break
 6.  Relativistic invariance. The Lorentz group\hfil\break
 7.  General representation of relativistic states\hfil\break
}
\vfill\eject
\phantom{x}
\vfill\eject
\booksection{Foreword}

\noindent
The following notes are the basis for a graduate course in the Universidad Aut\'onoma de Madrid. 
They are oriented towards the application of group theory to particle physics, 
although some of it can be used for general quantum mechanics. 
They have no pretense of mathematical rigour; but I hope no gross 
mathematical inaccuracy has got into them. 

The notes can be broadly split
 into three parts: from \sect~1 to sect~3, 
they deal with abstract mathematical concepts. Generally speaking, I
have not attempted to give 
proofs of the statements made. 
These sections I have mostly taken from some  lectures  I gave at the 
Menendez Pelayo University, in the summer of 1965.
In \sects~3 through 5, we consider specific groups, particularly
 the so-called {\sl classical groups}, which are the ones that
 have  wider application in particle physics. We then describe  practical methods to 
study their representations, which is the way that most applications of groups appear  in 
high energy physics. Finally, 
the last two sections 6 and 7 deal with properties and representations of 
the Lorentz group. It is really a shame that so many physicists, who show an astounding familiarity with 
$p$-dimensional noncommutative membranes, have only a vague idea of why the photon has two 
polarization states (although its spin is 1) or how to transform a particle 
to a moving reference frame.

There are few people with whom I have discussed about the contents of these notes, 
besides A.~Galindo in what respects the first sections, long time ago; 
but I would like to record here my gratefulness to Maria Herrero, whose enthusiasm 
decided me to give the lectures, and produce the text
 (besides providing a useful reference  for some of the matters treated in \sects~3, 4). 
\vfill\eject
\phantom{X}
\vfill\eject
{\abstracttype{CONTENTS}
\item{\fib 1. }{{\fib Generalities}\leaderfill 1}
\itemitem{1.1. }{Groups and subgroups. Homomorphisms\leaderfill 1}
\itemitem{1.2. }{Representations\leaderfill 2}
\itemitem{1.3. }{Finite groups. The permutation group. 
Cayley's theorem\leaderfill 4}
\itemitem{1.4. }{The classical groups\leaderfill 5}
\smallskip
\item{\fib 2. }{{\fib Lie groups and Lie algebras}\leaderfill 6}
\itemitem{2.1. }{Definitions\leaderfill 6}
\itemitem{2.2. }{Functions over the group; group integration;
 the regular\hb representation. Character of a representation\leaderfill 8}
\itemitem{2.3. }{Lie algebras\leaderfill 9}  
\itemitem{2.4. }{The universal covering group\leaderfill 10}
\itemitem{2.5. }{The adjoint representation. Cartan's tensor and\hb Cartan's basis\leaderfill 12}
\smallskip
\item{\fib 3. }{{\fib  The unitary groups}\leaderfill 14}
\itemitem{3.1. }{The SU(2) group and the Lie algebra ${\bf A}_1$ \leaderfill 14}
\itemitem{3.2. }{The groups SO(4) and SU(2)$\times$SU(2)\leaderfill 14}
\itemitem{3.3. }{The SU(3) group and the Lie algebra   ${\bf A}_2$\leaderfill 15}
\smallskip
\item{\fib 4. }{{\fib Representations of the SU(n) groups (and of their algebras)}\leaderfill  17} 
\itemitem{4.1. }{Representations of  ${\bf A}_1$ \leaderfill 17}
\itemitem{4.2. }{Representations of  ${\bf A}_2$ \leaderfill 18}
\itemitem{4.3. }{Products of representations. The Peter--Weyl theorem and\hb the 
Clebsch--Gordan coefficients.\hb Product of representations of SU(2)  \leaderfill 20}
\itemitem{4.4. }{Products of representations of  ${\bf A}_2$ \leaderfill 22}
\smallskip
\item{\fib 5. }{{\fib The tensor method for unitary groups, and\hb the permutation group}\leaderfill 22}
\itemitem{5.1. }{SU(n) tensors\leaderfill 22}
\itemitem{5.2. }{The tensor representations of the SU(n) group.\hb Young tableaux and patterns
\leaderfill 24}
\itemitem{5.3. }{Product of representations in terms of Young tableaux\leaderfill 27}
\itemitem{5.4. }{Product of representations in the  tensor formalism\leaderfill 29}
\itemitem{5.5. }{Representations of the permutation group\leaderfill 30}
\smallskip
\item{\fib 6. }{{\fib Relativistic invariance. The Lorentz group}\leaderfill 31}
\itemitem{6.1. }{Lorentz transformations. Normal parameters\leaderfill 31}
\itemitem{6.2. }{Minkowski space.  The full Lorentz group\leaderfill 33}
\itemitem{6.3. }{More on the Lorentz group\leaderfill 35}
\itemitem{6.4. }{Geometry of the Minkowski space\leaderfill 37}
\itemitem{6.5. }{Finite dimensional representations 
of the Lorentz group\leaderfill 40}
\itemitem{ }{i. The correspondence $\cal L\to$SL(2,C)\leaderfill 40}
\itemitem{ }{ii. Connection with the Dirac formalism\leaderfill 42}
\itemitem{ }{ii. The finite dimensional representations of 
the group SL(2,C)\leaderfill 43}
\smallskip
\item{\fib 7. }{{\fib General representation of relativistic states}\leaderfill 43}
\itemitem{7.1. }{Preliminaries\leaderfill 43}
\itemitem{7.2. }{Relativistic one-particle states: general description\leaderfill 45}
\itemitem{7.3. }{Relativistic states of  massive particles\leaderfill 48}
\itemitem{7.4. }{Massless particles\leaderfill 50}
\itemitem{7.5. }{Connection with the wave function formalism\leaderfill 53}
\itemitem{7.6. }{Two-Particle States. Separation of the Center of Mass Motion.\hb States
with Well-Defined Angular Momentum\leaderfill 56}
\item{}{References\leaderfill 59}
\itemitem{}{}
}
\vfill\eject
\phantom{x}
\vfill\eject

\brochureb{\smallsc  f. j.  yndur\'ain}{\smallsc 
elements of group theory}{1}



\booksection{\S1. Generalities}
\vskip-0.5truecm
\booksubsection{1.1. Groups and subgroups. Homomorphisms}

\noindent A set of elements, $G$, is said to form a group if there exists an 
associative operation, that we will call 
{\sl multiplication}, and an element, $e\in G$, called the identity or unity, 
with the following properties:
\item{1. }{For every $f,\;g\in G$ there exists the element $h$ in $G$  such that $fg=h$;}
\item{2. }{For all $g\in G$, $eg=ge=g$.} 
\item{3. }{For every element $g\in G$ there exists an element $g^{-1}$,
 also in $G$, called the {\sl inverse}, such that $g^{-1}g=gg^{-1}=e$.}

\noindent
In general, $fg\neq gf$. If one has $fg=gf$ for all $f,\;g\in G$, we say that the group is {\sl abelian}, 
or commutative. For abelian groups, the operation is at times called {\sl sum} 
and denoted by $f+g$.

A {\sl subgroup}, $H$ of $G$, is a subset of $G$ which is itself a group. Given a subgroup $H$ 
of $G$ we say that it is {\sl invariant} if, for every $h\in H$, and all $g\in G$, the element 
$ghg^{-1}$ is in  $H$. The element $e$ by itself, and the whole group $G$, are  invariant subgroups; 
they are called the {\sl trivial} subgroups. 
If a group has no invariant subgroup other than the trivial ones, then we say that the group is 
{\sl simple}. If a group has no {\sl abelian} invariant subgroup 
(apart from the identity) we say that the group is 
{\sl semisimple}.
\smallskip
\ejes{The $n$-dimensional Euclidean space, ${\bbbr}^n=\{{\bf v}\}$, with
$${\bf v}=\pmatrix{v_1\cr\vdots\cr v_n},$$
 the $v_i$ real numbers, is an abelian group with 
the vector law of composition: if
$${\bf u}=\pmatrix{u_1\cr\vdots\cr u_n},\quad{\bf v}=\pmatrix{v_1\cr\vdots\cr v_n}$$
then
$${\bf u}+{\bf v}=\pmatrix{u_1+v_1\cr\vdots\cr u_n+v_n}.$$
The same is true for the complex euclidean space, ${\bbbc}^n$,
 where the vector components are complex numbers.\hb 
The set ${\bbbr}_+$ of {\sl positive} real numbers is an abelian group with the operation of ordinary 
multiplication. \hb
The set ${\cal T}_n$ of translations in $\bbbr^n$ is an abelian group.\hb
The set of rotations defined by a three-dimensional vector, $\ybf{\theta}$, 
by angle $\theta=|\ybf{\theta}|$,
 around the (fixed) direction of $\ybf{\theta}$ in the sense of a corkscrew that advances with
  $\ybf{\theta}$ is an abelian group. If we do not fix the direction,
 then we get the group of three-dimensional rotations, 
which is {\sl not} abelian.}
\smallskip

Let $G$, $G'$ be groups. Let $f$ be an application of $G$ in $G'$. 
We say that it is a {\sl homomorphism} if it preserves the group operations, i.e., if
for all $a,\,b\in G$,
$$f(a)=a',\;f(b)=b'\;\hbox{implies}\; f(ab^{-1})=a'b'^{-1}.$$
If the image of $G$ is all of $G'$, and the inverse application also exists and is a 
homomorphism, we say that we have an {\sl isomorphism}.
If $G=G'$, and the image of $G$ is the whole of $G$,
 we say that the homomorphism is an {\sl automorphism}.

The various groups $G'$, $G''$,\tdots\ isomorphic to a group $G$, and the group $G$ itself, 
may be thought of as {\sl realizations} of a single abstract group, $\cal G$.

The set $K_f\subset G$ of elements such that $k\in K_f$ implies $f(k)=e'$ 
($e'$ is the unit of $G'$) is called the {\sl kernel} 
of the homomorphism.
If $K_f=G$, we say that $f$ is {\sl trivial}; if  $K_f=\{e\}$ and the image of 
$G$ is all of $G'$, then $f$ is an isomorphism. 

\teorema{$K_f$ is an invariant subgroup of $G$. Hence, if $G$ is simple, every homomorphism of 
$G$ is an isomorphism.}

If an automorphism $f$ of $G$ is induced by the formula
$$f(g)=aga^{-1},\quad\hbox{with}\;a\in G$$
we say that the automorphism is {\sl internal}; if no such $a$ exists, we say that it 
is external.
\smallskip
\eje{The application
$$\exp:\xi\in {\bbbr}^1\to\ee^{\ii \alpha\xi}\quad\hbox{$\alpha\neq0$\ fixed}$$
is a homomorphism; its kernel is $K_{\exp}=\left\{\xi:\xi=2n\pi/\alpha\right\}$, $n$ an arbitrary integer.}
\smallskip
\eje{Consider the group SL(n,C) consisting of $n\times n$ matrices, $n\geq2$, with complex elements, and 
unit determinant. The transformation $g\to g^*$, where the star means the complex conjugate, is 
an external automorphism.} 
\smallskip
\eje{Let us characterize a rotation of angle $\theta$ around the origin 
in two (real) dimensions by $R(\theta)$. The set of all $R(\theta)$ forms a group, 
that we may call SO(2). The application $D$ of SO(2) on $2\times2$ matrices
$${ D}(R(\theta))=\pmatrix{\cos\theta&\sin\theta\cr-\sin\theta&\cos\theta\cr}$$
is an isomorphism.}
\smallskip
Given two groups, $G_1$, $G_2$, we define their {\sl direct product},  $G=G_1\times G_2$
as the set of elements $(g_1,g_2)$ with $g_i\in G_i$, $g\in G$, that we will write in the form $g=g_1g_2$ 
when there is no danger of confusion, with the product law
$$gh\equiv(g_1g_2)(h_1h_2)=(g_1h_1)(g_2h_2).$$

Let  $G$ be a group with $I$ and $H$  subgroups of it,  $I$ being invariant. 
 If every element $g\in G$ may be written as 
$$g=hi,\quad h\in H,\;i\in I,$$
then we say that $G$ is the {\sl semidirect} product of $H$ and $I$, written as 
$$G=H\widetilde{\times}I.$$
\smallskip
\eje{Consider the {\sl euclidean group} in $n$ dimensions, ${\cal E}_n$,
consisting of the rotations (SO(n)) and translations ${\cal T}_n$  in 
${\bbbr}^n$. Then, ${\cal E}_n={\rm SO(n)}\widetilde{\times}{\cal T}_n$.
If $R$ is a general element in SO(n) and $\bf a$ one in ${\cal T}_n$, 
a general element $g$ in  ${\cal E}_n$ can be written as
$g=({\bf a},R)$;
it acts on an arbitrary vector  $\bf r$ in ${\bbbr}^n$ 
by
$$({\bf a},R):\;{\bf r}\to R{\bf r}+{\bf a}.$$
 The unit element is $e=({\bf 0},1)$ and the product law is
$$({\bf a},R)({\bf b},S)=({\bf a}+R{\bf b},RS). 
$$
}
\smallskip
\sports{Verify that  ${\cal T}_n$ is invariant. Evaluate the inverse of $({\bf a},R)$.}

\booksubsection{1.2.  Representations}

\noindent
A {\sl representation} $D$ of the group $G$ is a homomorphism
$$D: g\in G\to D(g)\in{\cal O}(\goth{H}),$$
where ${\cal O}(\goth{H})$ is the set of linear operators in the Hilbert space  $\goth{H}$, 
 over the complex numbers. To avoid inessential complications we will assume that,
as happens in physical applications, 
both $D$, $D^{-1}$ are bounded operators.
We
will generally write the scalar product in $\goth{H}$ as $\langle\phi|\psi\rangle$ for any pair 
$\phi,\;\psi\in \goth{H}$.

We say that $D$ is finite if the Hilbert space has finite dimension; hence, it is 
equivalent to the space ${\bbbc}^n$ and the $D(g)$ are 
equivalent to $n\times n$ complex matrices.

If we have two representations $D_1$, $D_2$ acting into the same  ${\cal O}(\goth{H})$, 
and there exists the (bounded) linear operator $S$ in  ${\cal O}(\goth{H})$ such that, 
for all $g$,
$$ D_1(g)=SD_2(g)S^{-1}$$
then we say that  $D_1$ and $D_2$ are equivalent; indeed,
 they can be deduced one from the other by the change of basis in $\goth H$ induced by $S$.

If all the $D(g)$ are unitary, $D(g)^{\dag}=D(g)^{-1}$, 
 we say that $D$ is a {\sl unitary representation}; if $D$ is an isomorphism, 
we say that $D$ is {\sl faithful}; if, 
for all $g$, $D(g)=1$, we say that $D$ is {\sl trivial}.

If the (nontrivial\fnote{The trivial subspaces are  $\goth{H}$ itself,
 and that subspace formed by just the  zero vector.}) subspace $\goth{K}$
 of $\goth{H}$ is invariant under all the 
$D(g)$, then we say that $D$ is {\sl partially reducible}. 
If also the complementary\fnote{The complementary, 
$\goth{H}\ominus\goth{K}$, is defined as the set of vectors orthogonal to $\goth K$.}
  $\goth{H}\ominus\goth{K}$ is invariant, 
we say that the representation is (fully) reducible.

As an example of a representation which is reducible, but not fully reducible, 
consider the euclidean group in two dimensions, with rotations $R(\theta)$ and translations by 
the vectors ${\bf a}=(a_1,a_2)$; we write its elements 
as $({\bf a},R(\theta))$. The group can be represented by the matrices
$$D({\bf a},R(\theta))\to\pmatrix{\ee^{\ii \theta/2}&\ee^{-\ii \theta/2}(a+\ii b)\cr
0&\ee^{-\ii \theta/2}\cr}.
$$
These leave invariant the subspace of vectors of the form $\pmatrix{\alpha\cr0}$, 
but not its orthogonal,  $\pmatrix{0\cr\beta}$.  
\smallskip
\sport{Prove that a unitary representation that is partially reducible is always fully reducible.}
\smallskip
Given two representations,   $D_1$ and $D_2$, acting on ${\cal O}(\goth{H}_1)$ 
and ${\cal O}(\goth{H}_2)$,  we can form two new representations 
$D_1\oplus D_2$ and $D_1\otimes D_2$ called, respectively, their 
{\sl direct sum} and {\sl direct product} as follows. First we define the direct sum of Hilbert spaces
$\goth{H}_1$, $\goth{H}_2$, denoted by $\goth{H}\equiv\goth{H}_1\oplus \goth{H}_2$ as the set of pairs 
$$\phi=\pmatrix{\phi_1\cr\phi_2},\quad \hbox{with}\;\phi_i\in \goth{H}_i,
$$
with the natural definitions of linear combinations and scalar products; 
e.g., $\langle\phi|\psi\rangle=\langle\phi_1|\psi_1\rangle+\langle\phi_2|\psi_2\rangle$.
We then define $D\equiv D_1\oplus D_2$, acting on $\goth H$, by
$$D(g)=\pmatrix{D_1(g)&0\cr
0&D_2(g)\cr}.
$$
Clearly, $D$ is reducible; its invariant subspaces $\goth{K}_i$ are formed by vectors of the form
$$\goth{K}_1=\left\{\pmatrix{\phi_1\cr0\cr}\right\}\quad {\rm and}\quad
\goth{K}_2=\left\{\pmatrix{0\cr\phi_2\cr}\right\}.
$$

As for the direct product, we start by defining the direct product of two Hilbert spaces,
 ${\cal O}(\goth{H}_1)$ 
and ${\cal O}(\goth{H}_2)$, assumed to be separable. Hence, they have numerable orthonormal bases, 
that we denote by $\{\epsilon^{(1)}_n\}$,  $\{\epsilon^{(2)}_n\}$ respectively. 
We now form a new Hilbert space, $\goth{H}\equiv\goth{H}_1\otimes \goth{H}_2$, 
as that generated by the basis $(\{\epsilon^{(1)}_i,\epsilon^{(2)}_j\})$,
that we will simply write 
$(\{\epsilon^{(1)}_i,\epsilon^{(2)}_j\})\to\{\epsilon^{(1)}_i\epsilon^{(2)}_j\}$.
 Its vectors are thus of the form
$$\phi=\sum_{ij}\alpha_{ij}\epsilon^{(1)}_i\epsilon^{(2)}_j$$
and the operations of linear combination and scalar product are defined in the natural manner; 
for e.g. the second, if we have 
$$\phi=\sum_{ij}\alpha_{ij}\epsilon^{(1)}_i\epsilon^{(2)}_j,\quad
\psi=\sum_{ij}\beta_{ij}\epsilon^{(1)}_i\epsilon^{(2)}_j$$ 
then
$$\langle\phi|\psi\rangle\equiv\sum_{ij}\alpha^{*}_{ij}\beta_{ij}.$$
The direct product
 $D\equiv D_1\otimes D_2$  is then defined as follows: if 
$\phi=\sum_{ij}\alpha_{ij}\epsilon^{(1)}_i\epsilon^{(2)}_j$; and if 
$D_1\epsilon^{(1)}_i=\sum_{i'}d^{(1)}_{ii'}\epsilon^{(1)}_{i'}$, 
$D_2\epsilon^{(2)}_j=\sum_{j'}d^{(2)}_{jj'}\epsilon^{(2)}_{j'}$, 
then
$$D\phi=\sum_{ij}\sum_{i'j'}\alpha_{ij}d^{(1)}_{ii'}d^{(2)}_{jj'}\epsilon^{(1)}_{i'}\epsilon^{(2)}_{j'}.
$$
\smallskip
\sports{Check that direct sum and product are commutative. 
Check that, for the finite dimensional case, direct sum and product
 agree with the ordinary direct sum and product of 
matrices. Check that the dimension of the direct sum is the sum of the dimensions, 
and the dimension of the direct product is the product of the dimensions.}
\smallskip
In the finite dimensional case, with dimensions $\mu$, $\nu$, 
 if $D_1(g)=(a_{nm})$ and  $D_2(g)=(b_{nm})$, 
then   $D\equiv D_1\otimes D_2$ is the matrix
$$D\equiv
\pmatrix{a_{11}\pmatrix{b_{11}&\cdots& b_{1\nu}\cr&\cdots&\cr b_{\nu1}&\cdots&b_{\nu\nu}}&\cdots\cr
&\cdots&\cr
\cdots&a_{\mu\mu}\pmatrix{b_{11}&\cdots& b_{1\nu}\cr&\cdots&\cr b_{\nu1}&\cdots&b_{\nu\nu}}\cr}.
$$

A  representation that cannot be split in the sum of two or more representations is called 
{\sl irreducible}. A useful criterion for reducibility is the following:
\smallskip
\noindent {\sc Lemma} ({\sc Schur}).
\hb
{\it If an operator $F$ commutes with all the representatives of a group representation,
$$[F,D(g)]=0,$$
then either the representation is reducible, or $F$ is a multiple of the identity operator.}

A second related lemma, also due to Schur, is the following:
\smallskip
\noindent{\sc Lemma}.\hb
{\it If the representations $D$, $D'$ are irreducible; and if the operator 
$A$ verifies $AD(g)=D'(g)A$, for all $g$ (if the dimensions of  $D$, $D'$ are different, 
$A$ would be a square matrix) then either  $D$, $D'$ are equivalent, or $A=0$.}

\booksubsection{1.3.  Finite groups. The permutation group. Cayley's theorem}

\noindent
If the number of elements in a group is finite, it is said to be a finite group. 
Important finite groups (that, however, we will not study here; see e.g. Lyubarskii, 1960; 
Hamermesh, 1963)
 are the crystallographic groups. Another important group 
 is the group $\piv_n$ of permutations of $n$ 
elements, called the {\sl permutation} or {\sl symmetric} group.
 It is defined as follows. Let the $n$ elements be labeled $v_i$, $i=1,\dots n$. 
Let us consider two arrays of these elements,
$$v_{i_1},\dots v_{i_n};\quad  v_{j_1},\dots v_{j_n}.$$
A permutation $P$ is the application of the first array over the second; we will denote it by
$$P\equiv P(\{v_{i_1},\dots v_{i_n}\}\to \{ v_{j_1},\dots v_{j_n}\}).$$
We will denote permutations by the letters $P$, $Q$, $R$\tdots.
We have the product law
$$P(\{v_{i_1},\dots v_{i_n}\}\to \{ v_{j_1},\dots v_{j_n}\})
Q(\{ v_{j_1},\dots v_{j_n}\}\to\{v_{k_1},\dots v_{k_n}\})=
R(\{v_{i_1},\dots v_{i_n}\}\to \{ v_{k_1},\dots v_{k_n}\}).$$
The inverse $P^{-1}$ of $P$ is given by
$$P^{-1}\equiv 
\left[P(\{v_{i_1},\dots v_{i_n}\}\to \{ v_{j_1},\dots v_{j_n}\})\right]^{-1}=
Q(\{ v_{j_1},\dots v_{j_n}\}\to \{ v_{i_1},\dots v_{i_n}\}).$$
Clearly, the permutation group is not abelian.

A {\sl transposition}, $T(v_i\leftrightarrow v_j)$ is a permutation that only changes 
$v_i$ into $v_j$, and $v_j$ into $v_i$.
Any permutation may be written as a product of transpositions. 
The quantity $\delta_P\equiv (-1)^{\nu_P}$, where $\nu_P$ is the number  of such transpositions, 
is called the {\sl parity} of $P$. Although the decomposition
 in transpositions is not unique, and hence neither is $\nu_P$,
 the parity only depends on the permutation  $P$ and not on how it was decomposed in transpositions.

The permutation group is also important because it exhausts the set of {\sl all} finite groups, 
in the following sense: 
\smallskip
\teoreman{(\sc Cayley)}{Any finite subgroup is isomorphic with a subgroup of the permutation group. 
That is to say, given a finite group $G$, there exists an $n$, and a subgroup 
$G_n$ of $\piv_n$, such that  $G_n$ is isomorphic to $G$.} 
\smallskip
For more details, see Hamermesh~(1963).

\booksubsection{1.4.  The classical groups}

\noindent
Among the more important groups are those defined in terms of matrices, often called 
{\sl classical groups}. We here describe a number of these; 
several among  them will be studied in more detail later on.
\smallskip
\item{}{GL(n,C). }{(General complex linear group). This is the group of complex
 $n\times n$ matrices with nonzero determinant.} 
\item{}{GL(n,R). }{(General real linear group). This is the group of real $n\times n$
 matrices with determinant $\neq0$.}
\item{}{O(n,C). }{(Complex orthogonal group). This is the group of complex orthogonal
 $n\times n$ matrices, i.e., such that 
if $M\in$O(n,C), then $MM^{\rm T}=1$ where $M^{\rm T}$ is the transpose of $M$.}
\item{}{O(n). }{(Orthogonal group). This is the group of real orthogonal
 $n\times n$ matrices, i.e., such that 
if $M\in{\rm O(n)}$, then $MM^{\rm T}=1$ where $M^{\rm T}$ is the transpose of $M$.}
\item{}{U(n). }{(Unitary group). The group of unitary complex $n\times n$ matrices.}
\item{}{Sp(2k). }{(Simplectic group). The group that leaves invariant the simplectic 
form in the $2k$-dimensional euclidean 
space.}
\smallskip
\sport{Which of these groups is {\sl not} simple? Find abelian invariant subgroups.}
\smallskip

The definitions of these groups are all well known and elementary except, perhaps, that of the 
simplectic group. It is the group of real transformations in the $2k$-dimensional space 
that leave invariant the skew-symmetric quadratic form $[{\bf x}{\bf y}]$ defined by
$$[{\bf x}{\bf y}]\equiv x_1y_1-x_2y_2+\cdots+x_{2k-1}y_{2k-1}-x_{2k}y_{2k}.$$

Important subgroups of these groups are those obtained requiring unit determinant; 
the corresponding matrices are called {\sl unimodular}. 
They are denoted by adding the letter S (and the calificative {\sl special}\/) to the name of the group, 
except for the first two which are called 
SL(n,C) and SL(n,R). 
Thus, SO(n) is the {\sl special orthogonal group} consisting of 
real orthogonal matrices in $n\times n$ dimensions, and with unit determinant. 
\smallskip
\sport{Prove that SO(n) coincides with the group of 
rotations in ${\bbbr}^n$.}
\smallskip
The standard text on the classical groups is that of Weyl~(1946); that of Hamermesh~(1963) 
is more oriented towards physical applications.

\vfill\eject 
\booksection{\S2. Lie groups and Lie algebras}
\vskip-0.5truecm
\booksubsection{2.1. Definitions}

\noindent
Many of the groups of interest in physics are {\sl Lie groups}.\fnote{The proof of 
the majority of result we will give on Lie groups, as well as
 a wealth of supplementary information on them, 
may be found in the classic treatise of Chevalley~(1946).}  A group 
$G$ is a Lie group, of dimension $d$ ($d$ finite) if every element 
$g\in G$ is specified by $d$ real parameters: $g\equiv g(\alpha_1,\dots,\alpha_d)$ 
in such a way that, if $\alpha_1,\dots,\alpha_d$ are the parameters of 
$g$, $\beta_1,\dots,\beta_d$ those of $h$ and $\gamma_1,\dots,\gamma_d$ 
those of $gh^{-1}$, then the $\gamma_n=\gamma_n(\alpha_1,\dots,\alpha_d;\beta_1,\dots,\beta_d)$
 are analytic functions of the 
$\alpha_i$ and $\beta_j$.
We will assume that the parameters are {\sl essential}; that is to say, 
$g(\alpha_1,\dots,\alpha_d)=h(\beta_1,\dots,\beta_d)$ only if 
$\alpha_1=\beta_1$, \tdots , $\alpha_d=\beta_d$.

For Lie groups we will narrow the definition of simple and semisimple groups 
as follows: we say that a Lie group is simple if it has
 no invariant subgroups that are also Lie groups; and 
we say that it is semisimple if it has no abelian invariant subgroups that are also Lie groups.
(However, {\sl simple} or {\sl semisimple} Lie groups may
 have invariant {\sl discrete} abelian subgroups.)
\smallskip
\eje{The ``special" groups SU(n), SL(n,C) and SL(n,R) are all simple 
{\sl as Lie groups} but, for $n=$even, 
the discrete subgroup $\{1,-1\}$ of SU(n) is invariant.}
\smallskip
\teorema{It is possible to reparametrize a Lie group in such a way that the 
parameters are {\rm normal}, that is to say, they verify 
$g(0,\dots,0)=e$ ($e$ being the unity) and, if the vectors $\ybf \alpha$ and $\ybf \beta$ are parallel, 
then 
$$g(\alpha_1,\dots,\alpha_d)h(\beta_1,\dots,\beta_d)=
f(\alpha_1+\beta_1,\dots,\alpha_d+\beta_d).
$$}
The interest of normal parameters is that one can reduce a finite
 transformation to powers of infinitesimal ones:
$$g(\ybf{\alpha})=\left[g(\ybf{\alpha}/N)\right]^N.$$
For groups whose elements are matrices (or, more generally, operators) 
this allows us to get finite group elements by exponentiation:
$$g(\ybf{\alpha})=\lim_{N\to\infty}\left[g(\ybf{\alpha}/N)\right]^N=
\exp\ybf{\alpha}{\bf L},\quad L_i\equiv
\left. \partial g(\ybf{\alpha})/\partial\alpha_i\right|_{\ybf{\alpha}=0}.
$$

Let $G$ be a Lie group, in normal coordinates. 
Let $g=g(\alpha_1,\dots,\alpha_d)$, $h=h(\beta_1,\dots,\beta_d)$ and define the {\sl Weyl commutator}
$c=g^{-1}h^{-1}gh\equiv c(\gamma_1,\dots,\gamma_d)$. Then, the quantities $C_{ik\nu}$ given by
$$ C_{ik\nu}\equiv\left.\dfrac{\partial^2\gamma_\nu(\alpha_1,\dots,\alpha_d;\beta_1,\dots,\beta_d)}
{\partial\alpha_i\partial \beta_k}\right|_{\alpha=\beta=0}
$$
are called the {\sl structure constants of the group}.

\smallskip
A fundamental theorem is the following:
\smallskip
\teorema{If the group $G$ is simple, the structure constants calculated 
for the group $G$, or for any nontrivial representation of $G$, are identical.}
\smallskip\noindent
It follows that we can evaluate the $C_{ik\nu}$ in whatever representation is convenient.
\goodbreak

We say that the Lie group $G$ is {\sl compact} if the subset of ${\bbbr}^n$ over which  the parameters
 $\alpha_1,\dots,\alpha_d$ vary when $g(\alpha_1,\dots,\alpha_d)$ ranges over the whole group is 
compact; for normal parameters, this 
essentially means that it is bounded.
 SO(n) and SU(n) are compact Lie groups; SL(n,C) and SL(n,R) are also Lie groups, 
but they are not compact.

\midinsert{
\setbox0=\vbox{\hsize9.5truecm{\epsfxsize=8.truecm\epsfbox{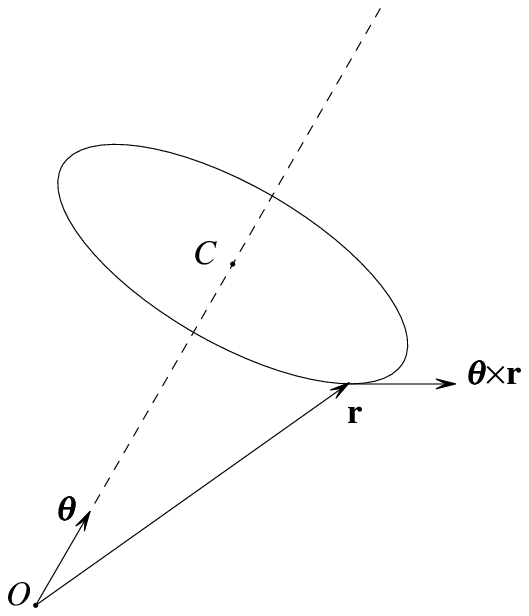}}}
\setbox1=\vbox{\hsize 6truecm{\petit The action of the rotation $R(\ybf{\theta})$.\hb
\phantom{X}\hb}}
\line{\otightboxit{\box0}\hfil\box1}
}\endinsert 

A simple and important example of Lie group is the rotation group, SO(3). 
We can parametrize the elements $R$ of SO(3) by three parameters, $\theta_i$, so that, on any vector 
$\bf r$ in three-dimensional space, $R(\ybf{\theta})$ acts as follows:
$$
{\bf r} \to {\bf r}' = R (\ybf{\theta}) {\bf r} =  (\cos \theta )
{\bf r} + (1 - \cos \theta ) \frac{\ybf{\theta} {\bf r}}{\theta ^2}
{\ybf {\theta}} + \frac{\sin \theta}{\theta} \ybf{\theta} \times
{\bf r} ; 
$$ 
see the figure. 
For $\ybf{\theta}$ infinitesimal,
$$R (\ybf{\theta}) {\bf r} = {\bf r} + \ybf{\theta} \times
{\bf r} + O (\theta ^2).
$$
A subtle point is that we must restrict $\ybf{\theta}$ to $|\ybf{\theta}|\leq2\pi$,
 and we have to identify the rotations $R(\ybf{\theta})$ for  $|\ybf{\theta}|=2\pi$ 
with the unity. 
\smallskip
\sports{Check that the matrix $R_{ij}$ is orthogonal and that $\det (R_{ij})=1$. 
Check that SO(3) is compact. {\sl Try} to draw the parameter space for 
SO(3).}
\smallskip
We finish this subsection with two important theorems:
\smallskip
\teorema{If the group $G$ is compact, then all its irreducible, finite dimensional representations, 
are equivalent to unitary representations (i.e., representations 
in which the matrices $D(g)$ are all unitary).}
\smallskip
\teorema{If the group $G$ is {\rm not} compact, then it does not have unitary finite dimensional 
representations.}

\break

\booksubsection{2.2. Functions over the group; group integration; the regular representation.\hb 
Character of a representation}

\noindent
Let $G$ be an arbitrary Lie group. We consider the space ${\cal F}(G)$ of functions, with complex values, 
and defined over the group,
$$\phi:\;g=g(\alpha_1,\dots,\alpha_d)\to \phi(g)\in {\bbbc}.
$$
Because $g$ is given by the parameters $\alpha_1,\dots,\alpha_d$, we can consider $\phi$
 as an ordinary function 
of $d$ variables, $\phi(g)=\phi(\alpha_1,\dots,\alpha_d)$.
\smallskip
\teoreman{{\sc(Haar integral)}}{ If $G$ is compact there exists a nonegative
 function $\mu(g)=\mu(\alpha_1,\dots,\alpha_d)$, unique up to normalization,
 called the {\sl Haar measure}, such that the integral 
$$\int_G\dd\mu(g) \phi(g)\equiv\int_{\{\alpha\}}\dd\mu(\alpha_1,\dots,\alpha_d)
\phi(\alpha_1,\dots,\alpha_d)$$
exists provided $\phi$ is bounded in all $G$. Moreover, $\mu$ is left and right invariant:
$\dd\mu(hg)=\dd\mu(gh)=\dd\mu(g)$.}
\noindent
If the group is not compact, but is semisimple, the result is still true 
but we have to restrict the function $\phi$ to decrease at infinity in parameter space. 
The proof of this theorem may be found in Naimark~(1956); cf. also Chevalley~(1946). 
An intuitive discussion may be seen in Wigner~(1959).

We may define a scalar product in the subset ${\cal C}(G)\subset{\cal F}(G)$ of continuous functions on 
$G$ (of fast decrease in parameter space, if the group is not compact); we write
$$\langle \phi|\psi\rangle\equiv\int_G\dd\mu(g) \phi(g)^*\psi(g).
$$
Then, ${\cal C}(G)$ can be extended to a Hilbert space, ${\rm L}^2(G)$.

For compact groups, the integral $\int_G\dd\mu(g)$ is finite. 
In this case one can, if so wished, normalize the Haar measure so that
 $\int_G\dd\mu(g)=1$ .

The Haar measure can be reduced to an ordinary integral by writing 
$$\dd\mu(\alpha_1,\dots,\alpha_d)=j(\alpha_1,\dots,\alpha_d)\dd\alpha_1\cdots\dd\alpha_d.$$
The functions $j$ can be found, for several important groups, 
in Hamermesh~(1963). 
\smallskip
\sport{Prove that, for SO(3), 
characterizing its elements as before by $R(\ybf{\theta})$,  one simply has
$\dd\mu=\dd\theta_1\dd\theta_2\dd\theta_3$.}

The notion of Haar integral can be extended to finite groups. If $G$ is a finite group 
with elements $g_i$, $i=1,\dots,n$ then the ``Haar integral" is simply the sum over all group elements:
$$\int\dd\mu\, \phi\equiv\sum_{i=1}^n\phi(g_i).$$

It is possible to construct a representation of the group $G$ over the set of functions 
 ${\rm L}^2(G)$, which is at times called the {\sl regular} representation. For an element 
$a\in G$, it is defined by
$${\rm reg}(a):\;\phi(g)\to\phi(ag).
$$
More on the important properties of the regular representation 
may 
be found in Naimark~(1959).
\smallskip
\sport{Prove that the regular representation is unitary.}
\smallskip

An important group function is what is called the {\sl character} of a (finite-dimensional) 
representation, $D(g)$. It is defined by $\chi_D(g)=\trace D(g)$.
 An important property of the character is that it is intrinsic to the representation, 
in the sense that, if $D$, $D'$ are equivalent, then  $\chi_D(g) =\chi_{D'}(g)$.
Moreover, if   $D$, $D'$ are {\sl not} equivalent, their characters are orthogonal:
$$\int\dd\mu(g)\,\chi_D^*(g) \chi_{D'}(g)=0.$$
This is a consequence of the Peter--Weyl theorem, that we will 
consider later.

The theory of characters is very important in the study of representations of {\sl finite} 
groups, in particular the permutation group or chrystalographic groups;
 see Lyubarskii~(1960) or Hamermesh~(1963).

\booksubsection{2.3. Lie algebras}

\noindent
Consider a linear space, $\bf L$, with elements $L$ that verify the following conditions:\fnote{A 
very comprehensive (and comprehensible) book on Lie algebras is Jacobson~(1962). 
In the present notes, we will only consider {\sl finite} Lie algebras, i.e.,
 such that the linear space $\cal L$ has finite dimension.}
\item{}{1. }{Any linear combination with real constants, $a L_1+bL_2$, 
$L_i\in {\bf L}$, is also in $\bf L$;}
\item{}{2. }{There exists a composition law, called the {\sl commutator}, 
$[L_1,L_2]=-[L_2,L_1]\in{\bf L}$ such that it is linear in both arguments;}
\item{}{3. }{For any three $L_i$, $i=1,\,2,\,3$ in $\bf L$ one has the 
{\sl Jacobi identity}\/
$$\sum_{\rm cyclic}[L_1,[L_2,L_3]]=0.$$}
Then we say that $\bf L$ is a {\sl Lie algebra}. 
If all commutators vanish we say that $\bf L$ is {\sl abelian}.

If $\bf H$ is a linear subspace in $\bf L$, which is in itself a Lie algebra, we say that 
it is {\sl invariant} if, for all $H\in{\bf H}$, $L\in{\bf L}$, 
the commutator $[H,L]$ belongs to $\bf H$. 
We say that $\bf L$ is {\sl simple} if it has no invariant subalgebra (except the trivial ones). 
We say that $\bf L$ is {\sl semisimple} if it has no abelian (nontrivial) invariant subalgebra.

If $\bf L$ is a Lie algebra and it has a basis $L_i$, $i=1,\dots,d$, then we can write
$$[L_i,L_j]=\sum_\nu C_{ij\nu} L_\nu.$$
The $C_{ij\nu}$ are called the {\sl structure constants} of the Lie algebra.

Given a Lie group, $G$, we can construct a corresponding Lie algebra as follows: 
consider the regular representation. 
Then the set $\bf G$ of operators $L$ of the form
$$L=\sum_i a_i\left. 
 \dfrac{\partial\,{\rm reg}(g(\alpha_1,\dots,\alpha_d))}{\partial\alpha_i}\right|_{\ybf{\alpha}=0},
\quad a_i\;{\rm real},
$$
is a Lie algebra. We say that $\bf G$ is the Lie algebra of $G$.
\smallskip
\sport{Check that the structure constants of the group $G$ are the same as those of its 
corresponding Lie algebra, $\bf G$.}

One has the following fundamental theorem:
\smallskip
\teoreman{(Lie and E. Cartan)}{To every (finite dimensional) Lie algebra $\bf L$ there corresponds 
at least a group, $G$, whose Lie algebra $\bf G$ is identical with $\bf L$, $\bf G=L$.}
\smallskip
\eje{The set ${\bf M}_n$, $n\geq2$, of real  $n\times n$ matrices $M$, with zero trace,
$\trace M=0$, is a Lie algebra. 
A basis of this algebra is formed by the matrices
$$L_{ij}=\pmatrix{0&\dots&\dots&\dots&0\cr
0&\dots&1\; (ij)&\dots&0\cr
0&\dots&\dots&\dots&0\cr
}\;{\rm for}\; i\neq j;\qquad
L_k=\pmatrix{0&\dots&\dots&\dots&\dots&0\cr
0&\dots& 1\;(k)&0\dots&\dots&0\cr
0&\dots&0&-1\;(k+1)&\dots&0\cr
0&\dots&\dots&\dots
&\dots&0}.
$$
The corresponding Lie group is SL(n,R).}
\smallskip
\sport{Evaluate the structure constants of  ${\bf M}_n$ for $n=2$ and $n=3$. 
What is the dimension of  ${\bf M}_n$?}
\smallskip
\sport{Consider the set ${\bf A}_{n-1}$ of complex   $n\times n$ matrices $A$
anti-hermitean (i.e., $A^{\dag}=-A$) and of zero trace, $\trace A=0$. 
Prove that it is a Lie algebra. Find a basis and the structure constants for 
 ${\bf A}_{n-1}$. What is the dimension of  ${\bf A}_{n-1}$?}
\smallskip
Given a Lie algebra $\bf L$, with generators $L_n$, we can form a new Lie algebra, 
over the complex numbers, that we call the {\sl complexification} of $\bf L$ 
and we denote by ${\bf L}^{\bbbc}$ (or by the same letter, $\bf L$, if there is no danger of confusion),
 by admitting linear combinations with 
{\sl complex} coefficients,
$$\sum_n\alpha_n L_n,\quad \alpha_n\in \bbbc.$$
 From any complex Lie algebra, ${\bf L}^{\bbbc}$, we can generate a new {\sl real} 
Lie algebra, 
$({\bf L}^{\bbbc})^{\bbbr}$  whose basis is formed by the set $\{L_n,\sqrt{-1}\,L_{m}\}$.
\smallskip
\sport{Prove that the complexification of ${\bf A}_{n-1}$ coincides with 
that of ${\bf M}_n$, and both with the Lie algebra of SL(n,C).}
\smallskip

The definitions of representations, direct product and direct sum for Lie algebras are similar to those
 for groups. 
Thus, a representation of $\bf L$ 
is an application into the set of operators in a Hilbert space,
$D(L)$, such that 
$$D(\alpha L+\beta L')=\alpha D(L)+\beta D(L');\quad D([L,L'])=[D(L),D(L')].$$

Likewise, we define {\sl reducible} representations of Lie algebras to 
be those that can be written as 
direct sum of nontrivial representations.

\booksubsection{2.4. The universal covering group}

\noindent
Consider two closed, oriented curves, $\ell$, $\ell'$, in a group $G$, such that 
both  $\ell$, $\ell'$ run through the identity $e$. 
We will say that $\ell$ is {\sl homotopic} to $\ell'$ if  $\ell$  can be 
continuously deformed 
into $\ell'$ ({\sl without} going out of  $G$).
 Let us define the product $\ell\ell'$ as the curve obtained joining
 $\ell$ and $\ell'$, and call a null curve to one that 
can be continuously deformed into the point $e$.
 If, moreover, we identify homotopic curves, we obtain a set $\cal P$ 
with a structure of abelian group, called the {\sl homotopy} or {\sl Poincar\'e} group.
\smallskip
\teorema{Given a Lie group, $G$, there exists a unique group $\hat{G}$, 
called the {\rm universal covering group of $G$} such that 
\item{}{i)\ \ }{ $\dim G=\dim \hat{G}$;}
\item{}{ii) }{ $\hat{G}/ {\cal P}=G$;}
\item{}{iii)}{ The Lie algebras of  $G$ and $\hat{G}$ are identical.}
\item{}{}}
\noindent If the number of elements of $\cal P$ is $N$, we say that $\hat{G}$ covers $G$ $N$ times.
\smallskip
\ejes{The homotopy group of SO(3) is isomorphic to 
the group $\{1,-1\}$ (with the ordinary multiplication law). The Lie algebra of  SO(3) is ${\bf A}_1$.
The covering group of  SO(3) is SU(2). The homotopy groups of SO(4), SO(6) 
or the (orthocronous, proper) Lorentz group, ${\cal L}\equiv{\cal L}_+^{\uparrow}$ are also 
 isomorphic to $\{1,-1\}$.
The covering group of SO(6) is SU(4). The covering group of  ${\cal L}$ is SL(2,C).} 
\smallskip
\sport{Consider the rotation group in two dimensions, SO(2), 
with elements characterized by the angle $\theta$, 
 $0\leq\theta<2\pi$. It can be mapped into the group of complex numbers 
of the form
$\ee^{\ii\theta}$. One can extend the group to include the 
rotation by $2\pi$ by identifying 
$\ee^{2\pi\ii}\equiv1$. Use this to find the homotopy group of SO(2) (it is isomorphic to the integers) 
and the covering group of   SO(2) (it is isomorphic to the set of real numbers).}
\smallskip
Because in quantum mechanics the vectors $|\phi\rangle$ and $\ee^{\ii \lambda}|\phi\rangle$  
represent the same state, covering groups play an important role there, as we will see later.

We next establish the correspondence SO(3)$\to$SU(2). 
We let $\sigma_i$ be the Pauli matrices,

$$\sigma_1=\pmatrix{0&1\cr1&0\cr},\quad\sigma_2=\pmatrix{0&-\ii\cr\ii&0\cr},\quad
\sigma_3=\pmatrix{1&0\cr0&-1\cr}.$$

\smallskip
\sport{Check that
$$\sigma_a\sigma_b=\ii\sum_c\epsilon_{abc}\sigma_c+2\delta_{ab}.$$
}
\smallskip
 To every three-vector, $\bf v$ we make correspond a
hermitean, traceless 2$\times$2 
matrix $\hat{v}$,
$$\hat{v}\equiv {\bf v}\ybf{\sigma}:\quad \hat{v}^{\dag}=\hat{v},\quad
\trace\hat{v}=0;\quad \det \hat{v}=-{\bf v}^2.$$
If $R$ is an element of SO(3) (a rotation), and ${\bf v}_R$ the image of $\bf v$
under $R$, $v_i=\sum_j R_{ij}v_j$, then the matrix 
$$\hat{v}_R\equiv {\bf v}_R\ybf{\sigma}$$
is still hermitean and traceless. It can be written as
$$\hat{v}_R=U\,\hat{v}\,U^{\dag}$$
with $U$ unitary and of unit determinant. 
In fact, the explicit form of $U$ is obtained as follows. Let $\ybf{\theta}$ be the parameters that 
determine $R$, $R=R(\ybf{\theta})$. 
Then,
$$U=\pm\exp(-\ii \ybf{\sigma}\ybf{\theta}/2).$$
The correspondence   SO(3)$\to$SU(2) is bi-valued; that of SU(2)$\to$SO(3) is single-valued.

\smallskip
\sport{Prove all this. {\rm Hint:} calculate for infinitesimal parameters $\ybf \theta$ and 
exponentiate.}
\smallskip
\sport{Calculate the $R(\ybf{\theta})$ that corresponds to a given 
 $U(\ybf{\theta})$. Hint: consider the quantity $\trace \sigma_a\hat{n}_R$, 
where 
$\bf n$ is a unitary vector along the $n$-th axis.}
\smallskip

If a Lie group is a matrix group, we may consider its Lie algebra to be a matrix 
algebra. The restriction to matrix groups is really no restriction as it can be proved that any 
Lie group has a faithful matrix representation. We have,
\smallskip
\teorema{If $G$ is a matrix Lie group, and $\bf G$ its matrix Lie algebra, 
with basis $\{L_n\}_1^d$,
then the set of elements of the form $\exp\sum_1^d \alpha_nL_n$, 
$\alpha_n$ real, generates the group $\hat{G}$.}
\noindent 
For this reason, the elements $L_n$ are also called the {\sl generators}
 of the group (or of the Lie algebra).
\smallskip
\teorema{If $\hat{G}$ is abelian, simple, semisimple then $\bf G$ 
is also abelian, simple, semisimple; and conversely.}
\smallskip
The proof of the last theorem is based on the relation, valid for small $L$, $L'$,
$$\ee^L\ee^{L'}\ee^{-L}\ee^{-L'}=[L,L']+\hbox{third order terms}.$$
$\ee^L\ee^{L'}\ee^{-L}\ee^{-L'}$ is called the {\sl Weyl commutator}.

There are two generalizations of the concept of (unitary) group representations which are important 
in physics. One are the {\sl representations up to a phase}, which are applications such that
$$D(g)D(h)=\ee^{\ii \lambda(g,h)}D(gh).$$
The other are multivalued representations, 
$$g\in G\to \ee^{\ii \lambda}D(g)$$
where the phase $\lambda$ may take several values; 
for example, one may have $g\to\pm D(g)$ as in the correspondence 
SO(3)$\to$SU(2) above.

With respect to the first, Wigner has shown that (for the groups of interest in physics) 
one can
 choose the phases of the vectors in the Hilbert spaces 
in which the $D(g)$ act so that $\phi(g,h)\equiv0$: 
that is to say, they can be reduced to ordinary representations. 
With respect to multivalued representations, one can show (see 
Chevalley~1946) that they correspond to 
{\sl single valued} representations of the covering group, $\hat{G}$. 

In the particular case of the rotation group, it follows that multiple-valued 
representations of SO(3) become single valued representations of SU(2). 
Likewise, multiple-valued representations of the Lorentz group, $\cal L$ (that we will discuss 
later) become single-valued representations of its covering group, SL(2,C).
Because SL(2,C) doubly covers $\cal L$, and SU(2) doubly covers SO(3), 
this implies that representations of  SO(3) or $\cal L$ can be at most double-valued.
Hence, in particular, spin can only be integer or half integer. 
For massive particles this follows also from the 
commutation relations of the generators of SO(3); for {\sl massless} particles, 
the proof based on the covering group is the only one known to the author.

\smallskip
\sport{From the fact that  that the covering group of the rotation group 
in two dimensions, SO(2), is isomorphic to the 
group of the real line deduce that, {\sl in two dimensions}, one  
can have any real value for the angular momentum; i.e., 
in two dimensions the angular momentum can vary continuously.}    

\booksubsection{2.5. The adjoint representation. Cartan's tensor and Cartan's basis}

\noindent
An important representation of Lie groups and Lie algebras is 
the so-called {\sl adjoint} representation. 
It represents the element $L_n$ in a Lie algebra $\bf G$ of dimension $d$ by the matrix 
${\rm ad}_{\bf G}(L_n)$ with components
$$\left({\rm ad}_{\bf G}(L_n)\right)_{ij}=C_{ijn};$$
the $C_{ijn}$ are the structure constants.
The dimension of this representation is, clearly, that of the Lie algebra, $d$.
This representation generates, by exponentiation, a representation of the 
covering group $\hat{G}$.
In turn, 
this representation induces a metric tensor $g_{ik}$, called the {Cartan tensor}
 (or also {\sl Killing form}),
as follows: 
$$g_{ik}=\trace L_iL_k=\sum_{nm}C_{nmi}C_{mnk}.$$
If $g_{ik}$ is negative-definite, we say that $\bf G$ is {\sl compact}.
\smallskip
\teoreman{(E. Cartan)}{The tensor $g_{ik}$ is non-degenerate if, and only if, $\bf G$ is semisimple.}
\smallskip
\teoreman{(H. Weyl)}{$\bf G$ is compact if, and only if, $G$ is compact.}

Given a semisimple, {\sl complex} Lie algebra, $\bf G$, consider all its abelian subalgebras 
(which cannot be invariant). Among these, that of maximum dimension,\fnote{There may 
exist several abelian subalgebras with the same maximum dimension; the results are independent of 
which one we choose as maximal abelian subalgebra.}  $\bf H$, 
is called the {\sl maximal abelian subalgebra}; if $l$ is its dimension, we also say that 
$l$ is its 
{\sl rank}. Consider now the maximal abelian subalgebra $\bf H$, and let us denote by 
$H_i$ to a basis of $\bf H$. We let the 
$E_\alpha$ be the remaining elements, obviously in ${\bf G}\ominus{\bf H}$, that complete 
a basis of $\bf G$. One has:\smallskip
\teoreman{(Killing and E. Cartan)}{There exists a basis of ${\bf G}^{\bbbc}$ 
(we will simply denote ${\bf G}^{\bbbc}$ by $\bf G$) such that all the
 ${\rm ad}_{\bf G}(H_i)$ are self-adjoint. Moreover, we can choose the 
$E_\alpha$ such that they are eigenvectors of the $H_i$,
$$[H_i,E_\alpha]=r_i(\alpha)E_\alpha;$$
for every $E_\alpha$ there exists $E_{-\alpha}$ with
$$[H_i,E_{-\alpha}]=-r_i(\alpha)E_{-\alpha}$$
and
$$[E_\alpha,E_{-\alpha}]=r^i(\alpha)H_i,
\quad r^i(\alpha)=\sum_j g_{ij}r_j(\alpha)$$
and, finally,
$$[E_\alpha,E_\beta]=n_{\alpha\beta}E_{\alpha+\beta}.$$
Here $n_{\alpha\beta}=C_{\alpha+\beta,\alpha\beta}$ if $E_{\alpha+\beta}$ exists; otherwise, 
 $n_{\alpha\beta}=0$.}
\smallskip
The $l$-dimensional vectors $\ybf{\alpha}$ with components 
$r_i(\alpha)$ are called {\sl roots} of $\bf G$. 
\smallskip
\teoreman{(Killing and E. Cartan)}{Apart from the so-called {\sl exceptional algebras},
 which we will not study here,\fnote{There are five such algbras, denoted by 
$G_2$, $F_4$, $E_6$, $E_7$ and $E_8$; the index is the rank. They may be found in Jacobson
 (1962).} the only possible compact algebras are those of the following 
table, where we also give the corresponding classical groups:
$$\matrix{
{\bf A}_l:\quad&{\rm SU(l+1)}\cr
{\bf B}_l:\quad&{\rm O(2l+1)}\cr
{\bf C}_l:\quad&{\rm Sp(2l)}\cr
{\bf D}_l:\quad&{\rm O(2l).}\cr
}
$$
}

We note that some of the lower dimensionality algebras are in fact isomorphic: 
${\bf B}_1$ and ${\bf A}_1$, ${\bf D}_2$ and  ${\bf A}_1\times{\bf A}_1$ 
and  ${\bf D}_3$ and  ${\bf A}_3$.

It is possible to give a concise characterization of all the compact Lie algebras in terms of the 
{\sl root} diagrams; we will give these in a few simple cases. 
An even more concise characterization is in terms of the so-called {\sl Dynkin diagrams}, 
which we will {\sl not} discuss here. 
We refer the reader to the text of Jacobson~(1962), where one can also find 
the proofs of many of the statements of this section, as well as the description
 of the so-called {\sl exceptional} groups (and algebras) of E.~Cartan.

\booksection{\S3. The unitary groups}

\noindent
The study of the unitary groups, SU(n), is equivalent to the study of the corresponding Lie 
algebras, ${\bf A}_{n-1}$. Because the groups SU(n) are their own covering groups, 
one can be obtained from the other by exponentiation or differentiation 
with respect to the parameters. We will in this, and the following sections, study
 in some detail the simplest groups corresponding to $n=2,\,3$, as well as their representations.
\smallskip
\sport{Prove that the automorphism $U\to U^*$ in SU(n) is external for $N\geq3$.  
Prove that it is internal for $n=2$. Hint: for the second, write 
$U=\exp\ii\ybf{\theta}\ybf{\sigma}/2$ and consider the 
transformation $U\to CUC^{-1}$ with $C=\ii\sigma_2$ ($\sigma_2$ the Pauli matrix) 
in SU(2).}

\booksubsection{3.1. The group SU(2) and the Lie algebra ${\bf A}_1$}

\noindent
By far the more important Lie groups are the unitary ones, SU(n). We will now construct explicitly 
their corresponding Lie algebras for $n=2,\;3$. 
\medskip
\noindent
${\bf A}_1$.\quad The (real) ${\bf A}_1$ algebra consists of traceless,
 antihermitean $2\times2$ matrices. 
A convenient basis for it are the $L_a=(-\ii/2)\sigma_a$, with $\sigma_a$ 
the Pauli matrices.
The commutation relations are
$$[L_a,L_b]=\sum_c\epsilon_{abc}L_c,$$
and $\epsilon_{abc}$ is the antisymmetric Levi-Civita tensor. 
Thus, the structure constants are $C_{abc}=\epsilon_{abc}$.
The adjoint representation is three-dimensional and has as basis the matrices with components
$$\left({\rm ad}(L_a)\right)_{ij}=\epsilon_{aij}.$$
The Cartan tensor is $g_{ij}=-2\delta_{ij}$.

The maximal abelian subalgebra consists of the multiples of a single generator, 
that we may take $T_3=\ii L_3$; 
we change somewhat the names and definitions to be in agreement with 
what is usual in physical applications. We will also work with the complexified algebra, 
${\bf A}_1^{\bbbc}$, that we will go on calling simply ${\bf A}_1$.
The Cartan basis of this (complex) algebra is completed with the elements
$$T_{\pm1}=\ii\left(L_1\pm\ii L_2\right),
$$
and one can easily check that
$$[T_3,T_{\pm1}]=\pm T_{\pm1},\quad [T_{+1},T_{-1}]=2 H.
$$
The root diagram of ${\bf A}_1$ is one dimensional, as shown in the figure.
\bigskip
\setbox1=\vbox{\hsize6.5truecm{\epsfxsize=5.truecm\epsfbox{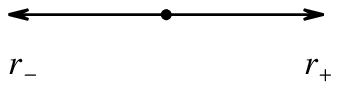}}}
\setbox2=\vbox{\hsize9 truecm{\petit The root diagram for  ${\bf A}_1$.\hb
\phantom{X}\hb}}
\line{\otightboxit{\box1}\hfil\box2}

\booksubsection{3.2. The groups SO(4) and SU(2)$\times$SU(2)}

\noindent We will here establish a correspondence between the groups SO(4) and SU(2)$\times$SU(2) 
(in fact, between the corresponding Lie algebras; we will work infinitesimally). 
For this, consider the set of matrices $\sigma_A$, $A=\,1,\,2,\,3,\,4$ with 
$\sigma_4=\ii$, and $\sigma_i$ the Pauli matrices for $i=1,\,2,\,3$.
 
For any {\sl real} four-dimensional vector, $v$ we will designate its components by 
$({\bf v},\,v_4)$. The scalar product in $\bbbr^4$ we then write as 
$$v\cdot w={\bf v}{\bf w}+v_4w_4.$$
For any vector $v$, we form the $2\times2$ matrix
$$\hat{v}=v\cdot \sigma={\bf v}\ybf{\sigma}+\ii v_4,$$
and we note that
$$\det \hat{v}=-v\cdot v.$$
We now consider the transformation
$$\hat{v}\to \hat{v'}=v'\cdot \sigma=V\hat{v} U^{\dag},\quad U,\,V\in SU(2).
\equn{(1)}$$
The set of such transformations builds the product group SU(2)$\times$SU(2). 
One can therefore write $U,\,V$ in all generality as
$$U=\ee^{-\ii\ybf{\alpha}\ybf{\sigma}},\quad V=\ee^{-\ii\ybf{\beta}\ybf{\sigma}}.$$

Eq. (1) establishes a correspondence between vectors in $\bbbr^4$,
$$v\to v'$$
which it is easy to check that it is linear and such that  $v\cdot v=v'\cdot v'$. 
It only remains to verify that $v'$ is {\sl real} to conclude that we can write 
$$v'_A=\sum_B R_{AB}v_B,\quad R\in {\rm SO(4)}.$$
We do this for infinitesimal $\ybf{\alpha},\,\ybf{\beta}$, that is to say, we take
$$U= 1-\ii \ybf{\alpha}\ybf{\sigma}+O(\alpha^2),\quad V= 1-\ii \ybf{\beta}\ybf{\sigma}+O(\beta^2);
$$
we will then neglect quadratic terms systematically.
It follows that, if  we  write
$$v'\cdot \sigma=V(v\cdot\sigma)U;\quad v'_A=\sum_B R_{AB} v_B$$
then, for infinitesimal transformations, the matrix elements $R_{AB}$ are given by
$$\eqalign{
{\bf v}'=&\,{\bf v}-(\ybf{\alpha}+\ybf{\beta})\times{\bf v}+v_4(\ybf{\alpha}-\ybf{\beta}),\cr
v'_4=&\,v_4-(\ybf{\alpha}-\ybf{\beta}){\bf v}.\cr
}
\equn{(2)}$$ 
This is clearly real, and therefore Eq.~(2) sets up the mapping 
$$(\pm V,\;\pm U)\in {\rm SU(2)}\times {\rm SU(2)}\to (R_{AB})\in{\rm SO(4)}
$$
for infinitesimal transformations.
\smallskip
\sport{Extend this to finite transformations.}

\booksubsection{3.3. The group SU(3) and the Lie algebra ${\bf A}_2$}

\noindent We now have $3\times3$ traceless, antihermitean matrices. 
For physical applications it is convenient to start with the 
 basis $L_a=-(\ii/2)\lambda_a$, $a=1,\dots,8$; 
$\lambda_a$ are the Gell-Mann matrices
$$\matrix{\lambda_j=\pmatrix{\sigma_j&0\cr0&0\cr},&
\lambda_4=\pmatrix{0&0&1\cr0&0&0\cr1&0&0\cr},&
\lambda_5=\pmatrix{0&0&-\ii\cr0&0&0\cr\ii&0&0\cr},\cr
&&\vphantom{(}&\cr
\lambda_6=\pmatrix{0&0&0\cr0&0&1\cr0&1&0\cr},&
\lambda_7=\pmatrix{0&0&0\cr0&0&-\ii\cr0&\ii&0\cr},&
\lambda_8=\tfrac{1}{\sqrt{3}}\pmatrix{1&0&0\cr0&1&0\cr0&0&-2\cr}.\cr}$$
The commutation relations are now
$$[L_a,L_b]=\sum_cf_{abc}L_c,
$$ so the structure constants are $C_{ikn}=f_{ikn},$
and only nonzero elements of the $f$, up to permutations, are as follows:
$$\eqalign{1=f_{123}=2f_{147}=2f_{246}=2f_{257}=2f_{345}\cr
=-2f_{156}=-2f_{367}=\dfrac{2}{\sqrt{3}}f_{458}=\dfrac{2}{\sqrt{3}}f_{678}.\cr}$$
For physical applications it is interesting to note that the $\lambda_a$ verify the 
anticommutation relations
$$\left\{\lambda_a,\lambda_b\right\}=2\sum d_{abc}\lambda_c+\tfrac{4}{3}\delta_{ab}$$
with the $d$ fully symmetric and all of them zero except for the following 
(and their permutations):
$$\eqalign{
\dfrac{1}{\sqrt{3}}=d_{118}=d_{228}=d_{338}=-d_{888},\quad
-\dfrac{1}{2\sqrt{3}}=d_{448}=d_{558}=d_{668}=d_{778},\cr
\tfrac{1}{2}=d_{146}=d_{157} =d_{247} =d_{256} =d_{344} =d_{355} =-d_{366} =-d_{377}. \cr}$$
\smallskip
\sport{Evaluate the Cartan tensor for SU(3).}
\smallskip

The maximal abelian subalgebra of SU(3) has now dimension 2; we may take as its basis the elements
$$T_3=\ii L_3,\quad Y=\dfrac{2}{\sqrt{3}}\ii L_8;$$
again here we use these names (instead of $H_1$, $H_2$) and definitions because they 
are the conventional ones in applications to particle physics.
With them the $T_3$, $Y$ are hermitean (instead of antihermitean).
Likewise, we will use names other than $E_{\alpha}$ for the remaining terms in a Cartan basis. 
To be precise, we define
$$T_{\pm}=\ii\left(L_1\pm\ii L_2\right);\quad U_{\pm}=\ii\left(L_6\pm\ii L_7\right);\quad
V_{\pm}=\ii\left(L_4\pm\ii L_5\right).
$$
In terms of these operators, the commutation relations are
$$\eqalign{
[T_3,Y]=&\,0,\quad [T_3,T_{\pm}]=\pm T_{\pm},\quad [T_+,T_-]=2T_3,\quad [Y,T_{\pm}]=0;\cr
[T_3,U_{\pm}]=&\,\mp\tfrac{1}{2}U_{\pm},\quad [T_3,V_{\pm}]=\pm\tfrac{1}{2}V_{\pm},\quad
 [Y,U_{\pm}]=\pm\tfrac{1}{2}U_{\pm},\quad [Y,V_{\pm}]=\pm\tfrac{1}{2}V_{\pm};\cr
[U_+,U_-]=&\,\tfrac{3}{2}Y-T_3\equiv2U_3,\quad [V_+,V_-]=\tfrac{3}{2}Y+T_3\equiv2V_3;\cr
[T_+,U_+]=&\,V_+,\quad [T_+,V_-]=-U_-,\quad [U_+,V_-]=T_-;\cr
[T_+,V_+]=&\,[T_+,U_-]=[U_+,V_+]=0.\cr}
$$
\smallskip
\sport{Prove that the three $T_\pm,\,T_3$ form the basis of a ${\bf A}_1$ subalgebra of 
${\bf A}_2$. Check that, with  the $U_3$, $V_3$ just defined, 
 the same is true for the three $U$s, $V$s.}
\smallskip
\sport{Verify that the root diagram of  ${\bf A}_2$ is as in the figure.}
\bigskip
\setbox1=\vbox{\hsize7.5truecm{\epsfxsize=6.truecm\epsfbox{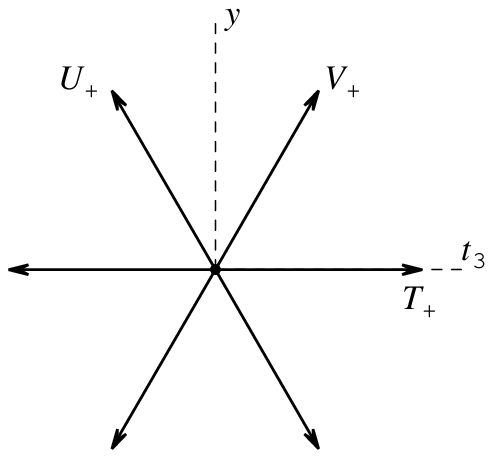}}}
\setbox2=\vbox{\hsize8 truecm{\petit The root diagram for  ${\bf A}_2$.\hb
\phantom{X}\hb}}
\line{\otightboxit{\box1}\hfil\box2}

\booksection{\S4. Representations of the SU(n) groups (and of their Lie algebras)}

\noindent Because the groups SU(n) are their own covering groups, it follows that their representations 
may be obtained from the representations of their (complex) Lie algebras,
${\bf A}_{n-1}$: a much simpler task. 
This task is further simplified because a representation of a {\sl real} Lie algebra, 
$\bf L$, can be extended to a representation of its complexification, 
${\bf L}^{\bbbc}$, 
by the simple expedient of allowing multiplication by complex numbers. 
We will use this trick systematically.

In the present section we will construct explicitly the representations of these 
Lie algebras for $l=n-1=1,\,2$; and, later on, of the groups for all $n$. 
There is a particularly important representation of the
 groups SU(n), namely that acting in a complex $n$-dimensional space in which the representatives of 
the elements in  SU(n) are the very unimodular, unitary $n\times n$ matrices in  SU(n).
It is called the {\sl fundamental} representation.
One has the important result that all the representations of  SU(n) can be generated by multiplying the 
fundamental representation by itself (Weyl,~1946).

A very understandable treatise on  representations of Lie groups, in particular of 
SU(n) and SL(n,C), is that
 of Hamermesh~(1963); for the rotation group, see Wigner~(1959).

\booksubsection{4.1.  The representations of ${\bf A}_1$}

\noindent The representations of the  ${\bf A}_1$ Lie algebra are 
well known from elementary quantum mechanics, but we will review them here because of their 
importance for more complicated cases. We work with the Cartan basis given above and look for irreducible, 
finite dimensional representations. Hence, in these representations the operators 
representing the $T_a,\;a=1,\,2,\,3$ [which we denote with the same letters, 
$D(T_a)\to T_a$] 
can be taken to be hermitean operators. Because of this, one has $T_+^{\dag}=T_-$. 
We construct an orthonormal basis of vectors $|t,t_3\rangle$ which are eigenvalues of 
$T_3$:
$$T_3|t,t_3\rangle=t_3|t,t_3\rangle;$$
the quantity $t$, that (as we will see) fully characterizes the representation is defined as the 
maximum of $t_3$; hence, there exists a state (that we assume to be {\sl unique}; see below) 
$|t,t\rangle$ with this maximum value of $t_3$.
 Because the transformation $T_3\to-T_3$ is a symmetry, it follows that, for each 
state $|t,t_3\rangle$, there exists the state $|t,-t_3\rangle$. It thus follows
 that the state with {\sl minimum} value of $t_3$ is $|t,-t\rangle$.

The commutation relations of the $T_3$, $T_\pm$ can be used to verify that the last act as 
rising/lowering operators for $t_3$. Hence the state
$$T_-^n|t,t\rangle\equiv C_{t,t-n}|t,t-n\rangle$$
is such that 
$$T_3|t,t-n\rangle=(t-n)|t,t-n\rangle.$$
The $C_{t,t-n}$ are constants introduced to make the states $|t,t-n\rangle$ 
normalized to unity; see below.
A first consequence of this is that one must necessarily have
$$T_+|t,t\rangle=T_-|t,-t\rangle=0.$$

It is easy to check that the operator $\sum_a T^2_a$ commutes with all the generators; 
hence, by virtue of the Schur Lemma, it has to be a multiple of the identity,
$\sum_a T^2_a=\lambda$. The number $\lambda$ is evaluated as follows. 
First, we note the identity
$$T_+T_-=\sum_a T^2_a-T^2_3+T_3;
\equn{(1)}$$
then we apply it to $|t,-t\rangle$. We find
$$0=T_+T_-|t,-t\rangle=\left(\sum_a T_a-T^2_3+T_3\right)|t,-t\rangle=
(\lambda-t^2-t)|t,-t\rangle$$
and hence
$$\sum_a T^2_a=t(t+1).
\equn{(2)}$$
An operator like $\sum_a T^2_a$ that commutes with all the 
generators is called a {\sl Casimir operator}.

Let us continue with the construction of the basis $|t,t_3\rangle$. 
When we apply $T_-^n$ to $|t,t\rangle$ with $n>2t$ we must find zero. Hence we have the
$2t+1$ basis vectors
$$|t,t\rangle,\;|t,t-1\rangle,\dots,\;|t,-t\rangle.$$
\smallskip
\sport{Prove that this implies that $t$ and the $t_3$ must be either integer or half-integer.}
\smallskip
We next have to find the coefficients $C_{tt_3}$. 
This is done by establishing a recursion relation as follows:
$$\eqalign{
1=&\,\langle t,t_3|t,t_3\rangle=\dfrac{1}{|C_{tt_3}|^2}\langle t,t|T^n_+T^n_-|t,t\rangle
=\dfrac{|C_{t,t_3+1}|^2}{|C_{tt_3}|^2}\langle t,t_3+1|T_+T_-|t,t_3+1\rangle\cr
=&\,\dfrac{|C_{t,t_3+1}|^2}{|C_{tt_3}|^2}\langle t,t_3+1|\Big(\sum_a T_a-T^2_3+T_3\Big)|t,t_3+1\rangle
=\dfrac{|C_{t,t_3+1}|^2}{|C_{tt_3}|^2}\left[t(t+1)-t_3(t_3+1)\right].
\cr}
$$
This implies the recursion formula
$$|C_{t,t_3+1}|=|C_{t,t_3}|/\sqrt{t(t+1)-t_3(t_3+1)}
$$
which, together with the requirement that $C_{tt}=1$ and that the $C_{tt_3}$ be positive gives all these 
coefficients. 
In particular, we find the action of the $T_\pm$ on our basis,
$$T_\pm|t,t_3\rangle=\sqrt{t(t+1)-t_3(t_3\pm1)}|t,t_3\pm1\rangle,
\equn{(3)}$$
which completely solves the problem.

\smallskip
\sport{Prove that, if there existed more than one state with maximum 
value of $t_3$, say, if one had $|t,t;\; {\rm I}\rangle$ and $|t,t;\; {\rm II}\rangle$, 
not proportional, then the representation would be reducible.}

\bigskip
\setbox1=\vbox{\hsize7.5truecm{\epsfxsize=6.truecm\epsfbox{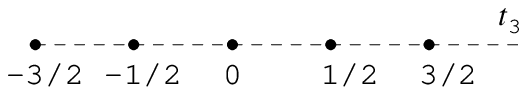}}}
\setbox2=\vbox{\hsize8.5 truecm{\petit The representation of ${\bf A}_1$ for 
$t=3/2$.\hb
\phantom{X}\hb}}
\line{\otightboxit{\box1}\hfil\box2}

\booksubsection{4.2. The representations of ${\bf A}_2$}

\noindent We have now two independent
 commuting operators, $T_3$ and $Y$. So,  we have to specify two eigenvalues, $t_3$ and $y$, 
and the diagrams for the representations of  ${\bf A}_2$ are two-dimensional.
Another thing in that the representations of SU(3) differ from 
those of SU(2) is that, if $D(g)$ is a representation of SU(3), the representation $D(g)^*$ may not 
be equivalent to it. When $D(g)^*$ is equivalent to $D(g)$, we say that the representation is {\sl real}. 
Thus, the 8-dimensional representation of SU(3) is real, but the 
3-, 6- or 10-dimensional representations are not: the representations 
$3^*$, $6^*$ or $10^*$ (with self-explanatory notation) are not equivalent to 
them.
 In the following figures we show the $t_3,\;y$ diagrams of 
the lowest dimensional representations of  ${\bf A}_2$ (the representations $6^*$, which is the 
up-down mirror image of the 6, and $10^*$, the mirror image of 10, are not shown).

\medskip
{
\setbox3=\vbox{\hsize5.2truecm{\epsfxsize=3.4truecm\epsfbox{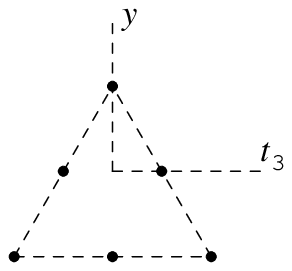}}}
\setbox5=\vbox{\hsize6.8truecm{\epsfxsize=5.4truecm\epsfbox{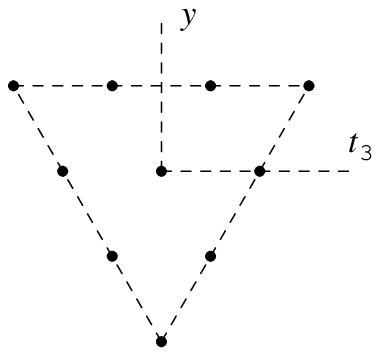}}}
\line{\qquad\qquad\notightboxit{\box3}\hfil\notightboxit{\box5}\qquad\qquad}
\bigskip
\centerline{\petit The representations 6 and 10.}
}

\topinsert{
\setbox1=\vbox{{\psfig{figure=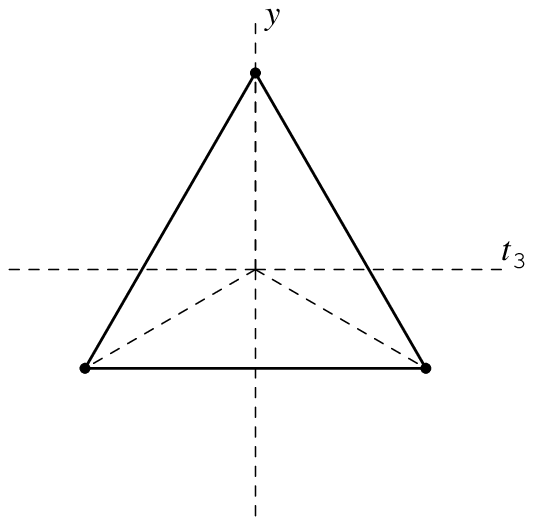,width=5truecm,angle=0}}} 
\setbox2=\vbox{{\psfig{figure=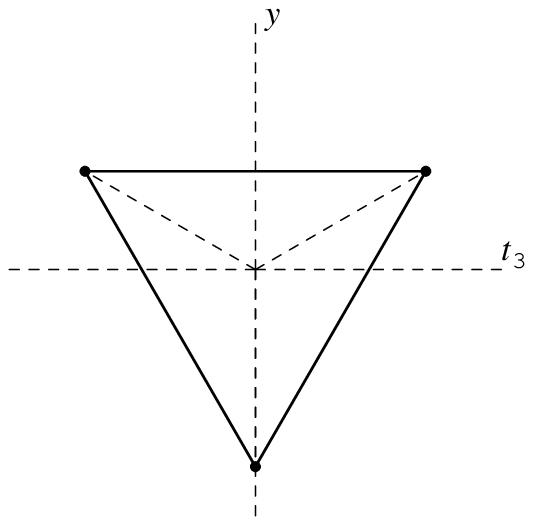,width=5truecm,angle=0}}}
\setbox4=\vbox{{\psfig{figure=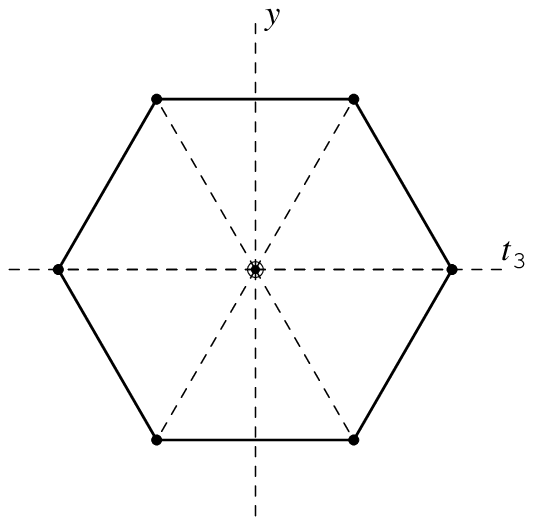,width=5truecm,angle=0}}}
\line{\notightboxit{\box1}\hfil\notightboxit{\box2}\hfil\notightboxit{\box4}}
\bigskip
\centerline{\petit The representations 3,  $3^*$ and 8.}
}\endinsert

\smallskip
\sport{Prove that the representations of 
SU(2) (that we deduced in the previous section) 
are all real. Hint: the matrix that does the trick is the representative of $\ii\sigma_2$.}
\smallskip

To describe the irreducible representations of  ${\bf A}_2$ we consider the plane $t_3\,y$ and put a dot 
for each state of said representation at the corresponding location on this plane. 
We then have a diagram that, as we shall see, fully characterizes the representation. 
On can move among the dots of the diagram with the operators\fnote{We also here 
denote with the same letters the elements of the Lie algebra and their representatives.} 
$T_\pm$, $U_\pm$ and $V_\pm$; 
in fact, using the commutation relations we can easily verify the following properties:
\item{}{$T_+$ raises $t_3$ by 1 unit, and leaves $y$ unchanged;}
\item{}{$U_+$ lowers $t_3$ by $\tfrac{1}{2}$ unit and raises $y$ by 1 unit (we note that the units of 
$y$ have a length $\sqrt{3}/2$ those of $t_3$).}
\item{}{$V_+$  raises $t_3$ by 1 unit  and raises $y$ by 1 unit.}
\itemitem{}{}

\noindent The $T_-$, $U_-$ and $V_-$ have the opposite effect. 
In view of this, it follows that by applying the  $T_\pm$, $U_\pm$ and $V_\pm$ we move in the 
diagram along lines forming angles multiple of   60\degrees, including 0\degrees. 

Another important property of the diagram of a representation is that its boundary 
forms a hexagon, in general irregular, symmetric around the $y$ axis, and 
where the length of the sides, equal to the number of states in such side minus 1,  
is given by just two integers, $p$ and $q$. Thus, the representation 8 (see figure) has 
$p=1$, $q=1$; the representations 3, 6 and 10 are degenerate hexagons, with $q=0$ and 
$p=1,\,2,\,3$ respectively.
For $p=q=0$ we have a single point, the trivial representation.

To construct all the points in a diagram , we start from the site with largest value of $t_3=t=
(p+q)/2$ 
(it can be proved that there is a single one), $|y,t\rangle$, and apply 
all operators $T_\pm$, $U_\pm$ and $V_\pm$ to $|y,t\rangle$, thereby generating the diagram.
We note that some of the points are multiple; thus, in diagrams 
3, 6, 8, 10 all points are simple, except for the central point in 8 which is double. 
We can separate the two points there by the value of the operator 
$\sum_aT_a^2$.
\smallskip
\sport{Reconstruct, from a single point with maximum $t_3$, the diagrams for the representations 
3, 6, 8, 10; $3^*$, $6^*$, $10^*$ shown in previous figures.}
\smallskip
\sport{Arrange the baryons with spin 1/2, $n,\,p,\,\Sigma$s, $\Lambda$ and $\Xi$s into 
an SU(3) octet; and the spin 3/2 resonances ($\Delta$s, etc.) into a decuplet.} 

\booksubsection{4.3. Products of representations. The Peter--Weyl theorem and 
the Clebsch--Gordan coefficients. Product of representations of SU(2)}

\noindent
Let us label the irreducible unitary representations of a compact group $G$ 
as $D^{(l)}(g)$. We then have:
\smallskip
\teoreman{(Peter--Weyl)}{The set of functions  $D^{(l)}_{ik}(g)$ forms a complete
 orthonormal basis in the 
space ${\rm L}^2(G)$ with respect to the Haar measure $\mu$, normalized to
$\int_G\dd \mu(g)=1$. That is to say, one has
$$\int_G\dd \mu(g)D^{(l)}_{ik}(g)^*D^{(l')}_{i'k'}(g)=\delta_{ll'}\delta_{ii'}\delta_{kk'}
$$
and any function $\phi(g)$ may be expanded in this basis.}
\noindent
For the proof, see Naimark~(1959) or Chevalley~(1946).
\smallskip

If we consider now the tensor product of two unitary, finite dimensional representations 
of ${\bf A}_1$,
$D^{(l_1)}\otimes D^{(l_2)}$, it will be reducible in general. 
The Peter--Weyl theorem guarantees that we can expand it as a direct sum of irreducible representations
$$D^{(l_1)}\otimes D^{(l_2)}={\bigoplus}_{l}^{ }D^{(l)}.$$
For the individual states we then find
$$|\psi^{(l_1)}\rangle\otimes|\psi^{(l_2)}\rangle=
\sum_{l,\phi^{(l)}}C(\phi^{(l)};\psi^{(l_1)},\psi^{(l_2)})\,|\phi^{(l)}\rangle.$$
The coefficients $C(\phi^{(l)};\psi^{(l_1)},\psi^{(l_2)})$ 
are called {\sl Clebsch--Gordan} coefficients and we will  
show how to calculate them in simple cases; here we start with SU(2) 
(actually, with ${\bf A}_1$). 

We consider two representations $D'$, $D''$, corresponding to the numbers $t'$, $t''$, and denote by 
$T'_a$, ${T''}_a$ to the operators that represent the Lie algebra in each of the two spaces. 
We will label the corresponding states as
$$|t',t'_3\rangle\otimes|t'',t_3''\rangle.$$
The operator $T_3$ corresponding to the product representation is obviously 
$$T_3=T'_3+ T_3''$$
hence its possible eigenvalues are $t'_3+t_3''$. It is also clear that there is only one 
state with maximum value of $T_3$, viz., $|t',t'\rangle\otimes|t'',{t''}\rangle$,
for which $t_3=t'+t''$.

Instead of considering the product $D'\otimes D''$,
 we could project it on the possible irreducible representations that it contains, 
$D^{(t)}$. We would than have a basis
$$|t,t_3\rangle.$$

By using the commutation relations one can verify the relations
$$\eqalign{
{\bf T}'{\bf T}''=&\,\Big\{t(t+1)-t'(t'+1)-t''(t''+1)\Big\}\cr
{\bf T}''{\bf T}'=&\,\Big\{t(t+1)+t'(t'+1)-t''(t''+1)\Big\}.
\cr}
\equn{(1)}$$

Let us now find the possible values of $t$, and the Clebsch--Gordan coefficients. 
First of all, we have that the maximum possible value of 
$t_3$ is $t'+t''$; hence the product $D'\otimes D''$ contains the
 representation characterized by such $t$.
Then, we start with the state
$$|t'+t'',t'+t''\rangle=|t',t'\rangle\otimes|t'',t''\rangle.$$
We then apply $T_-$ to this state. On one hand,
$${T}_-|t'+t'',t'+t''\rangle=\sqrt{t'+t''}|t'+t'',t'+t''-1\rangle,$$
and, on the other,
$$\eqalign{
{T}_-|t'+t'',t'+t''\rangle=&\,{T}'_-|t',t'\rangle\otimes|t'',t''\rangle
+|t',t'\rangle\otimes{T}''_-|t'',t''\rangle\cr
=&\,\sqrt{t'}|t',t'-1\rangle\otimes|t'',t''\rangle+\sqrt{t''}|t',t'\rangle\otimes|t'',t''-1\rangle\cr}
$$
and we have used Eq.~(3) in \sect~4.1. Equating,
$$|t'+t'',t'+t''-1\rangle=\sqrt{\dfrac{t'}{t'+t''}}|t',t'-1\rangle\otimes|t'',t''\rangle
+\sqrt{\dfrac{t''}{t'+t''}}|t',t'\rangle\otimes|t'',t''-1\rangle
\equn{(2)}$$
and, iterating the procedure, we would find all the states  
$$|t'+t'',t_3\rangle,\quad t_3=t'+t'',\,t'+t''-1,\,\dots,\,-(t'+t'').$$

The vector $|t'+t'',t'+t''-1\rangle$ is not the only one with $t_3=t'+t''-1$. 
In fact, this value of $t_3$ may be obtained adding $t'$ and $t''-1$ or $t'-1$ and $t''$:
we also have the  combination
$$|t'+t'',t'+t''-1\rangle_\perp=\sqrt{\dfrac{t''}{t'+t''}}|t',t'-1\rangle\otimes|t'',t''\rangle
-\sqrt{\dfrac{t'}{t'+t''}}|t',t'\rangle\otimes|t'',t''-1\rangle.
$$
which is orthogonal to the one above. 
[We have fixed the phases so that the corresponding  Clebsch--Gordan is real and, 
for the rest, followed the standard conventions of Condon and Shortley (1951).]

If we applied $T_+$ to this state we would get zero: which means that it corresponds to a representation 
with $t=t'+t''$: we can write above equality as
$$|t'+t'',t'+t''-1\rangle_\perp\equiv |t'+t'',t'+t''-1\rangle=
\sqrt{\dfrac{t''}{t'+t''}}|t',t'-1\rangle\otimes|t'',t''\rangle
-\sqrt{\dfrac{t'}{t'+t''}}|t',t'\rangle\otimes|t'',t''-1\rangle.$$
Applying repeatedly $T_-$ to this state, we would generate all the states
$$|t'+t''-1,t_3\rangle$$
in terms of the $|t',t'_3\rangle\otimes|t'',t''_3\rangle$.

We may then go to the states with $t_3=t'+t''-2$. They can be obtained in {\sl three} ways; two 
correspond to states already constructed. The third is obtained by taking a 
 combination orthogonal to the other two.
We can then continue the process (in which we evaluate all the  Clebsch--Gordan
coefficients) and find that
$$D'\otimes D''=\bigoplus_{t=|t'-t''|}^{t=t'+t''} D^{(t)}.
$$
The lower limit is obtained by remarking that, in the direct product basis we have 
$(2t'+1)(2t''+1)$, states while in the direct sum basis we have $\sum_{t_{\rm min}}^{t'+t''}(2t+1)$: 
 equality is only possible if $t_{\rm min}=|t'-t''|$.

Explicit expressions for the representations of SU(2) and for their 
Clebsch--Gordan coefficients may be found in Wigner~(1959); the book of  Condon and Shortley (1951)
contains a large number of properties and applications of products of 
representations of SU(2).

\booksubsection{4.4. Products of representations of ${\bf A}_2$}

\noindent
The most powerful method for multiplying (and, indeed, constructing)
 representations of the unitary groups is the tensor method; we will describe it below. 
Here we will follow a method similar to that used for SU(2). If we have two irreducible 
representations of ${\bf A}_2$, $D'$, $D''$, with diagrams $\cal D'$, $\cal D''$, 
the $t_3$ and $y$ quantum numbers\fnote{We will henceforth simplify the notation 
by using simple multiplication sign, $\times$, instead of the $\otimes$ one, for tensor products, and 
simple sum signs, $+$ instead of $\oplus$, when there is no danger of confusion.}
 of $D=D'\times D''$ must be such that they are obtained by
 adding the corresponding quantum numbers of $D'$, 
$D''$: $t_3=t'_3+t''_3$, $y=y'+y''$. 
Hence, the diagrams contained in the product representation must be contained in 
the diagram obtained by putting the center of the diagram $\cal D'$ on each of the points 
of $\cal D''$. 
The array of points so obtained may be resolved into the different diagrams for the 
irreducible representations that we have generated in a previous section. 
Thus, for example, multiplying $3\times 3^*$ one recognizes the superposition of the 
diagrams for 8 and 1; and multiplying $3\times3$ we get an array that can be resolved into the 
superposition of the diagrams for 6 and $3^*$ (see figure).

\bigskip
\setbox1=\vbox{\hsize6.truecm{\epsfxsize=4.6truecm\epsfbox{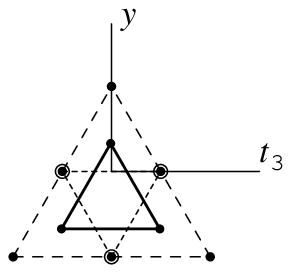}}}
\setbox2=\vbox{\hsize9 truecm{\petit $3\times3=3^*+6$.\hb
\phantom{X}\hb}}
\line{\otightboxit{\box1}\hfil\box2}
\smallskip
\sport{Verify that $3\times3\times3=1+8+8+10$. What is the result of $8\times8$?}
\smallskip
The values of the Clebsch--Gordan coefficients can be obtained as for products 
of representations of ${\bf A}_1$, starting with the state in $D'\times D''$ with largest $t_3$ 
and generating all the other states by applying the $T_\pm$, $U_\pm$, $V_\pm$.
This is a very cumbersome procedure; we will not give more details.
\smallskip
\sport{Assume that the particles in the 3 representation of SU(3) are the quarks $u,\,d,\,s$. 
Identify the mesons contained in the product  $3\times 3^*$ depending on the 
spin being 
0 or 1; consider that the quarks are in a relative S-wave.}
\smallskip
A detailed description of the representations of ${\bf A}_2$, and their 
Clebsch--Gordan coefficients, may be found in the treatise of Hamermesh~(1963)
 and, especially, in the review of de~Swart~(1963).

\booksection{\S5. The tensor method for unitary groups, and the permutation group}
\vskip-0.5truecm
\booksubsection{5.1. SU(n) tensors}

\noindent SU(n) tensors are the obvious generalization of ordinary tensors.\fnote{All 
the algebraic developments that we will 
give for SU(n) can be extended to SL(n,C) tensors in a straightforward manner. 
The tensor analysis of SL(n,C) [indeed, of GL(n,C)] may be found in 
Hamermesh~(1963).} 
A  SU(n) tensor of {\sl rank} $r$ is a set of complex numbers, with $r$ indices:
$\psi_{a_1,\dots,a_r}$, and the $a_i$ vary from 1 to $n$. 
They are assumed to transform, under unimodular unitary matrices $U$, as
$$U:\;\psi_{a_1,\dots,a_r}\to\psi_{U;a_1,\dots,a_r}\equiv\sum_{a'_1,\dots,a'_r}
U_{a_1,a'_1}\cdots U_{a_r,a'_r}\psi_{a'_1,\dots,a'_r}.
\equn{(1)}$$
We say that this is a {\sl covariant} tensor. 
If instead we had an object $\psi^{a_1,\dots,a_r}$ with the transformation law
$$U:\;\psi^{a_1,\dots,a_r}\to\psi^{U;a_1,\dots,a_r}\equiv\sum_{a'_1,\dots,a'_r}
U^*_{a_1,a'_1}\cdots U^*_{a_r,a'_r}\psi^{a'_1,\dots,a'_r}
\equn{(2)}$$
we would say that the tensor is {\sl contravariant}. We will write contravariant tensors with 
superindices. Another common notation is to put dots on contravariant indices, 
so we would have $\psi^{a_1,\dots,a_r}\equiv\psi_{\dot{a}_1,\dots,\dot{a}_r}$.
We will here use the upper indices notation.
It is also clear that tensors provide a representation of the group 
SU(n), in general reducible.

Because the $U$ are unitary, we obviously have
$$\sum_{a_1,\dots,a_r}\psi^{a_1,\dots,a_r}\psi_{a_1,\dots,a_r}=\hbox{scalar invariant}.
$$
More generally, we may define an {\sl invariant} scalar product of 
tensors $\psi$, $\phi$ with the same rank by
$$\langle\psi,\phi\rangle\equiv\sum_{a_1,\dots,a_r}\psi^*_{a_1,\dots,a_r}\phi_{a_1,\dots,a_r}.
$$
It is also easy to verify that the Levi-Civit\`a tensor in $n$ dimensions,
$\epsilon_{a_1,\dots,a_n}$ is an invariant tensor (of rank $n$).
It can also be considered a contravariant tensor, 
writing
$$\epsilon^{a_1,\dots,a_n}\equiv\epsilon_{a_1,\dots,a_n}.$$ 
It and the Kronecker delta $\delta_a^b$ (or products thereof)
 are the only invariant numerical tensors. The proof is left as 
an exercise.
\smallskip
\sport{Prove that, for any nonsingular matrix $S$,
$$\sum_{a'_1,\dots, a'_n}S_{a_1a'_1}\dots S_{a_na'_n}\epsilon_{a'_1,\dots,a'_n}=
(\det{S})\,\epsilon_{a_1,\dots,a_n}.$$
}
\smallskip

The unitarity of the $U$ can be used to prove the following result:
if $\psi_{a_1,\dots,a_r}$ is a covariant tensor of rank $r$, then
$$\psi^{a_{r+1},\dots,a_n}=\sum_{a_1,\dots,a_r}
\epsilon^{a_1,\dots,a_n}\psi_{a_1,\dots,a_r}
\equn{(3)}$$
is a contravariant tensor of rank $n-r$. 
\goodbreak

We could also construct mixed tensors (the Kronecker delta is one example) 
with $r$ subindices and $s$ superindices, $\psi_{a_1,\dots,a_r}^{a_{r+1},\dots,a_{r+s}}$;
 but this is not more general in the sense that we can use (3) to reduce them to
e.g. covariant tensors, which are the ones that we will (mostly) consider henceforth.

An important property of the tensor representations is that the {\sl permutations} 
of the indices commute with the SU(n) transformations. This occurs because all the 
$U$ in Eq.~(1) are the same. 
We can thus classify tensors according to their symmetric properties under the permutation 
group, and this classification will be SU(n) invariant: this will allow us to
explicitly construct all the 
irreducible representations of  SU(n). 
For example, consider a tensor of rank 2, $\psi_{ab}$. 
We may split it as
$$ \psi_{ab}=\tfrac{1}{2}\left\{\psi^{\rm S}_{ab}+\psi^{\rm A}_{ab}\right\}$$
where the symmetrized (S) or antisymmetrized (A) combinations are
$$\psi^{\rm S}_{ab}=\psi_{ab}+\psi_{ba},\quad  \psi^{\rm A}_{ab}=\psi_{ab}-\psi_{ba}.$$
Both $\psi^{\rm S,A}_{ab}$ are invariant under SU(n) transformations.

Because of this, the problem of constructing and multiplying tensor representations is related 
to that of constructing the irreducible representations of the permutation group, 
which we will discuss below.

\booksubsection{5.2. The tensor representations of the SU(n) group. Young tableaux and patterns}

\noindent The classification and product of representations of the SU(n) 
groups with the tensor method uses the technique of the so-called {\sl Young tableaux}. 
This technique was first developed for the permutation group;  it may be found
applied to it in 
Hamermesh~(1963). 
Here we will develop it directly for 
representations of SU(n). The results found are valid {\sl tels quels} for 
SL(n,C).

Let us consider a tensor $\psi_{i_1,\dots,i_r}$, where some of the indices may  be repeated, 
and we assume that there are $n$ different indices. This is what we would have if 
 $\psi_{i_1,\dots,i_r}$ was a general tensor under SU(n). 
We first define the {\sl Young frames} as arrays of $r$ equal squares (that we take of 
unit length) into rows, left justified. 
If there are $\rho$ rows and their lengths are $l_1,\dots,l_\rho$, 
then we require $l_1\geq l_2\geq\dots\geq l_\rho$.
Examples of Young frames for $r=2,\, 3$ and 4, and $n\geq4$, are shown in the figures below.
\medskip
\setbox2=\vbox{\hsize6.truecm{\epsfxsize=3.8truecm\epsfbox{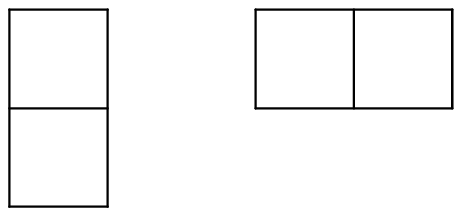}}}
\setbox3=\vbox{\hsize9.truecm{\epsfxsize=5.8truecm\epsfbox{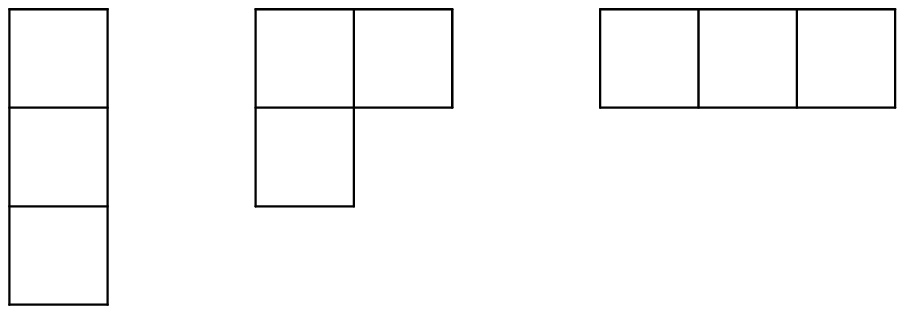}}}
\setbox4=\vbox{\hsize15.truecm{\epsfxsize=10.7truecm\epsfbox{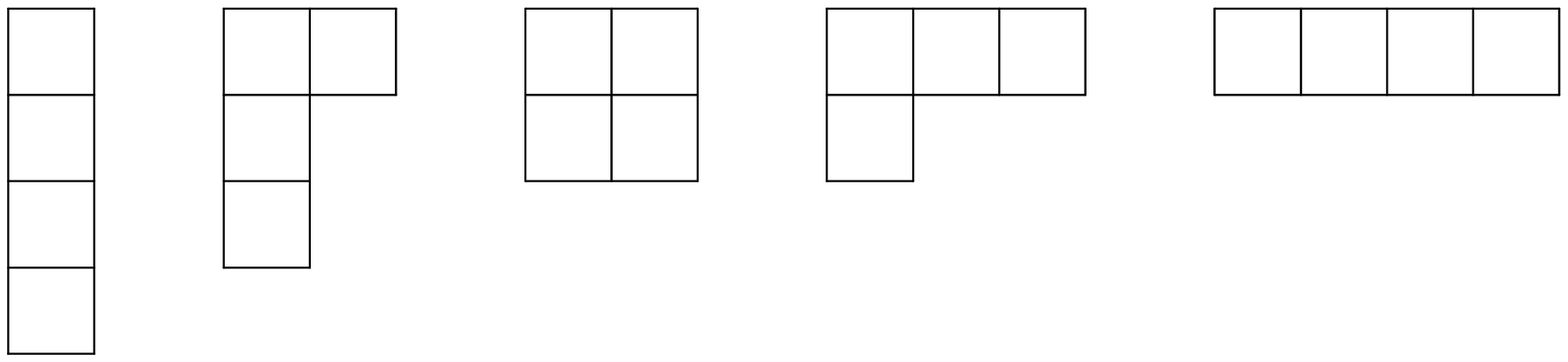}}}
\line{\otightboxit{\box2}\hfil\otightboxit{\box3}}
\medskip
\centerline{\otightboxit{\box4}}
\medskip
Once we have a Young frame, we define a Young tableau by 
putting an index among the $i_1,\dots,i_r$ into each frame.
Thus, from the frames in the second figure above we obtain the following tableaux:
\medskip
\setbox5=\vbox{\hsize10.truecm{\epsfxsize=7.7truecm\epsfbox{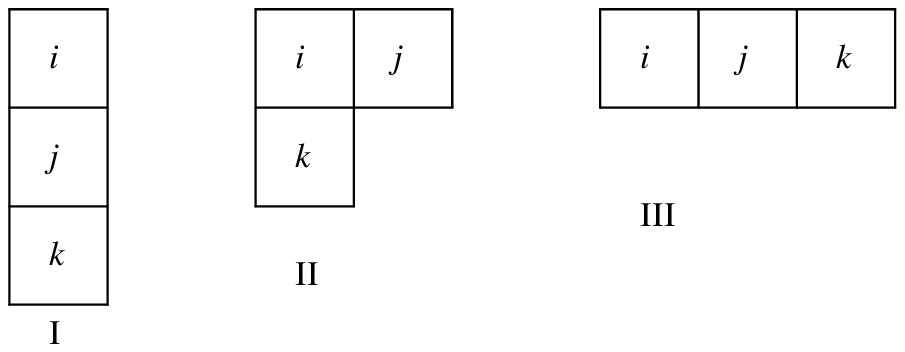}}}
\centerline{\notightboxit{\box5}\hfil}
\smallskip
\sport{Fill in the other two sets of frames to get the corresponding Young tableaux.}
\smallskip
When putting actual numbers (in lieu of the abstract indices $ijk$)
 in a Young tableau, we have a number of 
possibilities depending on which numbers we choose. 
We say that a tableau with actual numbers is a {\sl standard tableau} if 
the value of the indices does {\sl not} decrease as we go to the right along a row, for all rows,
and it does {\sl increase} as we go downwards along a column, for all columns.

For typographical reasons, as well as for ease when 
making hand drawings, one can replace the Young frames and 
tableaux by {\sl Young patterns}, as 
follows. Instead of the boxes of a Young frame, we put an array of dots. And, instead of the 
indices inside boxes in a tableau, we merely put the 
indices instead of the dots in the corresponding array. 
Thus, the pattern corresponding to the frame

\setbox1=\vbox{\psfig{figure=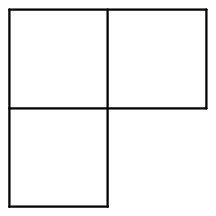,width=4truecm,angle=0}}
\centerline{\notightboxit{\box1}}

\noindent is the array
$\matrix{\bullet&\bullet\cr\bullet\cr}$
\smallskip
Likewise, to  the tableau

\setbox1=\vbox{\psfig{figure=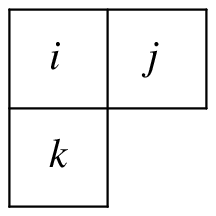,width=4truecm,angle=0}}
\centerline{\notightboxit{\box1}}

\noindent corresponds the pattern
$\matrix{i&j\cr k\cr}$\ \ .

With each Young tableau we associate the following operation on a tensor,  $\psi_{i_1,\dots,i_r}$:
\smallskip
\item{1.- }{Indices appearing in the same column 
of the tableau are {\sl antisymmetrized}. This gives a tensor, sum 
of the several tensors that are generated by  the symmetrization.}
\item{2.- }{Subsequently, in the sum just obtained, 
indices appearing in the same  row (of the tableau) are {\sl symmetrized}.}
\smallskip
Thus, from the three Young tableaux above we find the following tensors:
$$\eqalign{
{\cal Y}^{\rm I}\psi_{ijk}\equiv \psi^{\rm I}_{ijk}=&\,\psi_{ijk}-\psi_{ikj}-
\psi_{jik}+\psi_{jki}-\psi_{kij}+\psi_{kji};\cr
{\cal Y}^{\rm II}\psi_{ijk}\equiv 
\psi^{\rm II}_{ijk}=&\,\psi_{ijk}+\psi_{jik}-\psi_{kji}-\psi_{kij};\cr
{\cal Y}^{\rm III}\psi_{ijk}\equiv \psi^{\rm III}_{ijk}=&\,
\psi_{ijk}+\psi_{ikj}+\psi_{jik}+\psi_{jki}+\psi_{kij}+\psi_{kji}.
\cr}
\equn{(1)}$$
\smallskip
\sport{Show that, for A, B = I, II, III, 
$${\cal Y}^{\rm A}\left({\cal Y}^{\rm B}\psi_{ijk}\right)={\rm (Const.)}\times\delta_{\rm AB}
\left({\cal Y}^{\rm B}\psi_{ijk}\right),
$$
i.e., the operations ${\cal Y}^{\rm I}$, ${\cal Y}^{\rm II}$, ${\cal Y}^{\rm III}$ 
 are mutually {\sl orthogonal}.
Evaluate the constants above.}
\smallskip
\setbox1=\vbox{\hsize5truecm{\epsfxsize=3.5truecm\epsfbox{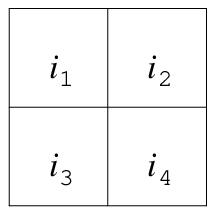}}}
\setbox2=\vbox{\hsize8 truecm{\petit The  tableau  $\cal Y$.\hb
\phantom{X}\hb
\phantom{X}\hb}}
\line{\qquad\notightboxit{\box1}\hfil\box2\qquad}

As a second example of  Young tableaux we apply the tableau of the figure above, 
that we denote by $\cal Y$, to the tensor
$\psi_{i_1i_2i_3i_4}$. 

First we antisymmetrize $i_1$, $i_3$, and $i_2$, $i_4$, {\sl and} 
 $i_1$, $i_3$ {\sl plus} $i_2$, $i_4$ getting
$$\psi_{i_1i_2i_3i_4}-\psi_{i_3i_2i_1i_4}-\psi_{i_1i_4i_3i_2}+\psi_{i_3i_4i_1i_2}.
$$
Then, we symmetrize the result in $i_1$, $i_2$, and $i_3$, $i_4$ {\sl and}
 $i_1$, $i_2$ {\sl plus} $i_3$, $i_4$. The final result is then
$$\eqalign{
{\cal Y}\psi_{i_1i_2i_3i_4}
=&\,\psi_{i_1i_2i_3i_4}-\psi_{i_3i_2i_1i_4}-\psi_{i_1i_4i_3i_2}+\psi_{i_3i_4i_1i_2}\cr
+&\,\psi_{i_2i_1i_3i_4}-\psi_{i_3i_1i_2i_4}-\psi_{i_2i_4i_3i_1}+\psi_{i_3i_4i_2i_1}\cr
+&\,\psi_{i_1i_2i_4i_3}-\psi_{i_4i_2i_1i_3}-\psi_{i_1i_3i_4i_2}+\psi_{i_4i_3i_1i_2}\cr
+&\,\psi_{i_2i_1i_4i_3}-\psi_{i_4i_1i_2i_3}-\psi_{i_2i_3i_4i_1}+\psi_{i_4i_3i_2i_1}.\cr
}$$
 
\sport{Show that, if $n\geq3$, the three tensors above are irreducible under SU(n).}
\smallskip
\sport{Show that, for SU(3), the only rank four Young tableaux 
have the frames shown in the figure:
\setbox4=\vbox{\hsize12.truecm{\epsfxsize=10.truecm\epsfbox{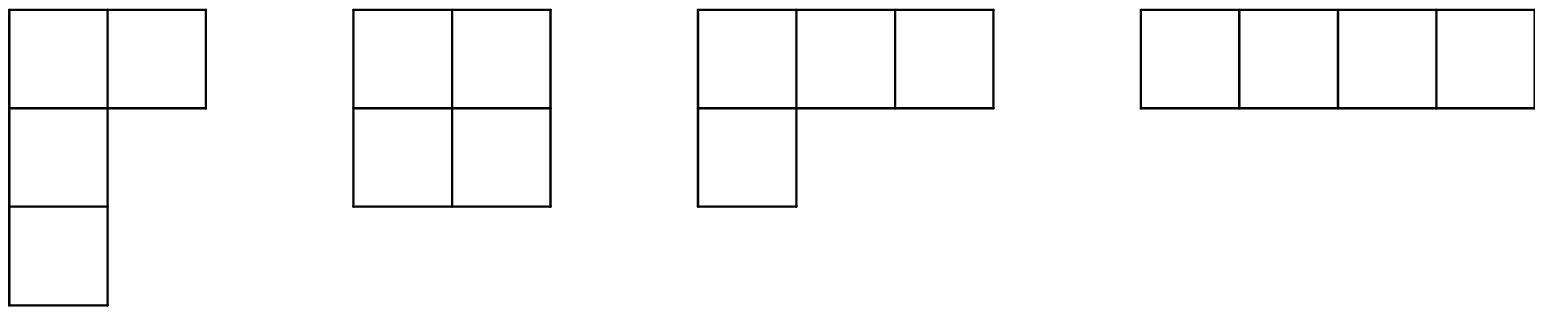}}}
\medskip
\centerline{\notightboxit{\box4}}
\smallskip
\noindent There is no vertical tableau with 4 or more rows for SU(3).}
Let us return to the example (1). When substituting actual numbers in lieu of the 
$ijk$, we need only 
do so with numbers that would lead to a standard tableau. If they formed a nonstandard 
tableau, the result would be (after appropriate symmetrization)
 either zero or a combination of the $\psi^{\rm I,II,III}$. 
We then find the following standard tableaux: for the case (I), there is only one, that of the figure.

\bigskip
\setbox1=\vbox{\hsize3.5truecm{\epsfxsize=2.truecm\epsfbox{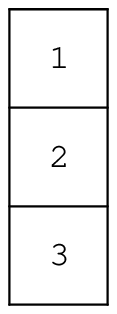}}}
\setbox2=\vbox{\hsize11 truecm{\petit The only standard tableau corresponding
 to the tensor $\psi^{\rm I}_{ijk}$.\hb
\phantom{X}\hb}}
\line{\notightboxit{\box1}\hfil\box2}
\medskip
\noindent
For the case (II), we have 8 standard tableaux, as shown below.
\bigskip
\midinsert{
\setbox0=\vbox{\hsize16.truecm{\epsfxsize=13.truecm\epsfbox{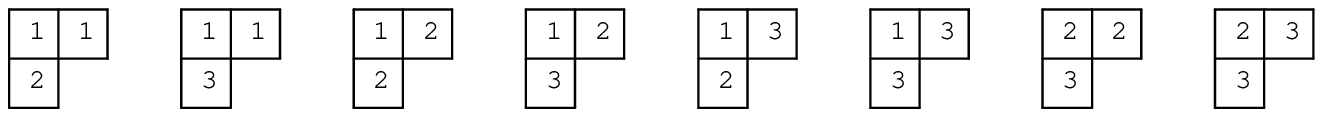}}}
\centerline{\notightboxit{\box0}}
\medskip
\centerline{\petit The eight standard tableaux corresponding to the tensor $\psi^{\rm II}_{ijk}$.}
}\endinsert
\smallskip
\sport{Construct the 10 standard tableaux corresponding to $\psi^{\rm III}_{ijk}$.}
\smallskip
In view of these results, it follows that the tensor corresponding to (I)
 has a single component, i.e., it is an invariant singlet; that corresponding to (II) has 8 
components (and thus  the tensor is a realization of the adjoint representation) 
and the tensor corresponding to (III) is a decuplet.
The (rather cumbersome) general formula for the dimension of the 
representation associated to a Young tableau may be found in Hamermesh~(1963), 
pp.~384~ff.
It is obtained by calculating how many standard tableaux exist for a given Young frame.

\booksubsection{5.3. Product of representations in terms of Young tableaux}

\noindent
Consider two representations of SU(n), corresponding to the Young tableaux 
$\cal Y$ and ${\cal Y}'$. The product of the two representations may be decomposed
 into irreducible representations, with corresponding Young tableaux ${\cal Y}^{(l)}$, 
$l=1,\,2,\dots$; we remind the reader that the product is commutative. 
We will write this symbolically as
$${\cal Y}\times{\cal Y}'={\cal Y}^{(1)}+{\cal Y}^{(2)}+\cdots
\equn{(1)}$$
We now give a procedure to find the tableaux ${\cal Y}^{(l)}$.
We do this in steps.

\item{}{Step 1. }{Label the boxes of tableau  ${\cal Y}'$ by putting the same index, 
$a$ in all the boxes in the first row; the same index, $b$, in all the boxes in the second row; 
the same index $c$ in all the boxes of the third row, etc. Note that we assume the 
tableau ${\cal Y}'$ to be standard, so we must have 
$a<b<c,\cdots$}
\item{}{Step 2. }{Glue all boxes labeled $a$ to the tableau $\cal Y$, in all possible combinations, 
in such a way that you form Young tableaux, 
but so that two identical letters do {\sl not} appear in the same column. 
In this way one finds a set of tableaux,
$${\cal Y}_{1},\;{\cal Y}_{2},\;\dots,{\cal Y}_{J_1}.
\equn{(2)}$$}
\item{}{Step 3. }{Glue the boxes labeled $b$ to the tableaux in (2),
with the same conditions as in Step~2, to get a second set of tableaux,
$$\eqalign{
{\cal Y}_{1,1},&\;{\cal Y}_{1,2},\;\dots,{\cal Y}_{1,J_2}\cr
&\cdots\quad \cdots\cr
{\cal Y}_{J_1,1},&\;{\cal Y}_{J_1,2},\;\dots,{\cal Y}_{J_1,J_2}.\cr
}
\equn{(3)}$$}
\item{}{Step 4. }{Do the same with the boxes labeled $c$, etc.} 
\item{}{Step 5. }{Once finished the process, consider each of the 
ensuing tableaux. For a given one, form the sequence of symbols $a$, $b$,\tdots\ by starting,
 {\sl from right to left}, from the upper row, then continuing along the second row, etc. 
This will give a sequence $aabcc...$. If the sequence is such that, to the left of any 
of its symbols, there are more $a$ than $b$, of $b$ than $c$, etc.,\fnote{For
 reasons that escape the present author, such a sequence is said {\sl not} to form 
a {\sl lattice permutation}; cf. Hamermesh~(1963), p.~198.}  then the tableau is to be rejected.
}
\item{}{Step 6. }{Remove the symbols $a$, $b$, $c$,\tdots\ from the remaining tableaux 
(keeping the boxes). 
These form the set 
$${\cal Y}_{1},\;{\cal Y}_{2},\;\dots,{\cal Y}_{J_1}.$$}

The whole procedure is best seen with an example. Consider the product
 of the tableau of the figure by itself.

\setbox1=\vbox{\psfig{figure=young_pat_blank.eps,width=4truecm,angle=0}}
\centerline{\notightboxit{\box1}}

According to the rules laid before, we must form the tableaux of the figure below:

\setbox1=\vbox{\psfig{figure=young_pat_blank.eps,width=4truecm,angle=0}}
\setbox2=\vbox{\psfig{figure=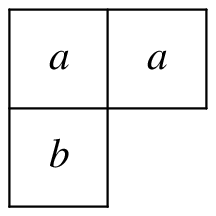,width=4truecm,angle=0}}
\line{\qquad\hfil\box1\hfil\box2\hfil\qquad}

Instead, we will 
 use the pattern representation and thus have the two following patterns:
$$\matrix{\bullet&\bullet\cr \bullet\cr}\quad\quad\quad\matrix{a&a\cr b\cr}$$
By glueing the ``boxes" with $a$ to the first pattern, we get the equivalent of (2),
$$\eqalign{
[1]:&\,\quad\matrix{\bullet&\bullet&a&a\cr\bullet&\cr}\qquad\qquad
[2]:\quad\matrix{\bullet&\bullet&a\cr\bullet&a&\cr}\cr
\phantom{c}\cr
[3]:&\,\quad\matrix{\bullet&\bullet&a\cr\bullet&\cr a&\cr}\qquad\qquad\quad\;\;
[4]:\quad\matrix{\bullet&\bullet\cr\bullet&a\cr a&\cr}
\cr}
$$
Note that the array 
$$\matrix{\bullet&\bullet\cr\bullet&\cr a&\cr a&\cr}
$$
need not be considered, as it vanishes under antisymmetrization.

We then glue the box containing $b$ to [1] in all (consistent) possible manners, 
finding
$$[1,1]:\quad\matrix{\bullet&\bullet&a&a\cr\bullet&b\cr}\qquad\qquad
[1,2]:\quad\matrix{\bullet&\bullet&a&a\cr\bullet&\cr b&\cr}
\equn{(4i)}
$$ 
Likewise, we glue the box containing $b$ to [2] and get the patterns
$$[2,1]:\quad\matrix{\bullet&\bullet&a\cr\bullet&a&b\cr}\qquad\qquad
[2,2]:\quad\matrix{\bullet&\bullet&a\cr\bullet&a\cr b&\cr}
\equn{(4ii)}$$
With [3], we have
$$[3,1]:\quad\matrix{\bullet&\bullet&a\cr\bullet&b\cr a&\cr}\qquad\qquad
[3,2]:\quad\matrix{\bullet&\bullet&a\cr\bullet&\cr a&\cr b&\cr}
\equn{(4iii)}$$
Finally, from [4],
$$[4,1]:\quad\matrix{\bullet&\bullet\cr\bullet&a\cr a&b\cr}\qquad\qquad
[4,2]:\quad\matrix{\bullet&\bullet\cr\bullet&a\cr a&\cr b&\cr}
\equn{(4iv)}$$ 
Among the patterns so obtained, there appear some that we rejected because 
they do not form a ``lattice permutation"; they are, for example, the patterns
$$\matrix{\bullet&\bullet&a&a&b\cr\bullet&\cr}\qquad\qquad
\matrix{\bullet&\bullet&a&b\cr\bullet&a&\cr}
$$
In both cases, the procedure of Step~5 gives the sequence $baa$, which has too many 
$a$s to the right of $b$.

The set of tableaux obtained by replacing the letters in Eqs.~(4) by dots gives the 
full set of tableaux that appear in the decomposition (1). 
Note that the pattern
$$\matrix{\bullet&\bullet&\bullet\cr\bullet&\bullet\cr
\bullet\cr}
$$
appears {\sl twice}, as it can be reached by two independent paths, [2,1] and [3,1]. 
This indicates that the corresponding representation will also appear twice in the 
reduction of the product.

\booksubsection{5.4. Product of representations in the tensor formalism}

\noindent
We will consider in detail the case SU(3); this will indicate the generalization to 
higher groups. 

First of all, we will construct all representations by composing the fundamental representation with 
itself. We consider tensors made up of products of vectors $u^{(\alpha)}_i$ 
(the index $i$ denotes the components) in the 3-dimensional complex space, 
 $u^{(\alpha)}\in\bbbc^3$: 
thus, we have a rank 1 tensor, $u_i$;  rank two tensors, $u_iv_j$;  rank three tensors,
 $u_iv_jw_k$; rank four tensors, $u_iu_jv_kw_l$; \tdots; rank $r$
 tensors $u^{(1)}_{i_1}u^{(2)}_{i_2}\dots u^{(r)}_{i_r}$. 
It is not difficult to prove that forming linear combinations of these tensors we 
generate all the tensors, i.e., the tensors  $u^{(1)}_{i_1}u^{(2)}_{i_2}\dots u^{(r)}_{i_r}$ 
form a complete basis. In particular, putting them in Young tableaux we 
generate all the irreducible tensors. Thus we have:

\item{}{Rank 1: }{$T^{(3)}_i=u_i$\quad[3].}
\item{}{Rank 2: }{
$T^{(3^*)}_{ij}=\dfrac{1}{\sqrt{2}}\left(u_iv_j-u_jv_i\right)$\quad [$3^*$];\quad 
$T^{(6)}_{ij}=\dfrac{1}{\sqrt{2}}\left(u_iv_j+u_jv_i\right)$\quad [6].}
\item{}{Rank 3: }{\phantom{x}\hb
$T^{(1)}_{ijk}=\dfrac{1}{\sqrt{6}}\left(u_iv_jw_k-u_jv_iw_k-u_iv_kw_j+
u_kv_iw_j-u_kv_jw_i+u_jv_kw_i\right)$\quad [1];\hb 
$T^{(8)}_{ijk}=\dfrac{1}{\sqrt{4}}\left(u_iv_kw_j-u_kv_iw_j+u_kv_jw_i-u_jv_kw_i\right)$\quad [8];\hb
$T^{(10)}_{ijk}=\dfrac{1}{\sqrt{6}}\left(u_iv_jw_k+u_jv_iw_k+u_iv_kw_j+
u_kv_iw_j+u_kv_jw_i+u_jv_kw_i\right)$\quad [10].}

\noindent
etc. We have arranged the numerical factors so that, if the $u,\,v,\dots$ are of unit length, 
so are the higher rank tensors. In brackets we have put the dimensionality of each 
representation.
\smallskip
\sports{Identify these tensors with the corresponding Young tableaux. 
Check that, if we assume the $u,\,v,\,w$ to be an orthonormal set, so are the 
tensors $T^{(I)}$ above.}
\smallskip
Instead of multiplying abstract representations, it is much simpler to multiply these 
explicit repre\-sentations 
and merely project them in the ones we have. 
We show this with an explicit example. We start by multiplying 
$3\times3$ and find the tensor $u_iv_j$; it can be expanded into rank 2 tensors trivially,
$$u_iv_j=\dfrac{1}{\sqrt{2}}T^{(3^*)}_{ij}+\dfrac{1}{\sqrt{2}}T^{(6)}_{ij},$$
hence we recover (with Clebsch--Gordan coefficients included!) the result $3\times3=3^*+6$.
If we multiply again by a vector we find
$$T^{(3^*)}_{ij}w_k=\dfrac{1}{\sqrt{2}}\left(u_iv_j-u_jv_i\right)w_k$$
and it is easy to see that one has
$$T^{(3^*)}_{ij}w_k=\dfrac{1}{\sqrt{2}}\left(\sqrt{6}\,T_{ijk}^{(1)}+\sqrt{4}\,T_{ijk}^{(8)}
\right):$$
thus, we find $3^*\times 3=1+8$, again including the Clebsch--Gordan coefficients. 
This expansion can be done in a systematic manner by applying 
the Young tableaux of rank 3 to the tensor 
$T^{(3^*)}_{ij}w_k=\dfrac{1}{\sqrt{2}}\left(u_iv_j-u_jv_i\right)w_k$.
\smallskip
\sports{ 
i) Decompose the  product, $T^{(6)}_{ij}w_k$. ii) Form baryons from the $u,\,d,\,s$ 
quarks, taking into account 
the colour quantum number (which generates a SU(3) invariance), including the requirement of 
colour singlet for ``physical" hadrons.}
\smallskip 

The book of Cheng and Li~(1984) contains a readable elementary description of the SU(n) groups, their 
representations and their 
multiplication, which the reader may find 
sufficient for most physical applications (although, 
 of course, the basic reference is the text of Hamermesh,~1963).
\smallskip
\sport{By going to Lie algebras, and then to the complexified Lie algebras, show that 
everything that has been said for the Young tableaux-tensor
 formalism of SU(n) holds also for GL(n,C).}

\booksubsection{5.5. Representations of the permutation group}

\noindent
The method of Young tableaux allows us also to find the representations of the permutation group.
We will here only give a few results, without proofs;
 a detailed treatment may be found in the books of 
Weyl~(1946) and Hamermesh~(1963). 

Consider the permutation group of $n$ elements, $\piv_n$, and take all the Young tableaux 
of rank $n$. We may interpret the permutations as acting on the indices in the 
Young tableaux. For each Young tableau, $\cal Y$, we assign a representation of 
 $\piv_n$ as follows. Denote by $p$ to the subgroup of all permutations 
that leave each box in the same row (but not necessarily in the same column) that it occupied before 
applying the permutation; and denote by $q$ to the subgroup of permutations which move the 
boxes only inside the same column. 
It is evident that the sets $p$, $q$ will be different for different tableaux.  
We then introduce the function $\phi(P)$, $P\in \piv_n$ by requiring
$$\phi(P)=\cases{0,\;\hbox{when $P$ is not contained in the product $pq$};\cr
\delta_P\;\hbox{if $P=P_pQ_q$ with $P_p\in p,\;Q_q\in q$}.
\cr
}$$
Here $\delta_P$ is the parity of the permutation $P$. The functions of the form
 $$f(Q)=\sum_P a_P\phi(QP)$$
with $a_P$ real numbers  
generate a linear space, that we may call $\goth{H}({\cal Y})$, associated with the given 
Young tableau.
We finally define the operator $D(S)$ 
that represents the permutation $P$ on the functions  $\goth{H}({\cal Y})$ by
$$D(S):\;f(P)\to f(SP).$$
It is easy to verify that these operators form a representation of $\piv_n$. 
Although it is more difficult, it can also be shown that the representation
is irreducible, that the representations corresponding to different tableaux are inequivalent, 
and that they exhaust the set of all representations of $\piv_n$.  

A more detailed discussion of representations of the permutation group
 may be found in the treatises of Weyl~(1946), Hammermesh~(1963) or Lyubarskii~(1960).

\vfill\eject
\booksection{6. Relativistic invariance. The Lorentz group}
\vskip-0.5truecm
\booksubsection{6.1. Lorentz transformations. Normal parameters}

\noindent
In relativity theory the passage from one inertial system to another one,
moving with respect to it with speed ${\bf v}$, is given by the {\sl Lorentz
boosts} (or {\sl accelerations}). Starting with the case where ${\bf v}$ is
parallel to the $OZ$ axis, these boosts are given by\fnote{The contents of 
this and the following sections is adapted from the author's textbook
 on relativistic quantum mechanics, Yndur\'ain~(1996).}
$$
\eqalign{ x&\, \to x, \ \ y \to y, \cr
z&\, \to \dfrac{1}{\sqrt{1-v^2 /c^2}} (z + vt), \cr
t&\, \to \dfrac{1}{\sqrt{1 - v^2 /c^2}} (t + \dfrac{v}{c^2}
z).\cr}
$$
Here and henceforth $c$ will denote the speed of light.

We also write this with shorthand notation
$$
{\bf r} \to L ({\bf v} _z ) {\bf r}, \ \ t \to L
({\bf v} _z ) t.
$$
(This really {\sl is} shorthand: $L ({\bf v} ) {\bf r}$ depends also on $t$,
and not only on ${\bf v}, {\bf r}$; likewise, $L ({\bf v} ) t$ depends also on
${\bf r}$.) For ${\bf v}$ directed in an arbitrary way, we use the following
trick. Let $R ({\bf  z} \to {\bf v})$ be a rotation carrying the $OZ$
axis over ${\bf v}$. For example, we may choose
$$
R ({\bf  z} \to {\bf v} ) = R ({\ybf{\alpha}}),\ R ({\ybf{\alpha}} )
{\bf  z} = {\bf v} / |{\bf v} |,
$$
with ${\bf  z}$ the unit vector along $OZ$
and
$$
\cos \alpha = v_3 /v, \quad {\ybf{\alpha}} = ( \alpha / v)(\sin \alpha ) {\bf  z}
\times {\bf v}.
$$
Denoting by $L ({\bf v} )$ the Lorentz boost with velocity ${\bf v}$, we define
$$
L({\bf v} ) = R ({\bf  z} \to {\bf v} ) L ({\bf v} _z) R^{-1} (
{\bf  z} \to {\bf v} ),
$$
where ${\bf v} _z$ is a vector of length $v$
along $OZ$. Using the explicit formulas  for $L ({\bf v} _z)$
and $R$ we find that
$$
\eqalign{{\bf r}&\, \to L ({\bf v} ) {\bf r} =
{\bf r} -
\frac{{\bf v} {\bf r}}{v^2} {\bf v} + \left( 1 - \frac{{\bf v} ^2}{c^2}
\right) ^{-1/2} \left( \frac{1}{v^2} {\bf r} {\bf v} + t \right) {\bf v}, \cr
t&\, \to L ({\bf v} ) t = \left( 1 - \frac{{\bf v} ^2}{c^2}
\right) ^{-1/2} \left( t + \frac{{\bf v} {\bf r}}{c^2} \right) .\cr}
$$

\sport{Verify that, for $t$, $t'$, ${\bf r}$, ${\bf r} '$, ${\bf v}$ arbitrary,
$$
c^2 (L ({\bf v}) t ) (L ({\bf v} ) t' ) - (L ({\bf v} ) {\bf r} ) (L ({\bf v}
) {\bf r}' ) = c^2 tt' - {\bf r} {\bf r}',
$$
i.e., that under Lorentz boosts one has 
$$
c^2tt' - {\bf r} {\bf r}' = \hbox{invariant}.
$$
}

The parameters
${\bf v}$ are now {\sl not} normal; it is not true that the product of boosts
by ${\bf v}$, ${\bf v} '$ is the boost by ${\bf v} + {\bf v} '$ (which does not
even exist if $| {\bf v} + {\bf v} ' | \geq c$). It is then convenient to use
other parameters, which will be denoted by ${\ybf{\xi}}, {\ybf{\eta}}, \ldots$ such
that, whenever ${\ybf{\xi}}$ and ${\ybf{\eta}}$ are parallel,
$$
L ({\ybf{\xi}} ) L ({\ybf{\eta}} ) = L ( {\ybf{\xi}} + {\ybf{\eta}} ).
$$
Note that we use the same notation for $L ({\bf v} )$ and $L ({\ybf{\xi}} )$;
the context, and the latin/greek characters should be enough to indicate whether we
are using velocities or the new normal parameters.

Let us choose ${\ybf{\xi}}$ along $OZ$. If we write
$$
L ({\ybf{\xi}} ) z = A (\xi ) z + B (\xi ) ct,\quad L ({\ybf{\xi}} ) t = \frac{1}{c}
C (\xi ) z + D (\xi ) t,
$$
where $A$, $B$, $C$, $D$ are functions to be
determined, we get the consistency conditions
$$
AB = CD, \ \ A^2 -C^2 = D^2 - B^2 = 1,
$$
so that we can find $\varphi
({\ybf{\xi}} )$ verifying
$$
A = D = \cosh \varphi ({\ybf{\xi}} ), \ \ B = C = \sinh \varphi ({\ybf{\xi}} ).
$$
This relation  implies that
$$
\cosh (\varphi ({\ybf{\xi}} ) + \varphi ({\ybf{\eta}} ) ) = \cosh \varphi
({\ybf{\xi}} ) \cosh \varphi ({\ybf{\eta}} ) + \sinh \varphi ({\ybf{\xi}} )
\sinh \varphi ({\ybf{\eta}} )
$$
$$
\sinh ( \varphi ({\ybf{\xi}} ) + \varphi
({\ybf{\eta}}) ) = \cosh \varphi ({\ybf{\xi}} ) \sinh \varphi ({\ybf{\eta}} ) +
\sinh \varphi ({\ybf{\xi}} )
\cosh \varphi ({\ybf{\eta}} ) ,
$$
and we can thus choose $\varphi ({\ybf{\xi}} ) =
\xi\equiv|\ybf{\xi}|$. Finally

$$\eqalign{
x&\, \to x, \ \ y \to y, \cr
z&\, \to (\cosh \xi ) z + (\sinh \xi ) ct, \cr
t&\, \to \frac{1}{c} (\sinh \xi ) z + (\cosh \xi ) t,\quad
 {\ybf{\xi}} \parallel OZ.\cr} 
$$
The relation between the ${\ybf{\xi}}$ and ${\bf v}$ is found by comparison 
of these relations:
$$
\cos \xi = \frac{1}{\sqrt{1- {\bf v} ^2/c^2}} , \quad \sinh \xi = \frac{|
{\bf v} |}{c} \, \frac{1}{\sqrt{1- {\bf v} ^2 /c^2}} , \quad {\ybf{\xi}}
\parallel {\bf v} .
$$
${\ybf{\xi}}$ is sometimes called the {\sl rapidity}. For a boost along an arbitrary
${\ybf{\xi}}$,  we find
$$
\eqalign{{\bf r}&\, \to L ({\ybf{\xi}} ) {\bf r} =
{\bf r} -
\dfrac{{\ybf{\xi}}
{\bf r}}{\xi ^2} {\ybf{\xi}} + \frac{1}{\xi} \left\{ ( \cosh \xi )
\frac{{\ybf{\xi}} {\bf r}}{\xi} {\ybf{\xi}} + c (\sinh \xi ) t {\ybf{\xi}}
\right\}, \cr
t &\,\to L ({\ybf{\xi}} ) t = ( \cosh \xi ) t + \frac{1}{c}
\,
\frac{\sinh \xi}{\xi} {\ybf{\xi}} {\bf r} .\cr}
$$

For speeds small compared with $c$,
$$
{\ybf{\xi}} \simeq {\bf v} /c ,
$$
and a Lorentz boost coincides with a
Galilean boost.

The transformations $\lambdav$ of the set $({\bf r} ,t)$ obtained
by
applying rotations and Lorentz boosts as a product,
$$
\lambdav = LR,
$$
are called {\sl Lorentz transformations}. As we will see in the next sections,
they form a group, called the {\sl Lorentz group}, or, sometimes, and for
reasons that will be apparent presently, the {\sl orthochronous, proper Lorentz
group}. 

If we include possible products by space, $I_s$, and time, $I_t$,
reversals,
$$
I_s \ :  {\bf r} \to - {\bf r},  t \to t;  I_t
\ : \
{\bf r} \to {\bf r} ,  t \to - t,
$$
we obtain a set (which is also a group) called the {\sl full Lorentz group}.
Its elements are of one of the following forms:
$$
LR,\ I_s LR,  I_t LR,  I_s I_t LR.
$$

\booksubsection{6.2. Minkowski Space. The Full Lorentz Group}

\noindent
As we saw in the previous section, Lorentz boosts mix space and
time. A unified treatment of relativistic transformations demands that
we work
in a set that contains both. This is {\sl Minkowskian spacetime}
(or
just {\sl Minkowski space}). Its elements, or points, which will
be
denoted\fnote{Our conventions are not universal, although they
are certainly quite common.} by
letters $x$, $y$, $\ldots$, are called {\sl four-vectors}, and
are
determined by {\sl four} coordinates, $x_{\mu}, \mu= 0,1,2,3,$
$$
x \sim \pmatrix{x_0\cr x_1\cr x_2\cr x_3} ,
$$
where $x_0 = ct$ corresponds to a time coordinate and $x_j =
r_j$,
$j=1,2,3$ are purely spatial coordinates.\fnote{For 
the sake of definiteness, we work here with the space-time Minkowski space; 
the considerations are of course also valid for the energy-momentum Minkowski space of
vectors $p$, with $\bf p$ the momentum and $p_0=E/c$, $E$ the energy.}

We will consistently tag Minkowskian coordinates with Greek indices $\mu , \nu
, \dots$ varying from 0 to 3; latin indices $i, j, \dots$ will be
restricted to varying from 1 to 3. We will also denote by ${\bf r}$ the
spatial part of $x$, and $x$ may thus also be written as
$$
x \sim \pmatrix{ct\cr {\bf r}} . 
$$
At times a horizontal notation is convenient, and we write  $x \sim (ct,{\bf r})$.

Lorentz boosts may be represented by $4 \times 4$ matrices $L$,
$x
\to Lx$, with elements $L_{\mu \nu}$, so that
$$
(Lx)_{\mu} = \sum ^{3}_{\nu =0} L_{\mu \nu} x_{\nu} ; 
$$
explicitly, we have
$$
 (Lx)_0 = ( \cosh \xi ) x_0 + \frac{\sinh \xi}{\xi}
\sum^{3}_{j=1} \xi_j x_j,
$$
$$
(Lx)_i = x_i - \frac{1}{\xi ^2}\! \left(\sum_j \xi_j x_j \right)\! \xi_i +
\dfrac{1}{\xi} \!\left( \dfrac{\cosh \xi}{\xi}
\sum_j \xi_j x_j + x_0 \sinh \xi \right)\! \xi_i .\;
$$

Rotations can also be defined as transformations in Minkowski space: $x
\to Rx$, with
$$
 (Rx)_{\mu} = \sum_{\nu} R_{\mu \nu} x_{\nu} , 
$$
and 
$$
\matrix{(Rx)_0 = x_0 , \cr
(Rx)_i = (\cos \theta ) x_i + \dfrac{1- \cos \theta}{\theta ^2}
\left(\sum_j \theta_j x_j \right) \theta_i + \dfrac{\sin \theta}{\theta}
\sum_{kl} \epsilon_{ikl} \theta_k x_l . \cr}
$$
Here $\epsilon_{ikl}$ is the Levi--Civit\`a symbol.

The transformations $L$, $R$ leave invariant the quadratic form
$x \cdot
y$ defined by
$$
x \cdot y \equiv x_0 y_0 - \sum^{3}_{j=1} x_j y_j .
$$
This form is known as the {\sl Minkowski} (pseudo) {\sl scalar
product},
and can be also written in terms of the (pseudo) {\sl metric
tensor}
$G$, with components $g_{\mu \nu}$,
$$
g_{\mu \nu} =  0, \mu \neq \nu , \quad g_{\mu \nu} = 1, \mu
= \nu = 0 , \quad g_{\mu \nu}=  -1,  \mu = \nu \neq 0 . 
$$
Indeed,
$$
x \cdot y = \sum_{\mu \nu} g_{\mu \nu} x_{\mu} y_{\nu} =
\sum_{\mu}
g_{\mu \mu} x_{\mu} y_{\mu} = x^{\rm T} Gy.
$$
In the last expression, $x$, $y$ are taken to be matrices.
The {\sl Minkowski square}, denoted by $x^2$ if there is no
danger of
confusion, is defined as $x^2 \equiv x \cdot x$.

As stated above, one can verify, by direct computation, that, when $\lambdav =
LR$ for
any $L$, $R$, then, for every pair $x$, $y$,
$$
(\lambdav x) \cdot ( \lambdav y ) = x \cdot y .
$$
In terms of the metric tensor, 
$$
\lambdav^{\rm T} G \lambdav = G.
$$
These relations suggest that we define a group, called the {\sl full Lorentz
group}, and denoted by $\overline{\cal{L}}$, to be the set of all matrices
$\overline{\lambdav}$ such that

$$
\overline{\lambdav}^{\rm T} G \overline{\lambdav} = G.
$$
It is obvious that such $\overline{\lambdav}$ form a group, and it is easy 
to verify that one also has
$$
\overline{\lambdav} G \overline{\lambdav}^{\rm T} = G.
$$

Let us take determinants in $\lambdav^{\rm T} G \lambdav = G$. We find that $(\det
\overline{\lambdav} )^2 = 1$, and hence $\det \overline{\lambdav} = \pm
1$.
Consider space reversal, acting in Minkowski space by $(I_s x)_0
=x_0$,
$(I_s x)_i = -x_i$. Clearly, $I_s$ is in $\overline{\cal{L}}$ and
moreover
$\det I_s = -1$. If $\overline{\lambdav}$ belongs to
$\overline{\cal{L}}$
and $\det \overline{\lambdav} = -1$, then we can write identically

$$
\overline{\lambdav} = I_s (I_s \overline{\lambdav} ) ,
$$
and now $\det (I_s \overline{\lambdav} ) = +1$. If we denote by
$\cal{L}_+$ to the subgroup of $\overline{\cal{L}}$ consisting of matrices with
determinant unity, we have just shown that
$\overline{\cal{L}}$ consists of matrices either in $\cal{L} _+$ or products
of $I_s$ time matrices in $\cal{L} _+$.

Consider next the four-vector $n_t$, a unit vector along the time
axis,
with components $n_{t \mu} = \delta_{\mu 0}$. Given
$\overline{\lambdav}$
in $\overline{\cal{L}}$, we may have either $(\overline{\lambdav}
n_t )_0 >
0$ or $(\overline{\lambdav} n_t)_0 < 0$; it is not possible to
have
$(\overline{\lambdav} n_t )_0=0$. Moreover, if
$(\overline{\lambdav}
n_t )_0 > 0$ and $(\overline{\lambdav}' n_t) _0 > 0$, then $(\overline{\lambdav}^{-1}n_t)_0>0$ and 
$(\overline{\lambdav} \overline{\lambdav}' n_t )_0 > 0$. (The
proofs
of these statements are left as exercises.) It
then
follows that the subset of $\overline{\cal{L}}$ consisting of
transformations $\overline{\lambdav}$ with $(\overline{\lambdav}
n_t)_0
>0$ forms a group, called the {\sl orthochronous Lorentz group},
and
denoted by $\cal{L}^{\uparrow}$; the corresponding
transformations
preserve the arrow of time. If the matrix $\overline{\lambdav}$
in
$\overline{\cal{L}}$ is such that $(\overline{\lambdav} n_t)_0 <
0$,
then we can write identically
$$
\overline{\lambdav} = I (I \overline{\lambdav} ) ,
$$
where $I$ is the {\sl total reversal}, $I = I_t I_s$: $Ix
\equiv -x$.
Clearly, $(I \overline{\lambdav} n_t)_0$ is now positive. We have
proved
that any element of $\overline{\cal{L}}$ is either an element of
$\cal{L}^{\uparrow}$ or a product $I \lambdav$ with $\lambdav$ in
$\cal{L}^{\uparrow}$.

Finally, the {\sl proper, orthochronous Lorentz group}
$\cal{L}^{\uparrow}_+$ (which we simply call, if there is no
danger of
confusion, the {\sl Lorentz group}, $\cal{L}$) is the group of
matrices
$\lambdav$ such that
$$
\lambdav^{\rm T} G \lambdav = G, \quad \det \lambdav = 1, \quad \lambdav_{00}
> 0.
$$

As we have just shown,  we have that any element in
$\overline{\cal{L}}$, $\overline{\lambdav}$ is of one of the forms
$$
I_s \lambdav, \ I_t \lambdav, \ I_s I_t \lambdav, \ \lambdav
$$
with $\lambdav$ in $\cal{L}^{\uparrow}_+$.

The transformations $I_s$, $I_t$, $I$ are at times called {\sl
improper}
transformations.
\smallskip

\sport{Prove that $\lambdav^{\rm T} G\lambdav = G$ implies
that $\lambdav n_t
\neq 0$. \quad Solution: Consider the $00$ components of $\lambdav^{\rm T} G \lambdav
= G$, and
$\lambdav G \lambdav^{\rm T} = G$; then,
$$
 \lambdav_{00}^2 - \sum_i \lambdav^2_{i0} = 1 ; \quad \lambdav_{00}^2 -
\sum_i \lambdav^2_{0i} = 1 . 
$$
From any of these, $|\lambdav_{00}| \geq 1$ so
$|(\lambdav n_t)_0 | \geq 1$.}
\smallskip
\sport{Show that $\lambdav_{00} >0$, $\lambdav'_{00} >0$
imply that $(\lambdav \lambdav ')_{00} >0$. \quad Solution: Using the
evaluations of the previous problem and Schwartz's inequality,
$$
\left| \sum_i \lambdav_{0i} \lambdav '_{i0} \right| \leq \sqrt{\sum
\lambdav_{0i}
\lambdav_{0i}} \sqrt{\sum \lambdav '_{i0} \lambdav '_{i0}} <
\lambdav_{00}
\lambdav '_{00} . 
$$
Hence,
$$
 (\lambdav \lambdav ')_{00} = \lambdav_{00} \lambdav '_{00} +
\sum_i
\lambdav_{0i} \lambdav '_{i0} > \lambdav_{00} \lambdav '_{00} - \left|
\sum
\lambdav_{0i} \lambdav '_{i0} \right| > 0. 
$$
}
\smallskip
\sport{Show that $\lambdav_{00} > 0$ implies that
$(\lambdav^{-1})_{00} > 0$.
}

\booksubsection{6.3. More on the Lorentz Group}

\noindent
In this section we further characterize the (orthochronous, proper) Lorentz
group. We start by proving a simple, but basic, theorem.

\teoreman{1}{If $R$ is in ${\cal L}$ and $Rn_t =
n_t$, then $R$ is a rotation.}

To prove this, we note that the condition $Rn_t = n_t$ implies
that $R$
is of the form
$$
 R = \pmatrix{ 1 & 0 & 0 & 0 \cr
0&&&&\cr 
 0&&\hat{R}&\cr
0&&&}, 
$$
with $\hat{R}$ a $ 3 \times 3$ matrix. The condition $R^{\rm T} GR = G$ implies
that $\hat{R}^{\rm T} \hat{R} = 1$; and $\det R = +1$ implies that also $\det
\hat{R} = +1$. Therefore, $\hat{R}\in{\rm SO(3)}$, i.e., it is
 a three-dimensional rotation. From now on we will
denote by the same symbol $R$ the Minkowski space transformation and the
restriction $(\hat{R})$ to ordinary three-space.

Now let $\lambdav$ be an arbitrary transformation in $\cal{L}$,
and let
$u \equiv \lambdav n_t$. We have $u_0 > 0$ and $u \cdot u =1$.
Consider
the vector ${\ybf{\xi}}$ such that $u_0 = \cosh | {\ybf{\xi}} |$, $|
{\bf u}
| = \sinh | {\ybf{\xi}} |$; this is possible because
$$
 1 = u \cdot u = ( u_0 )^2 - | {\bf u} |^2 = \cosh ^2 \xi -
\sinh ^2
\xi . 
$$
We choose $ {\ybf{\xi}}$ directed along ${\bf u}$,
$$
 {\ybf{\xi}} / | {\ybf{\xi}} | = {\bf u} / | {\bf u} | , 
$$
so that
$$
u_0 = \cosh \xi, \quad  u_i = \frac{1}{\xi} (\sinh \xi ) \xi _i .
$$

Using the explicit expressions for $L ({\ybf{\xi}} )$, we
see that $L ({\ybf{\xi}}) n_t = u$. It follows that the transformation $L^{-1}
({\ybf{\xi}} ) \lambdav$ is such that
$$
 L^{-1} ({\ybf{\xi}} ) \lambdav n_t = n_t , 
$$
so by Theorem 1, $L^{-1}
({\ybf{\xi}} ) \lambdav \equiv R $ has to be a
rotation, characterized by some ${\ybf {\theta}}$. We have therefore
proved
the following theorem:

\teoreman{2}{Any (proper, orthochronous) Lorentz transformation,
$\lambdav$, can be written as
$$
\lambdav = L ({\ybf{\xi}}) R ({\ybf{\theta}}) ,
$$ where $R$ is a rotation and $L$ a Lorentz boost (the
decomposition is {\sl not} unique).} In particular it follows from this that the Lorentz
group is a six-dimensional Lie group (three parameters from ${\ybf {\theta}}$ and
three from ${\ybf{\xi}}$).
 It is clearly non-compact (the parameters $\ybf{\xi}$ can take arbitrarily large values) 
and it is also simple and doubly connected; later we will find its covering group, 
which coincides with 
SL(2,C).

We may recall that the Lorentz boost $L ({\ybf{\xi}} )$ can be
written
as
$$
 R' L ({\ybf{\xi}}_z ) R'' , 
$$
with $R', R'' = R'^{\ -1}$ rotations and $L
({\ybf{\xi}}_z )$ an acceleration
along the $OZ$ axis. Thus, the general study of Lorentz
transformations
is reduced to that of rotations and pure accelerations, that may
be
taken to be along the $OZ$ axis.

\sport{Given two pure boosts $L({\ybf{\xi}} )$,
$L({\ybf{\eta}})$, find $L({\ybf{\zeta}})$, $R ({\ybf{\theta}})$ such
that
$$
 L ({\ybf{\xi}} ) L ({\ybf{\eta}} ) = L ({\ybf{\zeta}} ) R
({\ybf{\theta}} ).$$}
\smallskip\noindent
Note that in general (unless ${\ybf{\xi}}, {\ybf{\eta}}$ are parallel) the product
of two boosts is {\sl not} a pure boost

We finish the characterization by presenting two more theorems, and a
covariant parametrization of the Lorentz transformation $\lambdav$.

\teorema{A Lorentz transformation $\lambdav$
such that
$\lambdav n_t = u$ is a pure boost, times a rotation around
${\ybf{\xi}}$
(where ${\ybf{\xi}}$ is given in terms of $u$ by 
$\cosh \xi=u_0,\quad {\ybf{\xi}}/\xi={\bf u}/|{\bf u}|$)
 if, and
only if,
$\lambdav$ commutes with all rotations around ${\ybf{\xi}}$.
}

To prove this, we use  that a rotation around
${\ybf{\xi}}$,
which we denote by $R_{{\ybf{\xi}}}$, leaves ${\ybf{\xi}}$ invariant; hence, it 
follows that $L ({\ybf{\xi}} )$ and $R_{{\ybf{\xi}}}$ commute. [Use
that
${\ybf{\xi}} (R_{{\ybf{\xi}}} {\bf r} ) = (R_{\xi}^{-1} {\ybf{\xi}})
{\bf r}
= {\ybf{\xi}} {\bf r}$ for any ${\bf r}$]. The reciprocal is also
easy.
Given that $u = \lambdav n_t$, we construct  ${\ybf{\xi}}$ as before, and
then $L
({\ybf{\xi}} )$. Now, $L^{-1} ({\ybf{\xi}} ) \lambdav = R$ is a
rotation. As
we have just seen, $L ({\ybf{\xi}} )$ commutes with rotations
$R_{{\ybf{\xi}}}$; so does $\lambdav$, and hence $R$. But a rotation
that
commutes with all rotations around an axis ${\ybf{\xi}}$ is itself
a
rotation around that axis, so $\lambdav = L ({\ybf{\xi}} )
R_{{\ybf{\xi}}}$,
finishing the proof.

\teorema{We have, for any ${\ybf{\xi}}$ and any
rotation
$R$,
$$
R L ({\ybf{\xi}} ) R^{-1} = L (R {\ybf{\xi}} ),
$$
where $L (R {\ybf{\xi}} )$ is the boost characterized by
$R
{\ybf{\xi}}$.
}
\smallskip\noindent
The proof is straightforward  and is left as an exercise.

Instead of parametrizing a Lorentz transformation $\lambdav = L
({\ybf{\xi}} ) R ({\ybf {\theta}} )$ by the parameters ${\ybf{\xi}}$,
${\ybf {\theta}}$, it is at times convenient to use what is called
a {\sl
covariant parametrization}. We define the set of parameters
$\omega_{\mu
\nu}$ in terms of ${\ybf{\xi}}$, ${\ybf {\theta}}$ by
$$
\sum_{jk} \epsilon_{jkl} \omega_{jk} = \theta_l, \quad \omega_{j0}
=
\frac{1}{2} \xi_j; \quad \omega_{\alpha \beta} = - \omega_{\beta
\alpha} .
$$
For $\omega$ infinitesimal we write a Lorentz transformation as
$$
\lambdav = 1 - \sum \omega_{\alpha \beta} X ^{(\alpha \beta )} +
O
(\omega ^2 ) .
$$
Then, the matrices $X^{(\alpha \beta )}$ have components
$$
X_{\mu \nu}^{(\alpha \beta )} = - (\delta_{\mu \alpha} g_{\nu \beta}
-
\delta_{\mu \beta} g_{\nu \alpha} ).
$$
To prove this, we note that, on the one hand, and from the definition of
$X$,
$$
(\lambdav (\omega ) x)_{\mu} \simeq x_{\mu} - \sum_{\alpha \beta}
\sum_{\nu} \omega_{\alpha \beta} X_{\mu \nu}^{(\alpha \beta )}
x_{\nu} ;
$$
on the other, from the explicit formulas for $R$, $L$,
$$
 (R ({\ybf {\theta}} ) x)_0 = x_0,\quad  (R ({\ybf {\theta}} ) x)_i = x_i
- \sum
2 \omega_{ik} x_k; 
$$
$$
 (L ({\ybf{\xi}} )x)_0 \simeq x_0 + \sum 2 \omega_{j0}
x_j,\  (L ({\ybf{\xi}} ) x)_i \simeq x_i + 2 \omega_{i0} x_0, 
$$
so that letting
$\lambdav = LR$, we get
$$
 (\lambdav x)_0 \simeq x_0 - \sum 2 \omega_{0j} x_j,\ (\lambdav
x)_i
\simeq x_i + 2 \omega_{i0} x_0 - \sum 2 \omega_{ik} x_k  
$$
from which the desired result follows.

Beyond ${\cal L}^{\uparrow}_+$, the invariance group of relativity 
also includes space translations,
$$
 {\bf r} \to {\bf r} + {\bf a} ,
$$
and time translations,
$$
 ct \to ct + a_0; 
$$
in four-vector notation,
$$
x_{\mu} \to x_{\mu} + a_{\mu} .
$$

The group obtained by adjoining to ${\cal L}$ the translations will be called
the {\sl Poincar\'e}, or {\sl inhomogeneous Lorentz group}, written ${\cal
JL}$. Its elements are pairs $(a, \lambdav )$ with $a$ a four-vector and
$\lambdav$ in ${\cal L}$. They act on an arbitrary vector
$x$ by
$$
(a, \lambdav ) x = a + \lambdav x ,
$$
and satisfy the ensuing product and inverse law:
$$
\eqalign{
(a, \lambdav ) (a', \lambdav ') = (a + \lambdav a', \lambdav \lambdav
'),\cr
(a, \lambdav )^{-1} = (- \lambdav ^{-1} a, \lambdav ^{-1} ).
\cr}
$$
The unit element of the group is the transformation $(0,1)$.
At times we will simplify the notation writing $a$ instead of $(a,1)$ and $\lambdav$ 
instead of $(0,\lambdav)$. 
The mathematical structure of $\cal IL$ is 
$${\cal IL}={\cal L}\widetilde{\times}{\cal T}_4.$$

\booksubsection{6.4. Geometry of Minkowski Space}

The geometrical properties of spacetime present some peculiarities owing to
the indefinite character of the metric. A first peculiarity is that we can
classify vectors $v$ of a Minkowskian space, {\sl in a relativistically
invariant way}, in the following classes: timelike, lightlike, and
spacelike vectors. {\sl Timelike vectors} $v$ are such that $v \cdot v > 0$.
If
$v_0 > 0$, we say they are {\sl positive timelike}; if $v_0 < 0$, {\sl
negative} ($v_0 = 0$ is impossible). {\sl Lightlike} vectors $v$, which
satisfy $v \cdot v = 0$, are {\sl positive lightlike} if $v_0 > 0$, {\sl
negative} if $v_0 < 0$. $v_0 =0$ is only possible for the null vector, $v=0$.
Finally, we say that $v$ is {\sl spacelike} if $v \cdot v < 0$; the sign of
$v_0$ is {\sl not} invariant now.
\smallskip

\sports{i)~Prove that this classification is invariant under transformations in
${\cal L}^{\uparrow}_+$; in particular check invariance of sign $v_0$ if $v^2\geq0$.
ii)~Show that the trajectory of a particle with mass is given by a positive timelike
vector, and that of a light ray by a positive lightlike vector. Hint: Let ${\bf r}$
be the location of a particle (or signal) at time $t$. Form the four-vector $x, x_0=ct,
{\bf x} =
{\bf r}$. The velocity of the particle (assuming uniform motion) is ${\bf v} = {\bf r}
/t$}
\smallskip
The following lemma is very useful:
\smallskip
\noindent{\sc Lemma.}\hb{\it (i)~If $v$ is positive (negative) timelike, then there exists a vector
$v^{(0)}$ and a Lorentz transformation $\lambdav$ such that $v= \lambdav
v^{(0)}$, and $v_0^{(0)} = \pm m$, ${\bf v}^{(0)} =0$, $m>0$. (ii)~If $v$ is
positive (negative) lightlike there exists a $\overline{v}$ and $\lambdav$ with
$v=\lambdav \overline{v}$ and $\overline{v}_0= \pm 1$, $\overline{v}_1 =
\overline{v}_2 =0$, $\overline{v}_3 =1$. (Here and before the signs ($\pm$) are
correlated to positive--negative.) (iii)~If $v$ is spacelike, there exist a
$v^{(3)}$ and $\lambdav$ with $v=\lambdav v^{(3)}$, $v^{(3)}_{\mu} = \delta_{\mu
3} v^{(3)}_3$, $v_3^{(3)} >0$.
}
\smallskip
This means that, in an appropriate reference system, a positive lightlike
vector (e.g.) can be chosen to be of the form $\overline{v}$,
$$
\overline{v} = (1,  0 , 0 , 1) .
$$

\noindent The clumsy but simple proof of this lemma uses the explicit expression for the
Lorentz transformations to build explicit constructions.

The difference between an Euclidean space and Minkowski space is also apparent
in the two following results:

\teorema{If both $v$ and $v'$ are lightlike and they are orthogonal, i.e., $v \cdot v'
=0$, then they are parallel: $v' = \alpha v$.
}
The proof is left as a simple exercise, using the previous Lemma.

\teorema{If $v \cdot v \geq 0$ and $v \cdot u=0$,
either
$v$ and $u$ are proportional or necessarily $u$ is spacelike.
}
The proof is again left as an exercise, using the Lemma.

\teorema{The only invariant numerical tensors in Minkowski space are combinations of
the metric tensor, $g_{\mu \nu}$, and the {\sl Levi--Civit\`a} tensor
$\epsilon_{\mu \nu \rho \sigma}$,}
$$
\eqalign{\epsilon_{\mu \nu \rho \sigma}&\, =  1, \hbox{if $\mu
\nu \rho \sigma$ is an even permutation of $1230$}, \cr
\epsilon_{\mu \nu \rho \sigma}&\, -1, \hbox{if $\mu \nu
\rho \sigma$ is an odd permutation of $1230$}, \cr
\epsilon_{\mu \nu \rho \sigma}&\,= 0, \hbox{if two indices are
equal.}\cr
} 
$$
Note that $\epsilon_{ijk0}=\epsilon_{ijk}$, where
$\epsilon_{ijk}$ is
the Levi--Civit\`a tensor in ordinary three-space.

\teorema{Given a set of Minkowski vectors $v^{(a)}$, the only
invariants that are continuous and that can be formed with them are functions
of the scalar products $v^{(a)} \cdot v^{(b)}$ and, if there are four or more
vectors, of the quantities}
$$
 \sum 
\epsilon^{\mu \nu \rho \sigma} v^{(a)}_{\mu} v^{(b)}_{\nu}
v^{(c)}_{\rho}
v^{(d)}_{\sigma} .
$$ 

In spite of the fact that these theorems are similar to their analogues in
Euclidean space and also in spite of their apparent simplicity, proofs are
very complicated. For example, the later Theorem fails if we remove the requisite of
continuity: the functions (sign $v_0) \theta (v^2 )$ or $\delta_4 (v) \equiv
\delta (v_0) \delta ({\bf v} )$ are invariant: yet they {\sl cannot} be
written in terms of invariants. Proofs of the two Theorems can be found in,
for example, the treatise of Bogoliubov, Logunov and Todorov (1975).

Given a Minkowski vector, $v$, the set of Lorentz transformations
$\gammav$ that leave it invariant is called its {\sl little
group}\fnote{Little groups, first introduced by Wigner (1939), play a key
role in the study of relativistic particle states.} (or {\sl stabilizer}),
${\cal W} (v)$. The little group of a vector $v$ depends only upon the sign of
$v \cdot v$, in the sense that if, for example,
$v \cdot v >0$ and $u \cdot u >0$, then the little groups ${\cal W} (v)$,
${\cal W} (u)$ are isomorphic. To prove this, we first note that
${\cal W} (v)$ and ${\cal W} (\lambdav v)$ are isomorphic for any
$\lambdav$. Indeed, if $\gammav v =v$, then $\lambdav \gammav \lambdav^{-1}$ is in
${\cal W} (\lambdav v)$, and vice versa. Moreover, ${\cal W} (v)$ is identical
with ${\cal W} (\alpha v)$ for any number $\alpha \neq 0$. Using this in
conjunction with Lemma 1, we find that there are essentially only three little
groups. To be precise, we have that, if $v \cdot v >0$, the little group is
isomorphic to ${\cal W} (n_t)$; if $v \cdot v =0$, the little group is
isomorphic to ${\cal W} (\overline{v} )$, $\overline{v}_0 = \overline{v}_3$,
$\overline{v}_1 = \overline{v}_2 = 0$; and if $v \cdot v < 0$, the little group
is isomorphic to ${\cal W} (n^{(3)})$, $n_{\mu}^{(3)} = \delta_{\mu 3}$. This
greatly simplifies the study of the little groups.

\teorema{One has, (A)~${\cal W} (n_t) = SO (3)$, where by $SO (3)$ we denote
the group of ordinary rotations. (B)~${\cal W} (\overline{v} ) = SO (2) \times
{\cal T}_2$, where $SO_z (2)$ is the group of rotations around $OZ$, and
${\cal T}_2$ is defined below. (C)~${\cal W} (n^{(3)} ) = {\cal
L}^{\uparrow}_+ (3)$, where ${\cal L}^{\uparrow}_{+} (3)$ is identical to a
Lorentz-like group (in three dimensions) that acts only on time and the spatial plane $XOY$, but leaves
$OZ$ invariant. }

The result (A) is already known to us. Result (C) is
left as a simple exercise. We turn to the lightlike case (B). Let
$\gammav$ be an element of ${\cal W} (\overline{v} )$, and let $N$ be the
subspace of Minkowski space orthogonal to $\overline{v}$, that is, if
$u$ is in $N$, then $u \cdot \overline{v} =0$.

Clearly, the subspace $N$ is also invariant under $\gammav$. A basis of
$N$ is formed by the three vectors $v^{(a)}, a= 1,2,3$ with
$v^{(1)}= n^{(1)}$, $v^{(2)} = n^{(2)}$, $n^{(a)}_{\mu} = \delta_{a \mu}$, and
$v^{(3)} = \overline{v}$ : because $\overline{v}$ is lightlike the subspace
orthogonal to $\overline{v}$ contains $\overline{v}$ itself. If
$u$ is in $N$, we write $u = \sum_a \alpha_a v^{(a)}$. Because
$\gammav u$ is also in $N$, we can write
$$
\gammav u = \sum_{ab} \gammav_{ab} \alpha_{b} v^{(a)}; 
$$
thus the matrix elements $\gammav_{ab}$ determine $\gammav$, and vice versa. The
conditions $\gammav u \cdot \gammav u' = u \cdot u'$ and $\gammav \overline{v} =
\overline{v}$ imply that
$$
(\gammav_{ab} ) = \pmatrix{ \cos \theta & \sin \theta & 0
\cr - \sin \theta & \cos \theta & 0 \cr \gammav_{31} & \gammav_{32} & 1
\cr},
$$
with $\gammav_{31}$, $\gammav_{32}$ arbitrary. This set of matrices  has a
{\sl mathematical} structure like that of the Euclidean group of the plane, $SO_z
(2) \times {\cal T} _2$ where $SO_z (2)$ are  rotations around $OZ$,
$$
\pmatrix{\cos \theta & \sin \theta & 0 \cr
 -\sin \theta & \cos \theta & 0 \cr
 0  & 0  & 1  \cr} ,
$$
and the  ``translations" ${\cal T}_2$ are
$$
 \pmatrix{1  &  0  &  0 \cr 
0  & 1  &  0 \cr
\gammav_{31} & \gammav_{32}  & 1 \cr} .
$$

\topinsert{
\setbox0=\vbox{\hsize8.2truecm{\epsfxsize=6.5truecm\epsfbox{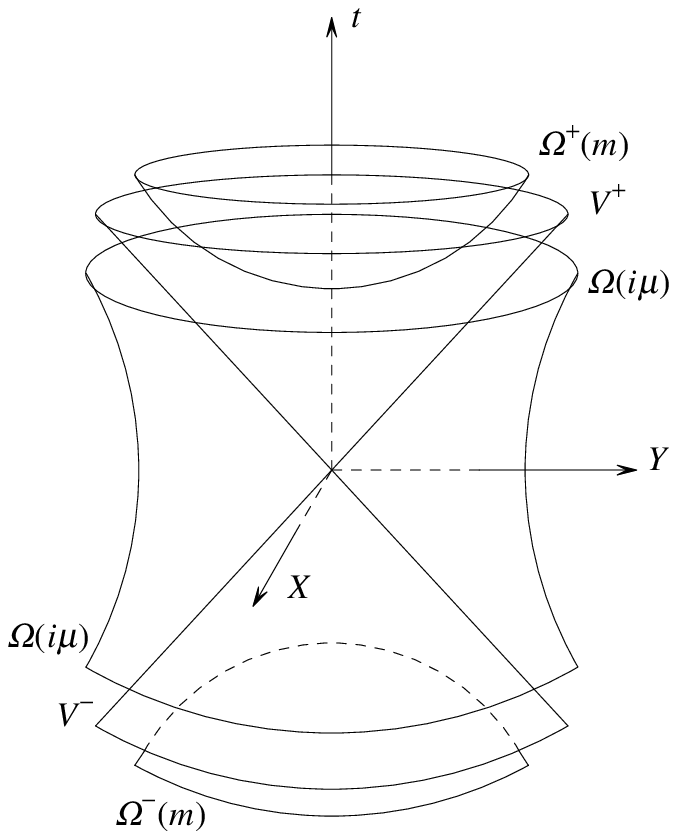}}}
\setbox1=\vbox{\hsize 6.8truecm{\petit Various regions in Minkowski space.\hb
\phantom{X}\hb}}
\line{\notightboxit{\box0}\hfil\box1}
}\endinsert

To finish this section we present a few more definitions (see the figure). The
{\sl light cone} is the set of vectors $v$ with $v^2 =0$. If, moreover,
$v_0>0$ ($v_0<0$), we speak of the {\sl future, forward or positive} ({\sl
past, backward or negative}) {\sl light cone}, denoted by $V^{+}$ ($V^-$). The
set of vectors $u$ with $u^2 = m^2 >0$ is denoted by $\Omega^{\pm} (m)$,
($\pm$) according to the sign of $u_0$, and is called the {\sl future, forward
or positive} ({\sl past, backward or negative}) {\sl mass hyperboloid},
for $u_0 >0$ ($u_0<0$). This name derives from ({\sl momentum}) {\sl Minkowski
space}. The set of $w$ with $w \cdot w = - \mu ^2$, $\mu ^2 >0$ is
called the {\sl imaginary mass hyperboloid}, $\Omega (i \mu)$.

\smallskip
\sport{Verify that the sets $V^{+}$, $V^-$, $\Omega^{+} (m)$, $\Omega^- (m)$, $\Omega
(i\mu )$ are invariant under ${\cal L} ^{\uparrow}_+$, and that each vector in
one of them can be reached by an appropriate transformation from any other one
in the same set.}

\booksubsection{6.5. Finite dimensional representations of the Lorentz group}

\noindent{\sl i. The correspondence $\cal L\to{\rm SL(2,C)}$}
\medskip\noindent
To every Minkowski vector $v$ with components $v_{\mu}$ we
associate the $2\times 2$ complex matrix
$$
\tilde{v} = v_0 + \ybf{\sigma} {\bf v} = \sum_{\mu \nu} g_{\mu
\nu} \tilde{\sigma}_{\mu} v_{\nu} = 
\pmatrix{v_0 + v_3 & v_1 - \ii v_2 \cr
v_1+\ii v_2 & v_0-v_3},
$$
$$\tilde{\sigma}_0 = \sigma_0 = 1, \ \tilde{\sigma}_i = -
\sigma_i . $$
We have
$$\eqalign{
\tilde{\sigma}_{\mu} = \sum_{\nu} g_{\mu \nu}
\sigma_{\nu} ; \quad {\rm Tr}\, \tilde{\sigma}_{\mu} \sigma_{\nu} =
2g_{\mu \nu} ;  \cr
\det \tilde{v} = v \cdot v , \quad v_{\mu} =
\tfrac{1}{2} {\rm Tr}\, \sigma_{\mu} \tilde{v} ;\quad \tilde{v}^{\dag} =
\tilde{v}, \cr}
$$
the last relation holding if the $v_{\mu}$ are real.

For every Lorentz transformation,
$$\lambdav \ : \ v \to \lambdav v \equiv v_{\lambdav} , $$
we have a corresponding matrix $A$, $A$ in SL(2,C). We {\sl
define} $A$ by
$$
A \tilde{v} A^{\dag} = \tilde{v}_{\lambdav} = \tilde{\sigma} \cdot
\lambdav v .
\eqno{(1)}
$$
Actually, both $\pm A$ correspond to the same $\lambdav$. An
explicit formula for the correspondence is obtained as follows.
Choose the vectors $v^{(\alpha )}$ with $v^{(\alpha )}_{\mu} =
\delta_{\alpha \mu}$. Applying (1) to these, we get
immediately
$$
\lambdav_{\beta \alpha} = \tfrac{1}{2} {\rm Tr}\, \sigma_{\beta} A
\sigma_{\alpha} A^{\dag}.
$$
The inverse is slightly more difficult to obtain. We will
consider separately accelerations $L (v)$ such that
$$L(v) n_t = v ; \ n_{t\mu} = \delta_{\mu 0} ,$$
and rotations, $R$. For the first, and because $\tilde{n}_t = 1$,
(1) gives
$$A (L (v) ) A^{\dag} (L (v) ) = \tilde{v} ,$$
with solution
$$
A (L (v)) = +\tilde{v}^{1/2} .
$$
Note that $\tilde{v} = L(v) n_t$ is positive definite. We choose the sign (+) for 
the square root for continuity.
For a pure boost, $A (L (v))^{\dag} = A (L(v) )$.
\smallskip
\sport{Prove this.}
\smallskip

For rotations, $R$, we have $R n_t = n_t$; hence (1) gives
$$A(R) A^{\dag} (R) = 1 ,$$
i.e., $A$ is {\sl unitary}. Let $\ybf{\theta}$ be the parameters
of $R$. For $\ybf{\theta}$ infinitesimal, and $v_0 = 0$,
$$\tilde{v} \equiv \ybf{\sigma}{\bf v} \to \ybf{\sigma}{\bf v} +
\sum \sigma_j \theta_k v_l \epsilon_{jkl} . 
$$
If we write
$$A (R) = \exp \ii \ybf{\theta} \ybf{\lambda} \simeq
1+i\ybf{\theta} \ybf{\lambda} ,
$$
we then get, from (1),
$$(1+\ii\ybf{\theta} \ybf{\lambda} ) \ybf{\sigma} {\bf v} ( 1 -
i \ybf{\theta} \ybf{\lambda} ) \simeq \ybf{\sigma} {\bf v} + \sum
\epsilon_{jkl} \sigma_j \theta_k v_l ,
$$
from which
$$[ \lambda_j , \sigma_k ] = -\ii \sum \epsilon_{jkl} \sigma_l ,
$$
and hence $\ybf{\lambda} = - \ybf{\sigma} /2$:
$$
A(R(\ybf{\theta})) = \exp\frac{-\ii}{2} \ybf{\theta}
\ybf{\sigma}.
\equn{(2)}
$$

If the four-vector $v$ is such that $v^2 = 1$, $v_0 > 0$, we
define $\ybf{\xi}$ by
$$\cosh \xi = v_0 , \ \sinh \xi = | {\bf v} | , \ \ybf{\xi} /
| \ybf{\xi} | = {\bf v} / | {\bf v} | . 
$$
Then,
$$\tilde{v}^{1/2} = \cosh \frac{\xi}{2} + \frac{1}{\xi}
\ybf{\xi} \ybf{\sigma} \sinh \frac{\xi}{2} = {\rm exp} \
\frac{1}{2} \ybf{\xi} \ybf{\sigma} ,$$
so that
$$
A(L(v)) = \exp\tfrac{1}{2}\ybf{\xi}\ybf{\sigma} .
\equn{(3)}$$
\smallskip
\sport{Prove that $\det A (L(v)) = \det A (R (\ybf{\theta} )) = 1$. Prove that the set
$A (L (v) ) A (R (\ybf{\theta}))$ exhausts the  group SL (2,C).
{\rm Hint.} Use the polar decomposition: any matrix $A$ may be written as
$$
A = H U
$$
with $H$ positive definite and $U$ unitary. If $\det A = 1$, $\det H$, $\det
U$ can also be taken to be so. Check that any such $H$ may be written as
(3), and any such $U$ as in (2).}

We next find the images of the little groups in SL(2,C). 
For the timelike case, this is accomplished by 
choosing the vector $n_t$, with $n_{t\mu}=\delta_{\mu0}$. 
Then, $\tilde{n}_t=1$ and the image $U$ of a rotation $R$ has to verify 
$UU^{\dag}=1$, i.e., the image of the SO(3) subgroup of $\cal L$ is the 
SU(2) subgroup of SL(2,C).

For the case of lightlike vectors, we choose $n=n_t+n^{(3)}$ with  $n_t$ as before and
$n^{(3)}_\mu=\delta_{\mu3}$. Then
$$\widetilde{n}=1+\sigma_3=\pmatrix{2&0\cr0&0}.$$
If $N$ is the image in SL(2,C) of the little group transformation $\gammav$, $\gammav n=n$,
 then it must satisfy the conditions
$$N\pmatrix{2&0\cr0&0}N^{\dag}=\pmatrix{2&0\cr0&0},\quad \det N=1$$
from which it follows that one can write
$$N=\pmatrix{\ee^{\ii \theta/2}&\ee^{-\ii \theta/2}(a+\ii b)\cr
0&\ee^{-\ii \theta/2}\cr}.$$
\smallskip
\sport{Find the image in SL(2,C) of the little group of a spacelike vector.} 
\smallskip
\noindent  {\sl ii. Connection with the Dirac formalism}
\smallskip
\noindent
Let us use the notation
$$
D^{(1/2)}_{\alpha \beta} (\lambdav ) \equiv A_{\alpha \beta}
(\lambdav ) ,
$$
$$
\tilde{D}^{(1/2)}_{\dot{\alpha} \dot{\beta}} (\lambdav ) \equiv
( A^{-1+} (\lambdav ) )_{\dot{\alpha} \dot{\beta}} .
$$
We also define
$$\eqalign{
\hat{v} \equiv v_0 - \ybf{\sigma} {\bf v} = \sigma
\cdot v ,  \cr
\hat{v}_{\lambdav} \equiv \sigma \cdot \lambdav v .
\cr}
$$
One may check by explicit verification that
$$
A^{-1+} \hat{v} A^{-1} = \hat{v}_{\lambdav} ,
\equn{(4)}$$
a formula which is the counterpart of (1) and which indeed
provides another representation of $\cal L$ into SL(2,C),
{\sl inequivalent} to that given by (1). 
(It is actually equivalent to the representation $\lambdav\to A^*$.)
\smallskip
\sport{prove that the representations $\lambdav\to A$ and $\lambdav\to (A^{\rm T})^{-1}$ 
are equivalent.\quad Hint: the matrix that does it is $C=\ii \sigma_2$.}
\smallskip 

We link this to the standard Dirac formalism  by noting that, in the Weyl realization 
of the gamma matrices,
$$\gamma_\mu=\pmatrix{0&\tilde{\sigma}_\mu\cr \sigma_\mu&0\cr},\quad \sigma_0=1
$$
one has
$$
\gamma \cdot v = \pmatrix{0 & \tilde{v} \cr
\hat{v} & 0}.
$$
We then define
$$\eqalign{
D (\lambdav ) = \pmatrix{D^{(1/2)}
(\lambdav )   & 0 \cr 0 & \tilde{D}^{(1/2)} (\lambdav)}  \cr
= \pmatrix{A_{\alpha \beta} (\lambdav )  & 0 \cr 0 &
(A^{-1+} (\lambdav ) )_{\dot{\alpha} \dot{\beta}}
}}.
$$
As an application we prove the transformation properties of the Dirac 
$\gamma$ matrices. In the Weyl realization, and
for an arbitrary four-vector $v$,
$$\eqalign{
D^{-1} (\lambdav ) \gamma \cdot v D (\lambdav ) = 
\pmatrix{A^{-1} & 0 \cr 0 &
A^{\dag}}  
\pmatrix{0 & \tilde{v} \cr \hat{v} & 0} 
\pmatrix{A & 0 \cr 0 & A^{-1\dag}} \cr
= \pmatrix{0 & A^{-1} \hat{v} A^{-1\dag} \cr
A^{\dag} \hat{v} A  & 0 }=
\pmatrix{0 & \tilde{\sigma} \cdot
\lambdav^{-1} v  \cr \sigma \cdot \lambdav^{-1} v 
& 0} \cr
= \pmatrix{0 & (\lambdav
\tilde{\sigma}) \cdot v \cr (\lambdav \sigma )
\cdot
v  & 0} = (\lambdav \gamma )
\cdot \sigma , \cr}
$$
and we have used (1), (4). Because $v$ is arbitrary, this gives
$$D^{-1} (\lambdav ) \gamma_{\mu} D (\lambdav ) = \sum
\lambdav_{\mu \nu} \gamma_{\nu} .$$

The similitude with the treatment of the group SO(4) in \sect~3.2 will be noted. 
In fact, the groups SO(4)  and $\cal L$ can be related one to the other 
through analytical continuation on the variable $v_0$ and the complexification of 
their Lie algebras coincide.
We will not delve into this question further.
\medskip\noindent
{\sl iii. The finite-dimensional representations of {\rm SL(2,C)}}
\medskip\noindent
The finite dimensional representations of  SL(2,C) are very easy to construct. Denoting by 
${\bf M}_2$ to the Lie algebra of  SL(2,C), it is easily seen to consist of 
$2\times2$ complex traceless matrices. It is obvious that, if we complexify the 
${\bf A}_1$ algebra corresponding to the SU(2) subgroup of 
 SL(2,C), it generates all of ${\bf M}_2$:   ${\bf A}_1^{\bbbc}={\bf M}_2$.
Therefore, we may generate in this way the representations of the
 Lorentz group from those of the rotation group. 
In particular, it follows that the Clebsch--Gordan coefficients of 
SU(2) and  SL(2,C) are the same. 
Thus, we may, by simple tensor product
$$A_{\alpha_1\beta_1}A_{\alpha_2\beta_2} \cdots A_{\alpha_j\beta_j}$$
construct a representation of   SL(2,C) which, when restricted to the rotation subgroup, 
corresponds to spin $j/2$.

More on the matters treated in this section may be found in
Bogoliubov, Logunov and Todorov (1975) or Wightman (1960).

\booksection{\S7. General Description of Relativistic States}
\vskip-0.5truecm
\booksubsection{7.1. Preliminaries}

It is in many applications convenient to introduce an
abstract characterization of relativistic states, freeing it from the
problems encountered in explicit realizations. We will thus describe the
states by ``safe'' observables: momentum ${\bf p}$ and another one that
we label $\zeta$ and that will be related to a spin component: our task
will then be to construct the states, $| {\bf p} , \zeta \rangle$, and
study their transformation properties under relativistic
transformations. This we will do from the next section onwards; in what
remains of the present section we will introduce some standard theorems on
group representations, without proofs, and, at the end, describe the
group of relativistic transformations, the Poincar\'e group.

The invariance group of relativity is the Poincar\'e group, also called
the inhomogeneous Lorentz group. Its elements are pairs $(a, \lambdav )$ with
$a$ a four-translation consisting of a spatial translation by ${\bf a}$,
and a time translation by $a_0/c$; and a (proper, orthochronous) Lorentz
transformation, $\lambdav$. 
The generators of the Poincar\'e group may be described as generators of
rotations, boosts and translations. Let us consider any representation,
$U (a, \lambdav )$ of the Poincar\'e group; then, for infinitesimal
transformations we write
$$
\eqalign{ U(0, R({\ybf {\theta}})) \simeq 1 - \frac{\ii}{\hbar}
{\ybf {\theta}} {\bf L} , \cr
U(0, L({\ybf{\xi}})) \simeq 1 - \dfrac{\ii}{\hbar} {\ybf{\xi}}
{\bf N} , \cr
U(a,1) \simeq 1 + \dfrac{\ii}{\hbar} a \cdot P .}
$$
The commutation relations may be evaluated in any (faithful)
representation; indeed, since these respect product and inverse rules,
commutators will also be respected. We may then choose the regular representation 
with the $U$  acting on scalar functions of $a,\lambdav$.
We can then take
$$ \eqalign{L_j = \ii \hbar \sum \epsilon_{jkl} x_k \partial_l ,
\cr
N_j = \ii \hbar (x_0 \partial_j - x_j \partial_0 ) ,\cr
P_j = \ii \hbar \partial_j , \ P_0 = \ii \hbar \partial_0 }
$$
and evaluate the commutators with these explicit expressions. That way
we find the relations, valid in any representation,
$$
\eqalign{
[ L_k , L_j ] = &\,\ii \hbar \sum \epsilon_{kjl} L_l ,
\cr
[ L_k , N_j ] = &\,\ii \hbar \sum \epsilon_{kjl} N_l ,
\cr
[ L_k , P_j ] = &\,\ii \hbar \sum \epsilon_{kjl} P_l ; 
\cr
[ L_k , P_0 ] = &0, \quad [ P_{\mu} , P_{\nu} ] = 0 ; 
\cr
[ N_k , N_j ] = &\,- \ii \hbar \sum \epsilon_{kjl} L_l , 
\cr
[ N_k , P_j ] =&\, - \ii \hbar \delta_{kj} P_0 , \cr
[ N_k , P_0 ] =&\, - \ii \hbar P_k . \cr
}
$$
We may also write them in covariant form. If we let
$$
U(\lambdav ) \simeq 1 - \frac{i}{\hbar} \omega^{\mu \nu} M_{\mu \nu} ,
$$
then a simple calculation, making use of the fact that
$$ [ \partial_{\mu} , x_{\nu} ] = g_{\mu \nu} $$
allows us to write the commutation relations in the form
$$\eqalign{
[ M_{\mu \nu} , P_{\alpha} ] = & \ii \hbar (g_{\nu \alpha} P_{\mu} -
g_{\mu \alpha} P_{\nu} ) , \cr
[ M_{\mu \nu} , M_{\alpha \beta} ] = & \ii \hbar (g_{\mu \alpha} M_{\beta
\nu} + g_{\mu \beta} M_{\nu \alpha} \cr
  & + g_{\nu \alpha} M_{\mu \beta} + g_{\nu \beta}
M_{\alpha \mu} ),\cr 
[ P_{\mu} , P_{\nu} ]&\,= 0.\cr
}
$$

Consider now a quantum system represented by the state $| \Psi \rangle$.
A Poincar\'e transformation $g$ will carry it over a new state, $| \Psi_g
\rangle$. According to the rules of quantum mechanics, we expect that
this will be implemented by a linear unitary operator,
$$\eqalign{
U(g) = U (a, \lambdav ) : \cr
| \Psi_g \rangle = U (a, \lambdav ) | \Psi \rangle .
}$$
We will require that this be a representation of the Poincar\'e group.
Actually, this is asking for too much; in principle, one could have,
more generally, a representation up to a phase:
$$ U(a, \lambdav ) U (a' , \lambdav ') = e^{i \varphi} U(a+\lambdav a' ,
\lambdav \lambdav ') . $$
In the following sections we will give an {\sl explicit} construction with
$\varphi = 0$; the proof that the result is general is fairly
complicated and will not be given here (see Wigner, 1939).

We will then consider unitary representations of the Poincar\'e group.
Since a {\sl reducible} representation can be decomposed into orthogonal
{\sl irreducible} ones, we need only consider the latter, which may be
identified as those describing elementary systems that we will call {\sl
particles}. Note that here ``elementarity'' is not used in a dynamical
sense; it only means that the corresponding {\sl isolated} system cannot
be described as two or more systems, {\sl also isolated}\fnote{Our treatment will not
be mathematically rigorous. Mathematical rigour can be provided by
consulting the treatises of Bogoliubov, Logunov and Todorov (1975) or
Wightman (1960). The problem of giving the general description of
relativistically invariant systems was first fully solved by Wigner
(1939), whose paper we will essentially follow.}.

\booksubsection{7.2. Relativistic one-particle states: general description}

Let us denote by ${\goth H}$ the Hilbert space for {\sl free}
one-particle states. We will construct a basis of ${\goth H}$, working in
the Heisenberg picture, the simplest one to use for our analysis.

Consider the operators that represent translations, $U(a,1) \equiv
U(a)$. If we write them in exponential form,
$$
U(a) = \exp \ii a \cdot P ,
$$
then unitarity of $U$ implies Hermiticity of the $P_{\mu}$. We will
identify $P_0$ with the energy\fnote{Unless otherwise explicitly stated, 
we will use natural units with
$\hbar = c = 1$.} operator (the Hamiltonian), and ${\bf P}$ the
ordinary momentum operator; the four $P_{\mu}$ form the four-momentum
operator.

From the commutation relations,  it follows that the
operator $P^2 =P \cdot P$ commutes with all the generators of the
Poincar\'e group, and hence also with all the $U(a, \lambdav )$. Schur's lemma
then implies that it is a constant, which we identify with the square of
the mass (which can be zero):
$$
m^2 = P \cdot P .
$$
Because of this, it follows that, for free particles, the operator $P_0$
is actually a function of the ${\bf P}$:
$$
P_0 = + (m^2 + {\bf P}^{\,2} )^{1/2} ,
$$
where we have chosen the positive square root to get positive energies. If
${\bf p}$ are the eigenvalues of the ${\bf P}$, and $p_0$ those of
$P_0$, we thus have
$$
p_0 = + \sqrt{m^2 + {\bf p}^2} ,
$$
as was to be expected for a relativistic particle.

As we know, the $P_{\mu}$ commute among themselves. We can then
diagonalize them simultaneously, and consider the corresponding
eigenvectors as the desired base of ${\goth H}$, which we denote by $| p,
\zeta \rangle$, with $\zeta$ being whatever extra quantum numbers necessary to
specify the states; as we will see, the $\zeta$ will be essentially a
spin component. Note that the notation $| p, \zeta \rangle$, although
convenient, is redundant; we could also write $|p, \zeta \rangle = |
{\bf p} , \zeta \rangle$, since $p_0$ is fixed  once ${\bf p}$
is given.

Because $|p, \zeta \rangle$ are eigensates of the $P_{\mu}$, we have
$$
P_{\mu} |p, \zeta \rangle = p_{\mu} | p, \zeta \rangle ,
$$
and, exponentiating, and writing $U(a)$ for $U(a,1)$,
$$
U(a) | p, \zeta \rangle = \ee^{\ii a \cdot P} | p , \zeta \rangle = \ee^{\ii a
\cdot p} | p, \zeta \rangle .
$$
Let us select a fixed momentum, $\overline{p}$, with $\overline{p} \cdot
\overline{p} = m^2$, $\overline{p}_0 > 0$. This means that we are
choosing a fixed reference system. Any admissible four-vector for the
particle, $p$, may be written as
$$ p = \lambdav (p) \overline{p} ,$$
where $\lambdav (p)$ is a (not unique) Lorentz transformation. We then {\sl choose} a
family of such Lorentz transformations, $\lambdav (p)$, one for each $p$.
The basis we will find will depend on the family of $\lambdav (p)$ we
choose; but the choice will be left unspecified for the moment. Then, we
{\sl define} the basis $| \lambdav (p) , \zeta \rangle$ by\fnote{The 
notation $|\lambdav(p),\zeta\rangle$ is  shorthand. 
A more precise notation for this state would be  $|p,\zeta; \lambdav(p)\rangle$, 
i.e., a state with momentum $p$, other quantum number $\zeta$, and obtained with 
the Lorentz transformation $\lambdav(p)$. Our notation is simpler and, 
hopefully, transparent enough.}
$$
| \lambdav (p) , \zeta \rangle \equiv U (\lambdav (p) ) | \overline{p} ,
\zeta \rangle ,
$$
i.e., by accelerating via $\lambdav (p)$ to momentum $p$; to simplify 
the notation, we write $U(\lambdav )$ for $U(0, \lambdav )$.

Let us first prove that the state $|\lambdav (p), \zeta \rangle$
corresponds to four-momentum $p$. To see this, we evaluate
$$ U(a) | \lambdav (p) , \zeta \rangle = U(a) U(\lambdav (p) ) |
\overline{p} , \zeta \rangle . $$
Using the identity
$$ U(a) U(\lambdav (p)) = U (a, \lambdav (p) ) = U (\lambdav (p) ) U
(\lambdav (p)^{-1} a), $$
we obtain
$$ U(a) | \lambdav (p) , \zeta \rangle = U(\lambdav (p) ) U(\lambdav
(p)^{-1} a) | \overline{p} , \zeta \rangle . $$
Taking into account that
$$ (\lambdav (p)^{-1} a) \cdot \overline{p} = a \cdot \lambdav (p)
\overline{p} = a \cdot p , $$
we get
$$ \eqalign{
U(\lambdav (p) ) U(\lambdav (p)^{-1} a) | \overline{p} , \zeta \rangle
\cr
= U ( \lambdav (p) ) \ee^{\ii (\lambdav (p)^{-1} a) \cdot \overline{p}} |
\overline{p} , \zeta \rangle \cr
= \ee^{\ii p \cdot a} U ( \lambdav (p) ) | \overline{p} , \zeta \rangle
\cr
= \ee^{\ii p \cdot a} | \lambdav (p) , \zeta \rangle .
\cr} $$
We have thus shown that

$$
U(a) | \lambdav (p) , \zeta \rangle = \ee^{\ii a \cdot p} | \lambdav (p) ,
\zeta \rangle ,
$$
and (for example, by differentiating with respect to $a_{\mu}$ at $a=0$)
that
$|\lambdav (p) , \zeta \rangle$ is a state with momentum $p$, as claimed
above:
$$
P_{\mu} | \lambdav (p) , \zeta \rangle = p_{\mu} | \lambdav (p) , \zeta
\rangle .
$$

These equation  tell us how the translations act upon our basis of state
vectors, $| \lambdav (p) , \zeta \rangle$. We will now deduce
corresponding formulas for Lorentz transformations. To do so, we start
by considering transformations, which we will denote by $\gammav , \gammav', \ldots$,
 contained in the little group of $\overline{p}$, ${\cal W}
(\overline{p} )$; and we will let
these
transformations act on $| \overline{p} , \zeta \rangle \equiv | \lambdav
(\overline{p} ) , \zeta \rangle$ itself. Because the $\gammav$ leave
$\overline{p}$ invariant, it follows that the state vector $U(\gammav ) |
\overline{p} , \zeta \rangle$ still corresponds to momentum
$\overline{p}$. Therefore, it will have to be a linear combination of
vectors $| \overline{p} , \zeta ' \rangle$:

$$
U (\gammav ) | \overline{p} , \zeta \rangle = \sum_{\zeta '} D_{\zeta '
\zeta} (\gammav ) | \overline{p} , \zeta ' \rangle ,
$$
where the $D_{\zeta ' \zeta}$ are certain coefficients. So, 
in the case of massive particles of 
spin 1/2, the parameter
$\zeta$ will, for example, represent the third component of spin. Thus, 
 we can have\fnote{In some cases it may
be convenient to label the matrix elements not with the indices $\pm 1/2$, but
with indices 1, 2. We thus identify
$$ \pmatrix{ D_{1/2, 1/2} & D_{1/2,-1/2} \cr D_{-1/2, 1/2}
& D_{-1/2, -1/2}}  \equiv \pmatrix{
D_{11} & D_{12} \cr D_{21} & D_{22}} $$ 
 that we
may take to be the components of a matrix $D$:
$$ D(\gammav )^{\rm T} = (D_{\zeta ' \zeta} (\gammav ) ), \ {\rm i.e.}, \
D(\gammav ) = (D_{\zeta \zeta '} (\gammav ) ) . $$} $\zeta = \pm 1/2$.
It is easy to verify that the conditions
$$ U(\gammav ) U (\gammav ') = U (\gammav \gammav ' ), \ U(\gammav^{-1} ) =
U^{-1} (\gammav ) , \ U^{\dag} (\gammav ) = U^{-1} ( \gammav ) $$
imply that
$$
\eqalign{
D(\gammav ) D(\gammav ') = D(\gammav \gammav ') , \cr
D(\gammav^{-1} ) = D (\gammav )^{-1} , \cr
D^{\dag} (\gammav ) = D( \gammav )^{-1} ;
\cr}
$$
it follows that the matrices $D$ build up a {\sl unitary representation
of the little group}, ${\cal W} (\overline{p} )$. From the
``elementarity'' of the system, that is to say, from the fact that $U(a,
\lambdav )$ is irreducible, we can deduce that the representation $D$ {\sl
must also be irreducible}.
\smallskip
\sport{Prove this.}
\smallskip
    
The specific form of the $D$ will be given in the next two sections. For
the moment we will assume that we have such a representation, so that we
know the values of the coefficients $D_{\zeta ' \zeta} (\gammav )$; with
their help we will be able to solve in full generality the problem of
finding how arbitrary Lorentz transformations act. In fact, we have,

$$
\eqalign{
U(\lambdav) | \lambdav (p) , \zeta \rangle = U (\lambdav ) U(\lambdav (p)) |
\overline{p} , \zeta \rangle \cr
= U(\lambdav (\lambdav p)) U( \lambdav (\lambdav p))^{-1} U( \lambdav \lambdav
(p)) | \overline{p} , \zeta \rangle \cr
= U (\lambdav (\lambdav p)) U(\lambdav (\lambdav p)^{-1} \lambdav \lambdav
(p)) | \overline{p} , \zeta \rangle ,
\cr}
$$
where $\lambdav (\lambdav p) \overline{p} = \lambdav p$, and we have
introduced a term $U (\lambdav (\lambdav p)) U(\lambdav (\lambdav p))^{-1} =
1$ and used the group properties of the $U$. Now,
$$ (\lambdav (\lambdav p ))^{-1} \lambdav \lambdav (p) \overline{p} =
(\lambdav ( \lambdav p))^{-1} \lambdav p = \overline{p} , $$
so that the transformation $(\lambdav (\lambdav p))^{-1} \lambdav \lambdav
(p)$, which we will write as $\gammav (p, \lambdav )$, is in ${\cal W}
(\overline{p} )$, since it leaves $\overline{p}$ invariant. We  thus
find
$$
U ( \gammav (p, \lambdav )) | \overline{p} , \zeta \rangle = \sum_{\zeta
'} D_{\zeta ' \zeta} (\gammav (p, \lambdav )) |\overline{p} , \zeta '
\rangle ;
$$
substituting this  we get the explicit formula

$$
\eqalign{
U(\lambdav ) | \lambdav (p) , \zeta \rangle = \sum_{\zeta '} D_{\zeta '
\zeta} (\gammav (p, \lambdav )) | \lambdav (\lambdav p) , \zeta ' \rangle ,
\cr
\gammav (p, \lambdav ) \equiv (\lambdav (\lambdav p))^{-1} \lambdav \lambdav
(p) .
}
$$

Besides choosing the family of $\lambdav (p)$, and finding the explicit
values of the $D_{\zeta ' \zeta}$, the only thing that we need to have
the problem totally solved is to find the normalization of the states
$|\lambdav (p) , \zeta \rangle$ such that relativistic transformations
leave it invariant, i.e., such that the $U(a, \lambdav )$ are unitary. 

The
$U(a)$ are unitary by construction. If we assume the $\zeta$ to be
eigenvalues of an observable, we will have

$$
\langle \lambdav (p) , \zeta | \lambdav (p') , \zeta ' \rangle = N(p)
\delta ({\bf p} - {\bf p} \ ') \delta_{\zeta \zeta '} ,
$$
where $N$ is a factor to be determined by the requirement that, for any
$\lambdav$,
$$
\eqalign{
\langle U(\lambdav ) (\lambdav (p) , \zeta ) | U(\lambdav ) (\lambdav (p') ,
\zeta ' ) \rangle \cr
= \langle \lambdav (p) , \zeta | \lambdav (p') , \zeta ' \rangle
}
$$
({\sl unitarity}). Substituting  and recalling
that the matrix $D=(D_{\zeta ' \zeta} )$ is unitary, we find the
condition

$$
N(\lambdav p) \delta (\lambdav {\bf p} - \lambdav {\bf p} \ ') = N(p)
\delta ({\bf p} - {\bf p} \ ') .
$$
If $\lambdav$ is a rotation $R$, and since $\delta (R{\bf p}) = \delta
({\bf p})$, it follows that $N$ can only depend on $|{\bf p} |$, or,
equivalently, on $p_0$, $N=N(p_0)$. Considering next a boost along {\sl
OZ}, $L_z$, with parameter $\xi$,
$$  \eqalign{
L_z : & p_0 \to (\cosh \xi ) p_0 + (\sinh \xi ) p_3 , \cr
      & p_3 \to (\cosh \xi ) p_3 + (\sinh \xi ) p_0 , \cr
      & p_1 \to p_1 , \ p_2 \to p_2:
} $$
we find
$$ N ( (\cosh \xi ) p_0 ) \frac{1}{(\cosh \xi ) p_0} \delta ({\bf p} -
{\bf p} \ ') = N (p_0 ) \delta ({\bf p} - {\bf p} \ ') , $$
for any $\xi$, so that we get $N (p_0 ) = {\rm constant}\times p_0$. We will
follow custom in choosing this constant equal to 2, so the
invariant form of the scalar product is finally
$$
\langle \lambdav (p) , \zeta | \lambdav (p') , \zeta ' \rangle = 2 p_0
\delta ({\bf p} - {\bf p} \ ') \delta_{\zeta \zeta '} , \quad
p_0 = + \sqrt{m^2 + {\bf p}^{2}}.
$$

Before moving on to the detailed analysis of the various different
cases, a few more words on general matters are in order. First of all we
again remark that the analysis of this section is valid for massive as
well as massless particles; for the latter it is sufficient to set $m=0$
in the appropriate formulas. Secondly, it may appear that our analysis
is dependent on the fixed vector (or reference system) $\overline{p}$,
from which we build the basis. This is not so; because the little groups
of two $\overline{p}, \overline{p} \ '$ are isomorphic, it follows that
substituting $\overline{p} \ '$ for $\overline{p}$ merely result in a
change of basis in ${\goth H}$. The same is true if we replace the family
$\lambdav (p)$ by another family, $\lambdav '(p)$.

\smallskip
\sport{Find the operators that implement the changes of basis (A) when replacing
$\overline{p}$ by $\overline{p} \ '$, and (B) when replacing $\lambdav (p)$ by
$\lambdav '(p)$.}

\smallskip

\sport{Suppose that, for a particle, there existed a
state $| \overline{p}_{\bot} \rangle$ {\sl different} from all the $p =
\lambdav
\overline{p}$. Prove then that $\langle \overline{p}_{\bot} | \lambdav
\overline{p} \rangle = 0$ for all $\lambdav$, and that the representation
turns out to be reducible.}
\smallskip
Finally, the analysis of this section may appear excessively abstract to the
reader. This could be overcome by returning to it {\sl after} having
gone over the next two sections.

\booksubsection{7.3. Relativistic states of massive  particles}

The idea behind Wigner's method is actually very simple, at least for
particles with mass. In this case, one chooses a reference system with
$\overline{p}_0 =m$, $\overline{p}_i =0$, that is to say, the
reference system in which the particle is at rest. Here, nonrelativistic
quantum mechanics is manifestly valid, which suggests to us that we take the
quantum numbers $\zeta$ to be the values of the third component of spin.
In this case, we will use the label $\lambda$ instead of $\zeta$. We
thus start by considering the states at rest,
$$ | \overline{p} , \lambda \rangle . $$

The little group of $\overline{p}$ consists of ordinary three-dimensional
rotations, which we denote by $R$ rather than $\gammav$. The matrices
$D(R)$ are just the standard $D^{(s)} (R({\ybf {\theta}}))$, for a particle
with total spin $s$. They are
$$ D^{(s)} (R ({\ybf {\theta}})) = \exp \frac{-\ii}{\hbar} {\ybf {\theta}}
{\bf S} , $$
where ${\bf S}$ are the familiar spin operators. For $s=1/2$,
$$ D^{(1/2)} (R({\ybf {\theta}})) = \ee^{-\ii \ybf{\sigma} {\ybf {\theta}}/2} . $$
For arbitrary $s$, the values of the matrix elements $D_{\lambda \lambda'}^{(s)} (R)$
 of $D^{(s)}$ can be found in Wigner~(1959). We then have
$$
U(R) | \overline{p} , \lambda \rangle = \sum_{\lambda'} D_{\lambda'
\lambda}^{(s)} (R) | \overline{p} , \lambda' \rangle .
$$

For states in an arbitrary reference system, with momentum $p$, we may
boost by a $L(p)$ 
 such that $L(p)\overline{p}=p$.

Then the states $| L(p), \lambda \rangle$ are {\sl defined} as
$$
| L(p) , \lambda \rangle \equiv U (L (p)) | \overline{p} , \lambda
\rangle ,
$$
and we normalize them to
$$
\langle L(p) , \lambda | L(p') , \lambda' \rangle = 2 p_0 \delta
({\bf p} - {\bf p} \ ') \delta_{\lambda \lambda'} .
$$
To find the transformation properties of the $| L (p) , \lambda \rangle$
under an arbitrary Lorentz transformation $\lambdav$, we proceed as
follows: $\lambdav$ will carry $p$ over $\lambdav p$. Therefore we (a) go
to the reference system where the particle is at rest decelerating by
$L^{-1} (p)$, (b) see how the state transforms there and (c) boost now by
$L(\lambdav p)$. In formulas,
$$ \eqalign{
U(\lambdav ) | L(p) , \lambda \rangle = U (\lambdav ) U (L (p)) |
\overline{p} , \lambda \rangle \cr
= U (L (\lambdav p)) U(L (\lambdav p)^{-1} ) U(\lambdav ) U (L (p)) |
\overline{p} , \lambda \rangle \cr
= U (L (\lambdav p)) U(R (p, \lambda )) | \overline{p} , \lambda \rangle
,} $$
where
$$ R(p, \lambdav ) = L(\lambdav p)^{-1} \lambdav L(p) $$
is called a {\sl Wigner rotation}; it {\sl is} a rotation since $R (p,
\lambdav ) \overline{p} = \overline{p}$. We obtain the result
$$ \eqalign{
U(\lambdav ) | L(p) , \lambda \rangle = U (L( \lambdav p)) U(R(p, \lambdav
)) | \overline{p} , \lambda \rangle \cr
= U (L(\lambdav p)) \sum_{\lambda '} D^{(s)}_{\lambda' \lambda} (R(p,
\lambdav )) | \overline{p} , \lambda' \rangle \cr
= \sum_{\lambda'} D^{(s)}_{\lambda' \lambda} (R(p, \lambdav )) |
L(\lambdav p) , \lambda' \rangle ,
} $$
so that
$$
\eqalign{
U(\lambdav ) | \lambdav (p) , \lambda \rangle = \sum_{\lambda'}
D^{(s)}_{\lambda' \lambda} (R (p, \lambdav )) | L(\lambdav p) , \lambda'
\rangle , \cr
R(p, \lambdav ) = L(\lambdav p)^{-1} \lambdav L (p) .
}
$$

Of course, we have already seen this in the previous section. The basis
$|L (p) , \lambda \rangle$ is sometimes called the {\sl covariant spin}
basis. Another useful basis is the {\sl helicity} basis. To build it, we
choose, instead of pure boosts $L(p)$, the transformations $H(p)$
defined as follows: first, take a pure boost $L(p^z)$ that carries
$\overline{p}$ over $p^z$ with $p_0^z=p_0$, $p_1^z=p_2^z=0$,
$p_3^z=p_3$. Then, let $R({\bf  z} \to {\bf p})$ be a rotation
around the axis ${\bf  z} \times {\bf p}$ that carries the {\sl OZ} axis
over ${\bf p}$. We define
$$ H(p) \equiv R ({\bf  z} \to {\bf p} ) L(p^z) , \ | H(p) , \
\eta= \zeta \rangle = U (H(p)) | \overline{p} , \zeta \rangle . $$

\noindent The corresponding states $|H (p), \ \eta = \zeta \rangle$ are the
helicity states, since $\eta$ is the projection of the spin on the vector
${\bf p}$.

The analysis is fairly straightforward for massive particles. The reason
why we gave the general discussion of the previous section is its usefulness
in studying the case of massless particles.

The nonrelativistic limit is obtained when $|{\bf p}| \ll m$, so that
$p_0 \simeq m$. The normalization becomes (taking the covariant spin
case for definiteness)
$$ \langle L(p) , \lambda | L(p') , \lambda' \rangle \simeqsub_{NR} 2m
\delta_{\lambda \lambda'} \delta ({\bf p} - {\bf p} \ ') , $$
so that
$$
|L(p) , \lambda \rangle = \sqrt{2p_0} | {\bf p} , \lambda
\rangle_{NR}
\simeqsub_{NR} \sqrt{2m} | {\bf p} , \lambda
\rangle_{NR} ,\quad  _{NR} \langle {\bf p} , \lambda | {\bf p}', \lambda' \rangle_{NR} 
= \delta_{\lambda \lambda'} \delta ({\bf p} - {\bf p}') .
$$
Because of this some authors define
$$ | L(p) , \lambda \rangle_{\rm I} =
\frac{1}{\sqrt{2m}} | L(p) , \lambda \rangle , $$
or
$$  | L(p) , \lambda \rangle_{\rm II} =
\dfrac{1}{\sqrt{2p_0}} | L(p) ,\lambda \rangle .  $$
Here we will stick to our conventions. Choice I presents the
problem of collapsing for massless particles; choice II is not
relativistically invariant. Our choice is valid for massless as
well as massive particles, and is relativistically invariant; the price
to pay is a factor $\sqrt{2p_0}$ between relativistic and NR 
normalization, a price that is quite justified.

Next we turn to the discrete symmetries ${\cal C}$, $\cal P$, $\cal T$.
$\cal C$ is defined trivially by setting
$$ {\cal C} | p, \lambda \rangle \equiv \eta_C | \overline{p, \lambda}
\rangle , $$
where $| \overline{p, \lambda} \rangle$ denotes the state of an
antiparticle with the same momentum $p$ and spin $\lambda$ as the
particle $| p, \lambda \rangle$. $\cal P$ and $\cal T$ are not given by
the previous analysis; but we can use the same method, with slight
modifications. Beginning with parity, we define the operator $\cal P$ by
considering that it is the representative of space reversal, $I_s$,
$(I_s x)_{\mu} = g_{\mu \mu} x_{\mu}$: ${\cal P} = U(I_s)$. We then
write
$$ \eqalign{
{\cal P} | L (p) , \lambda \rangle = U(I_s ) U(L (p)) | \overline{p} ,
\lambda \rangle \cr
= U(L (I_s p)) U(L(I_s p)^{-1} I_s L(p)) | \overline{p} , \lambda
\rangle .
} $$
Now, $L(I_s p)^{-1} I_s L(p)$ leaves $\overline{p}$ invariant. It is
{\sl not} a rotation, because its determinant is ($-1$); but then

$$
R(p, I_s) \equiv L(I_s p)^{-1} I_s L(p) I_s
$$
{\sl is} a rotation. In the nonrelativistic case,
$$ {\cal P} | \overline{p} , \lambda \rangle = \eta_P | \overline{p} ,
\lambda \rangle , $$
so that, finally,
$$
{\cal P} | L(p) , \lambda \rangle = \eta_P \sum_{\lambda'}
D^{(s)}_{\lambda' \lambda} (R(p, I_s )) | L(I_s p) , \lambda' \rangle .
$$

For time reversal we can repeat the analysis with the modifications due
to the {\sl antiunitary} character of $\cal T$. Using that
$$ {\cal T} P_{\mu} {\cal T}^{-1} = (I_s P)_{\mu} , $$
we find that

$$
{\cal T} | L(p) , \lambda \rangle = \eta_T \sum_{\lambda'}
D^{(s)}_{\lambda', -\lambda} (R(p, I_s)) (-\ii)^{2\lambda} | L(I_s p),
\lambda' \rangle .
$$
\smallskip
\sport{Evaluate ${\cal P} | H(p) , \zeta \rangle , {\cal T} | H(p), \zeta \rangle$.}

\booksubsection{7.4. Massless particles}

This case is {\sl essentially} different from the previous one, not
merely the limit as $m \to 0$, something that could already have
been imagined from what one finds for massless particles with the wave
function formalism. To begin with, since a particle without mass cannot
be at rest, the choice of $\overline{p}$ is less helpful than before.
What we do is merely define our spatial axes so that
$\overline{{\bf p}}$ points in a convenient direction, say, along
{\sl OZ}: we thus take
$$ \overline{p}_1 = \overline{p}_2 = 0, \ \overline{p}_3 =
\overline{p}_0 . $$
The particular value of $\overline{p}_0$ is (for systems with a single
particle) irrelevant; we may get $\overline{p}_0 =1$ by a boost, or by just taking
$\overline{p}_0$ as the unit of energy.

Let us now consider the little group of this $\overline{p}$, ${\cal W}
(\overline{p})$. If $\gammav$ is in ${\cal W} (\overline{p})$, we can
represent it as before. We then decompose $\gammav$ as

$$
\gammav = \lambdav_t R_z (\theta ) ,
$$
where $R_z (\theta )$ is a rotation around {\sl OZ} by an angle $\theta$,
so that the corresponding matrix $(\gammav )$ is
$$ \eqalign{
(\gammav ) = \pmatrix{1 & 0 & 0 \cr 0 & 1 & 0 \cr \xi &
\eta & 1} \pmatrix{ \cos \theta &
\sin \theta & 0 \cr - \sin \theta & \cos \theta & 0 \cr 0 & 0 & 1
}, \cr
\gammav_{31} = \xi \cos \theta - \eta \sin \theta , \quad \gammav_{32} = \xi
\sin \theta + \eta \cos \theta .
} $$

The first term in the expression for $(\gammav)$, viz.,
$$ \pmatrix{ 1 & 0 & 0 \cr 0 & 1 & 0 \cr \xi & \eta & 1
}, $$
corresponds to $\lambdav_t$; the second one to $R_z (\theta )$. Because
the product of two transformations $\gammav_1$, $\gammav_2$ in ${\cal W}
(\overline{p})$ lies in ${\cal W} (\overline{p})$, it follows that we can
write

$$
\gammav_i = \lambdav_{it} R_z (\theta_i ) , \ i=1,2,
$$
and
$$
\gammav_1 \gammav_2 = \lambdav_{12t} R_z (\theta_{12}) ,
$$
where the angle $\theta_{12}$ will depend on $\gammav_1$, $\gammav_2$:
$$ \theta_{12} = \theta_{12} (\gammav_1 , \gammav_2 ) . $$
\smallskip
\sport{Prove that, with self-explanatory notation,
$$\eqalign{
\theta_{12}(\gammav_1,\gammav_2)=&\,\theta_1+\theta_2,\cr
\xi_{12}(\gammav_1,\gammav_2)=&\,\xi_1+(\cos\theta_1)\xi_2-(\sin\theta_1)\eta_2,
\quad
\eta_{12}(\gammav_1,\gammav_2)=\eta_1+(\cos\theta_1)\eta_2+(\sin\theta_1)\xi_2.\cr}
$$}
\smallskip

To get a representation of the Poincar\'e group we require a
representation of this little group, ${\cal W} (\overline{p})$. This
little group  is actually isomorphic to the
Euclidean group in two dimensions, and its representations can be
studied by the same methods we
are using to find the representations of the Poincar\'e group. The details
may be found in Wigner (1939)\fnote{Or in Wightman (1960), Bogoliubov,
Logunov and Todorov (1975).}; we
will take from there, and without proof, the following result. If we want to have particles
with {\sl discrete} spin values, then the representation must be of the form

$$D (\gammav ) = D (R_z(\theta)),
\equn{(1)}
$$
i.e., we must have
$$
D(\lambdav_t)\equiv 1 .
\equn{(2)}$$
Moreover, the representation $D(R_z (\theta ))$ can be at most
double-valued, so that
$$
D(R_z (2\pi )) = \pm 1 .
$$
This is because the covering group of the Lorentz group, SL(2,C), is simply connected 
and covers twice  $\cal L$.

There is no physical reason for excluding particles with continuous spins
(which have been studied by Wigner, 1963); but it is a fact that all
particles found in nature have discrete spin values. We will therefore
{\sl require} (2).

With the help of this the analysis is easily completed. The {\sl
irreducible} representations of the $R_z (\theta )$, rotations around a
fixed {\sl (OZ)} axis, are trivial. Since the group is Abelian, Schur's
lemma implies that these representations must be one-dimensional. From
this
it follows that the index $\lambda$ in the classification of the states,

$$ | \overline{p} , \lambda \rangle , $$
can only take  {\sl one} value. The matrices $D_{\lambda' \lambda}
(\gammav )$ are therefore just numbers, equal to $\delta_{\lambda
\lambda'} d_{\lambda} (\theta )$.
Because the representation has to be unitary,
these numbers are of modulus unity and we can write
$$ d_{\lambda} (\theta ) = e^{- \ii \lambda \theta} . $$
The fact that the representation is at most two-valued,
implies
that the number $\lambda$ is integer or half integer. Its interpretation
is readily accomplished by comparing the expression for $d(\theta )$
with that for a rotation around the {\sl OZ} axis in terms of the $S_z$
component of the spin operator,
$$ U(R_z (\theta ) ) = \ee^{-\ii \theta S_z / \hbar} : $$
$\lambda$ is the spin component along {\sl OZ} (or along
${\overline{\bf p}}$, since it coincides with the {\sl OZ} axis).
This is the {\sl helicity}. Because there is only one possible value of
$\lambda$, it follows that, {\sl for massless particles, the helicity is
relativistically invariant}, something that can be seen in specific cases
with the wave function formalism.

Once the transformation properties of the states $| \overline{p} ,
\lambda \rangle$ under the little group ${\cal W} (\overline{p})$,
$$
U (\gammav ) | \overline{p} , \lambda \rangle = e^{-\ii \lambda \theta
(\gammav )} | \overline{p} , \lambda \rangle ,
$$
are known, we have to specify the family of transformations $\lambdav (p)$ with
$\lambdav (p) \overline{p} = p$ to extend the analysis to arbitrary
transformations. Choose $\overline{p}_0 = 1$; for an arbitrary $p$ we
set
$$ \eqalign{
\lambdav (p) = H (p) , \cr
H(p) = R ({\bf  z} \to {\bf p} ) L (p^z).
} $$
$L(p^z)$ is the pure boost along {\sl OZ} such that
$$ \eqalign{
L(p^z ) \overline{p} = p^z , \cr
p_0^z = p_0 , \ p_1^z = p_2^z = 0, \ p_3^z = p_0 ;
} $$
$R ({\bf  z} \to {\bf p} )$ is the rotation around the axis
${\bf  z} \times {\bf p}$ that carries {\sl OZ} over ${\bf p}$. We then define
$$
| p, \lambda \rangle \equiv U(H (p)) | \overline{p} , \lambda \rangle ,
$$
and we find that
$$
U(\lambdav ) | p, \lambda \rangle = e^{-\ii \lambda \theta (p, \lambdav )} |
\lambdav p, \lambda \rangle ;
$$
the angle $\theta (p, \lambdav )$ is the angle of the {\sl OZ} rotation
contained in
$$ \gammav (p , \lambdav ) = H(\lambdav p)^{-1} \lambdav H (p) , $$
when we decompose it as 
$$ \gammav (p, \lambdav ) = \lambdav_t R_z (\theta (p, \lambdav )) . $$
The normalization is
$$ \langle p, \lambda | p, \lambda \rangle = 2 p_0 \delta ({\bf p} ,
{\bf p}' ) . $$

Next we consider the discrete symmetries ${\cal P}$, ${\cal T}$.
Starting with parity, the corresponding operator should satisfy
$$ \eqalign{
{\cal P} P_0 {\cal P}^{-1} = P_0 , &\quad {\cal P} {\bf P} {\cal P}^{-1} = -
{\bf P} , \cr
{\cal P} {\bf L} {\cal P}^{-1} = {\bf L} , &\quad {\cal P} {\bf S} {\cal
P}^{-1} = {\bf S} ;
} $$
from this, and for the helicity operator
$$ S_{{\bf p}} = (1/|{\bf p}|)\, {\bf P}{\bf S}, $$
we obtain
$$ {\cal P} S_{{\bf p}} {\cal P}^{-1} = - S_{{\bf p}} . $$
Therefore we would have to postulate that
$$
{\cal P} | p,\lambda \rangle = \eta_P | I_s p, -\lambda \rangle .
$$
In general this will be impossible: because the value of $\lambda$ is now 
{\sl invariant}, this requires that there exist {\sl two independent}
states, a state with helicity $\lambda$ and another with $-\lambda$. In
nature we find two kinds of particle. In one class we have particles
like the photon, gluons or, presumably, the graviton, which can exist in
the two helicity states: $\pm 1$ for the first two, $\pm 2$ for the
last. In the second class we have particles,\fnote{We are here neglecting 
neutrino masses.} like the neutrinos, which
exist only with helicity $-1/2$; or the {\sl antineutrinos} which always
carry helicity $+1/2$. {\sl For these particles parity is not defined}
and indeed the interactions that involve them violate parity.

For neutrinos and antineutrinos we can define a combined operation, ${\cal
C} {\cal P}$, the product of parity and particle--antiparticle conjugation
that carries neutrinos (with helicity $-1/2$) into antineutrinos (with
helicity $+1/2$), and vice versa\fnote{One can prove quite generally
that the product ${\cal CPT}$ is always a symmetry for any relativistic
theory
of local fields. For the proof see, for example, the text of Bogoliubov,
Logunov and Todorov (1975).}. There is a third class, that of particles
with helicity $\lambda$ for which neither particles or antiparticles
with helicity $- \lambda$ existed, which is mathematically possible but
of which no representative has been found in nature.

For time reversal,
$$ {\cal T} {\bf S} {\cal T}^{-1} = - {\bf S} , \ {\cal T} {\bf P} {\cal
T}^{-1} = - {\bf P} , $$
so that
$$ {\cal T} S_{{\bf p}} {\cal T}^{-1} = S_{{\bf p}}, $$
and we can define the antiunitary operator ${\cal T}$ with
$$
{\cal T} | p , \lambda \rangle = \eta_T (-\ii)^{2\lambda} | I_s p, \lambda
\rangle ;
$$
the phase $(-\ii)^{2\lambda}$ is introduced for aesthetic reasons, to
make the massless case similar to the massive one.

Let us return to parity. If the state $|I_s p, - \lambda \rangle$
exists, we will have to double our Hilbert space of states to make room
for it. We define total spin as $s= {\rm max} \ |\lambda|$, and
chirality $\delta$ as $\delta = \lambda/s = \pm 1$. We may label the
states as
$$ | p, s, \delta \rangle , $$
and the transformation properties can then be written as
$$
\eqalign{
U(\lambdav ) | p, s , \delta \rangle = \ee^{-\ii \delta s \theta (p, \lambdav
)} | \lambdav p, s, \delta \rangle , \cr
{\cal P} | p, s, \delta \rangle = \eta_P | I_s p, s, - \delta \rangle .
}
$$
The representation is {\sl reducible} as a representation of the
Poincar\'e group because the subspaces with $\delta =1$ and $\delta = -1$
are separately invariant; it is {\sl irreducible} as a representation of
the orthochronous (but {\sl not} proper) group obtained adjoining space
reversal, $I_s$, with $U(I_s) \equiv {\cal P}$, to the orthochronous,
proper Poincar\'e group.

\booksubsection{7.5. Connection with the wave function formalism}

\noindent
The construction of relativistic
states with well-defined position, $|{\bf r} ,t, a \rangle$ ($t$ is the
time, and $a$ represents possible extra labels) does not make much
physical sense. Therefore, the connection between the abstract ket
formalism and the wave function formalism is now less straightforward
than in the nonrelativistic case, where we simply have $\Psi_a ({\bf r},t) =
 \langle {\bf r} , t,a | \Psi \rangle$. Now, we will connect with
the {\sl momentum space} wave functions; these can be then linked, via
the appropriate Fourier transformations, to $x$-space ones.

We then want to establish the correspondence between  ket states and (multicomponent) wave
functions $\psi^{({\bf k} , \lambda )}_a ({\bf p})$, corresponding to
momentum ${\bf k}$ and spin component $\lambda$ (note that here
${\bf p}$ is the variable). We will work in the Heisenberg
representation, so the $\psi$ are time independent. Time dependence can
be introduced, if so wished, by writing
$$
\Psi^{({\bf k} , \lambda )}_a ({\bf p} , t) = \ee^{-\ii k _0 t}
\psi_a^{({\bf k}, \lambda )} ({\bf p} ) , \ k_0 = \sqrt{m^2 + {\bf k}^{\ 2}} .
$$
Here we work in natural units, $\hbar = c = 1$.

The case of spinless particles is simple. We just have
$$
\varphi^{({\bf k})} ({\bf p}) = \langle p | k \rangle = 2k_0 \delta
({\bf p} - {\bf k} ) ,
$$
but spin poses nontrivial problems. We will only consider the spin
$1/2$ case; the generalization to higher spins is straightforward, for
$m \neq 0$, and can be found in Moussa and Stora (1968), Weinberg (1964)
and Zwanziger (1964a,b). (The latter also treat the massless case).

The wave function of a particle of spin $1/2$, with third component of
covariant spin $s_3$ and momentum ${\bf k}$ can be written 
 (extracting the time dependence) as 
$$
\psi^{(k, s_3)} ({\bf p}) = D(L(k))u(0, s_3) 2k_0 \delta ({\bf k}-
{\bf p}).
$$
Taking into account that
$$u (0, 1/2) = \pmatrix{ 1 \cr 0 \cr 0 \cr 0},\quad 
 u(0, -1/2) = \pmatrix{ 0 \cr 1 \cr 0 \cr 0}
$$
it becomes convenient for our calculations to change the labels $s_3 =
\pm 1/2$ to $\tau = 1,2$, so that $1/2 \to 1$, $-1/2 \to
2$. Then we may write $u_a (0, \tau ) = \delta_{a \tau}$, and (6.6.3)
adopts the simple form
$$
\psi^{({\bf k} , \tau )}_a ({\bf p}) = D_{a\tau} (L(k)) 2k_0 \delta
({\bf k} - {\bf p} ) ,
$$
and we then have the explicit expression
$$
u_a (k, \tau ) = D_{a \tau} (L(k)) .
$$
$D_{ab} (L(k))$ is the $ab$ matrix element of the matrix $D(L(k))$;  
we will here use the Weyl representation of the
$\gamma$ matrices, so that
$$\gamma_{\mu}^{\rm W} = \pmatrix{0 & \tilde{\sigma}_{\mu} \cr
\sigma_{\mu} & 0} ,\quad \ \tilde{\sigma}_i = - \sigma_i ,
\; \tilde{\sigma}_0 = \sigma_0 = 1. 
$$

We have
$$ D(L(k)) \equiv D(L({\bf k})) = \frac{1}{\sqrt{m}} (k_0 + {\bf k}
\ybf{\alpha}
)^{1/2} = \frac{1}{\sqrt{m}} (k \cdot \gamma \gamma_0 )^{1/2} ,
$$
a formula valid in any representation. In Weyl's, this becomes
$$
D^{\rm W} (L(k)) = \frac{1}{\sqrt{m}} \pmatrix{(k \cdot
\tilde{\sigma} )^{1/2} & 0  \cr 0 & (k \cdot \sigma )^{1/2}}.
\equn{(1)}$$
This is of course the reason why the Weyl representation is useful for
us: the matrix $D^{\rm W}$ is ``box-diagonal''. Taking into account 
that the matrix that leads from the Pauli to the Weyl representation is
$$\frac{1}{\sqrt{2}} (\gamma_0^{\rm P} + \gamma_5^{\rm P} ) = \dfrac{1}{\sqrt{2}}
\pmatrix{ 1 & 1 \cr 1 & -1},
$$
and the known expression for the spinors in the 
Pauli relization (see, e.g., Yndur\'ain, 1996)
 we find for the spinors $u(0, \tau )$, in the Weyl realization,
$$
u^{\rm W} (0,1) = \dfrac{1}{\sqrt{2}} 
\pmatrix{1 \cr 0 \cr 1 \cr 0},\quad \ u^{\rm W}(0,2) = \dfrac{1}{\sqrt{2}} 
\pmatrix{0 \cr 1 \cr 0 \cr 1}.
\equn{(2)}$$
In what follows we suppress the label ``$W$''.

We may rewrite the wave function  as
$$
\psi = \pmatrix{\varphi \cr \tilde{\varphi}}
\quad
\psi^{({\bf k}, \tau )}_a ({\bf p})  = \varphi_{\alpha}^{(k, \tau
)} ({\bf p}), \ a= \alpha
= 1,2 ; \quad
 \psi_b^{({\bf k}, \tau )} ({\bf p})  =
\tilde{\varphi}_{\dot{\beta}}^{(k, \tau )} ({\bf p}) , \ b= \dot{\beta} +2
= 3,4 ,
$$
with
$$
\eqalign{\varphi_{\alpha}^{(k, \tau )} ({\bf p}) =&\,
\dfrac{1}{\sqrt{2m}} ((k \cdot
\tilde{\sigma} )^{1/2} )_{\alpha \tau} 2k_0 \delta ({\bf p}-{\bf k}) , 
\cr
\tilde{\varphi}^{(k, \tau )}_{\dot{\beta}} ({\bf p}) =&\,
\frac{1}{\sqrt{2m}}
((k \cdot \sigma )^{1/2} )_{\dot{\beta} \tau} 2k_0 \delta ({\bf p}
-{\bf k})}
\equn{(3)}$$
(the notation with dotted indices, such as $\dot{\beta}$, for the components
$\tilde{\varphi}_{\dot{\beta}}$ is the traditional one).

Because $\psi$ satisfies the Dirac equation, it follows that we can get
$\tilde{\varphi}$ in terms of $\varphi$ (or vice versa). Indeed, we have
$$
\tilde{\varphi}_{\dot{\beta}}^{(k, \tau )} ({\bf p} ) = \sum_a \left(
\frac{k \cdot \sigma}{m} \right)_{\dot{\beta} \alpha}
\varphi_{\alpha}^{(k, \tau )} ({\bf p} ) .
\equn{(4)}$$

\smallskip
\sports{i) Prove (4) by verifying that the
identity $(k \cdot \sigma ) (k \cdot \tilde{\sigma} ) = k \cdot k$
implies that (3) is equivalent to the Dirac equation $( k\cdot \gamma
 - m) \psi^{({\bf k} , \tau )} ({\bf p} ) = 0$. ii) Check that
$$ (k \cdot \tilde{\sigma} )^{1/2} = [ 2 (k_0 + m )]^{-1/2} (m+k_0 +
{\bf k} \ybf{\sigma} ).$$}
\smallskip
Owing to this relation (4), it is sufficient to establish the
connection between the states $|k , \tau \rangle$ and the wave functions
$\varphi_{\alpha}^{(k , \tau )} ({\bf p} )$. This is achieved by
introducing the so-called {\sl spinorial states}, $|p, \alpha \rangle$,
defined to be such that
$$
\varphi_{\alpha}^{(k, \tau )} ({\bf p} ) \equiv \langle p, \alpha | k,
\tau \rangle .
$$
Taking into account the explicit form of the $\varphi$, 
 we obtain the formula that links the spinorial states to the
familiar states with given covariant spin $| k, \tau \rangle$: it is
$$
| p, \alpha \rangle = \sum_{\tau} \int \frac{d^3 k}{2k_0} \left( \left(
\frac{k \cdot \tilde{\sigma}}{2m} \right) ^{1/2} \right)_{\tau \alpha}
2k_0 \delta ({\bf p}- {\bf k} ) | k, \tau \rangle ,
$$
and we have used the Hermiticity of the matrix $(k \cdot \tilde{\sigma}
)^{1/2}$.

The matrix $(k \cdot \tilde{\sigma} /m)^{1/2}$ is {\sl not} unitary. The
basis
$|p, \alpha \rangle$ is therefore not orthogonal; rather one has
$$
\langle p', \alpha ' | p, \alpha \rangle = \frac{(p \cdot
\tilde{\sigma} )_{\alpha ' \alpha}}{2m} 2p_0 \delta ({\bf p} - {\bf p}' )
$$
The index $\alpha$ does not correspond to any quantum number.

\smallskip
\sport{Prove that $d^3 p /2p_0$, $2p_0 \delta ({\bf p} - {\bf p}')$
 are invariant by writing, for $p_0 > 0,$
$$\delta_4 (p-p') = \delta (p^2 - p'^2 ) 2p_0 \delta ({\bf p} - {\bf p}' ) . $$
}
\smallskip
\sport{Find $R(p, \lambdav )$ in the NR limit,
including corrections $O(v^2 /c^2)$.
}
\smallskip
\sport{Find $R(p, I_s )$ for $\lambdav (p) = L(p)$.
 Find $| H(p) , \lambda \rangle$ in terms of $|
L(p) , \eta \rangle$, and viceversa.
}
\smallskip
\sport{Let $W^{\mu}=
\epsilon^{\mu\nu\rho\sigma}P_{\nu}M_{\rho\sigma}$ 
({\sl Pauli-Lubanski} vector). Prove that $W^2={\rm invariant}=-m^2s(s+1)$, $s$ the spin.
}
\smallskip
\sport{Verify that, for any $\lambdav$,
$$\eqalign{
U(\lambdav ) : &\, \varphi_{\alpha}^{(k, \tau )} (p) \to
\sum_{\alpha '} D_{\alpha \alpha '}^{(1/2)} (\lambdav ) \varphi_{\alpha
'}^{(k, \tau )} (\lambdav^{-1} p) , \cr
U(\lambdav ) :&\, \ \varphi_{\alpha}^{(k, \tau )} (p) \to \sum_{\tau'}
 D_{\tau ' \tau}^{(1/2)} (R (k, \lambdav )) \varphi_{\alpha}^{(\lambdav
k, \tau ')} (p) .} 
$$
Here, $D (\lambdav ) = D (L) D(R)$, for $\lambdav = LR$, with
$$D^{(1/2)}_{\alpha \beta} (L(p)) = m^{-1}(p \cdot
\tilde{\sigma} )^{1/2}_{\alpha \beta} , \ D_{\alpha
\beta}^{(1/2)} (R(\ybf{\theta} )) = \left(\ee^{-\ii \ybf{\theta} \ybf{\sigma} /2}
\right)_{\alpha \beta} , \; {\rm etc.} $$}

\booksubsection{7.6. Two-Particle States. Separation of the Center of Mass Motion. States
with Well-Defined Angular Momentum}

\noindent Although the subject of this subsection has little to do with groups, we 
include it here for completeness.

Let us consider two free particles (which for simplicity we take
to be
distinguishable), $A$, $B$, with masses $m_A$, $m_B$. A state of
these
two particles can be specified by giving the momenta ${\bf p}_A$,
${\bf p}_B$ and spin quantum numbers (for example, the
helicities) to be
denoted by $\alpha , \beta$: we thus write it as
$$
| p_A , \alpha ; p_B , \beta \rangle,\quad  p_{A0} \equiv \sqrt{m^2_A
+
{\bf p}^2_A}, \quad p_{B0} \equiv \sqrt{m^2_B + {\bf p}^2_B}
$$
with normalization
$$
\langle p'_A , \alpha '; p'_B , \beta ' | p_A , \alpha ; p_B ,
\beta
\rangle   = \delta_{\alpha \alpha '} 2p_{A0} \delta ({\bf p}_A
-
{\bf p}'_A ) 
   \times \delta_{\beta \beta '} 2p_{B0} \delta ({\bf p}_B -
{\bf p}'_B ) .
$$
The same state can be specified by giving the total
four-momentum,
$p=p_A + p_B$, the direction of the relative three-momentum,
${\bf k} =
({\bf p}_A - {\bf p}_B ) /2$, and the spin labels $\alpha$,
$\beta$:
$$
| p_A , \alpha ; p_B , \beta \rangle = | p ; {\bf k} ; \alpha ,
\beta
\rangle ;
$$
we write ${\bf k}$, which is redundant (just as $p_{A0}$,
$p_{B0}$ were
redundant before) instead of $\Omega_{{\bf k}}$ (the
angular
variables of ${\bf k}$) for simplicity of notation.

\smallskip
\sport{Show that, given $p$, $\Omega_{{\bf k}}$ we can
reconstruct
$p_A$, $p_B$.}
\smallskip

The tensor product notation is at times convenient, and we will thus write
$$
| p_A , \alpha \rangle \otimes | p_B , \beta \rangle = | p_A , \alpha ; p_B ,
\beta \rangle 
= | p ; {\bf k} ; \alpha , \beta
\rangle 
= | p \rangle \otimes | {\bf k} ; \alpha , \beta \rangle .
$$

The scalar product  can be easily expressed in terms of
the new
variables: first,
$$ \delta ({\bf p}_A - {\bf p}'_A ) \delta ({\bf p}_B -
{\bf p}'_B ) = \delta ({\bf p} - {\bf p}') \delta ({\bf k} - {\bf k}' ) ;
$$
then, we can use the relation
$$
\delta ({\bf k} - {\bf k}') =
\frac{1}{{\bf k}^{2}}
\delta ( |{\bf k} | - | {\bf k}' |) \delta (\Omega_{{\bf k}} -
\Omega_{{\bf k}'} )
=\dfrac{1}{{\bf k}^{2}} J^{-1} \delta (p_0 - p'_0
)
\delta (\Omegav_{{\bf k}} - \Omegav_{{\bf k}'} ) , 
$$
where $J$ is the Jacobian $J= \partial | {\bf k} | / \partial
p_0$, to
get
$$\delta ({\bf p}_A - {\bf p}'_A) \delta ({\bf p}_B -
{\bf p}'_B)=
(1/J {\bf k}^2) \delta (p_0 - p'_0 ) \delta
(\Omegav_{{\bf k}} - \Omegav_{{\bf k}'} ) .  $$
We will only need the relative motion (described by ${\bf k}$)
in the
center of mass (c.m.) system, ${\bf p}=0$. Here, $p_0=p_{A0} +
p_{B0} =
(m^2_A + {\bf k}^2 )^{1/2} + (m_B^2 + {\bf k}^2 )^{1/2}$
so that
$$J = \partial | {\bf k} | / \partial p_0 = p_{A0} p_{B0} /p_0
|
{\bf k} | , $$
and finally we obtain
$$
\eqalign{\langle p'_A , \alpha ' ; p'_B , \beta ' | p_A ,
\alpha ;
p_B , \beta \rangle 
=&\,\langle p' ; {\bf k}'; \alpha ' , \beta ' |
p;
{\bf k} ; \alpha , \beta \rangle = \frac{4p_0}{|{\bf k} |}
\delta_4
(p-p') \delta (\Omegav_{{\bf k}} - \Omegav_{{\bf k}'} )
\delta_{\alpha
\alpha '} \delta_{\beta \beta '} ,\cr
\delta (\Omegav - \Omegav') \equiv&\, \delta ( \cos
\theta -
\cos \theta ') \delta (\phi - \phi ') ,
\cr}
$$
with $\theta$, $\phi$ the polar angles corresponding to the
solid
angle $\Omega$. We write this also as
$$
\langle p' | p \rangle = \delta_4 (p' -p) , \ \langle {\bf k}';
\alpha ' , \beta ' | {\bf k} ; \alpha , \beta \rangle =
\dfrac{4p_0}{|
{\bf k} |} \delta ( \Omega_{{\bf k}'} - \Omega_{{\bf k}} )
\delta_{\alpha \alpha '} \delta_{\beta \beta '} .
$$
This will allow us to introduce a completeness relation once we
ascertain the range of the variables $p_0$, ${\bf p}$.
Clearly, ${\bf p}$ varies over all space; but $p_0$ is limited
by
$$\eqalign{
p_0 =&\, p_{A0} + p_{B0} = \sqrt{m^2_A + {\bf p}_A^2} +
\sqrt{m^2_B +
{\bf p}^2_B} = \sqrt{p^2 + {\bf p}^2} , \cr
p^2 \geq&\, (m_A + m_B )^2 .
\cr}
$$
We can thus write the four-dimensional delta as
$$
\delta_4 (p-p')=2p_0 \delta ({\bf p} - {\bf p}') \delta
(p^2-p'^2) ,
$$
so that the completeness relation can be expressed separating the c.m. piece, which
behaves as a composite particle
with (variable) squared mass $p^2$ and momentum ${\bf p}$, and the
relative
motion, described by ${\bf k}$, as follows:
$$\eqalign{
1 &\,= \sum_{\alpha \beta} \int
\frac{\dd^3 p_A}{2p_{A0}}
\int \frac{d^3 p_B}{2p_{B0}} | p_A , \alpha ; p_B , \beta \rangle
\langle p_A , \alpha ; p_B , \beta |  \cr
  &\,= \sum_{\alpha \beta} \int \dd^4 p \int d \Omega_{{\bf k}}
\dfrac{|{\bf k} |}{4p_0} | p ; {\bf k} ; \alpha , \beta \rangle
\langle p
; {\bf k} ; \alpha , \beta | \cr
  & \,= \int_{(m_A + m_B)^2}^{\infty} \dd(p^2) \int
\dfrac{d^3
p}{2p_0} | p \rangle \langle p |  
\otimes \sum_{\alpha \beta} \int d
\Omega_{{\bf k}}
\dfrac{| {\bf k} |}{4p_0} | {\bf k} ; \alpha , \beta \rangle
\langle
{\bf k} ; \alpha , \beta |\cr
&\, =1_{\rm c.m.} \otimes 1_{\rm rel} . 
\cr}
$$

In the c.m. system one can construct states with well-defined
{\sl
orbital} angular momentum $l$, and third component $M$ as in the
nonrelativistic case: we have
$$
| l, M; \alpha , \beta \rangle = \int d \Omega_{{\bf k}} Y^l_M
(\Omega_{{\bf k}} ) | {\bf k} ; \alpha , \beta \rangle .
$$

The completeness relation  can again be expressed in terms
of the
states $| l , M; \alpha , \beta \rangle$: separating c.m. and
relative
motion, we get
$$\eqalign{
1 & = 1_{\rm c.m.} \otimes 1_{\rm rel} ;\cr
1_{\rm c.m.} & = \int \dd^4 p | p \rangle \langle p | ,
\cr
1_{\rm rel} & = \sum_{\alpha \beta} \int \dd
\Omega_{{\bf k}}
\dfrac{| {\bf k} |}{4p_0} | {\bf k} ; \alpha , \beta \rangle
\langle
{\bf k} ; \alpha , \beta | \cr
& = \frac{|{\bf k} |}{4p_0} \sum_{\alpha \beta}
\sum_{l M}
| l, M; \alpha , \beta \rangle \langle l,M; \alpha , \beta | .
\cr}
$$
One can, if so wished, compose the angular momentum 
and spins; we leave the subject here (see e.g. Yndur\'ain,~1996).

\vfill\eject
\phantom{x}
\vfill\eject
\booksection{References}

\noindent
Bargmann, V. and Wigner, E. P. (1948), {\sl Proc. Nat. Acad.
Sci. USA\/} {\bf 34}, 211.\hb
Bogoliubov (Bobolubov), N. N., Logunov, A. A. and Todorov, I.
T. (1975), {\sl Axiomatic Quantum Field Theory}, Benjamin.\hb
Cheng, T.-P. and Li, L.-F. (1984). {\sl Gauge theory of elementary particle physics}. Oxford.\hb
Chevalley, C. (1946). {\sl Theory of Lie groups}. Princeton U. Press.\hb
Condon, E. U. and Shortley, G. H. (1967), {\sl The Theory of
Atomic Spectra}, Cambridge.\hb
de Swart, J. J. (1963). {\sl Rev. Mod. Phys.} {\bf 35}, 916.\hb
Hamermesh, M. (1963). {\sl Group theory}. Addison-Wesley.\hb
Jacobson, N. (1962). {\sl Lie algebras}. Interscience.\hb
Lyubarskii, G. Ya. (1960). {\sl The application of group theory in physics}. Pergamon Press.\hb
Moussa, P. and Stora, R. (1968), in {\sl Analysis of Scattering
and Decay} (Nikolic, ed.), Gordon and Breach.\hb
Naimark, M. (1959). {\sl Normed rings}. Nordhoof.\hb
Weinberg, S. (1964), in {\sl Brandeis Lectures on Particles and
Field Theory}, Vol. 2 (Deser and Ford, eds.), Prentice Hall.\hb
Weyl, H. (1946). {\sl The classical groups}. Princeton U. Press.\hb
Wightman, A. S. (1960), in {\sl Dispersion Relations}, Les
Houches Lectures (de Witt and Omn\`es, eds.), Wiley.\hb
Wigner, E. P. (1939), {\sl Ann. Math.} {\bf 40}, No. 1.\hb
Wigner, E. P. (1959). {\sl Group theory}. Academic Press.\hb
Wigner, E. P. (1963). in {\sl Proc. 1962 Trieste Seminar}, IAEA, Vienna.\hb
Yndur\'ain, F. J. (1996). {\sl Relativistic quantum mechanics and introduction to 
field theory}. Springer-Verlag.\hb
Zwanziger, D. (1964a), {\sl Phys. Rev.} {\bf 113B}, 1036.\hb
Zwanziger, D. (1964b), in {\sl Lectures in Theoretical
Physics}, Vol. VIIa, University of Colorado Press.

\bye